\documentclass[graybox]{svmult}

\usepackage{bm}
\usepackage{mathptmx}       
\usepackage{helvet}         
\usepackage{courier}        
\usepackage{type1cm}        
\usepackage{makeidx}         
\usepackage{graphicx}        
\usepackage{multicol}        
\usepackage[bottom]{footmisc}
\usepackage{amsmath}
\usepackage{subfigure}

\makeindex             

\begin{document}
\title*{Quantum Transport Simulation of III-V TFETs with Reduced-Order $\bm{k}\cdot\bm{p}$ Method}
\author{Jun Z. Huang, Lining Zhang, Pengyu Long, Michael Povolotskyi, and Gerhard Klimeck}
\institute{J. Z. Huang, P. Long, M. Povolotskyi, and G. Klimeck \at Network for Computational Nanotechnology, School of Electrical and Computer Engineering, Purdue University, West Lafayette, IN 47907, USA. \email{huang599@purdue.edu}
\and L. Zhang \at Department of Electronic and Computer Engineering, The Hong Kong University of Science and Technology, Kowloon, Hong Kong.}
%
%
\maketitle

\abstract{III-V tunneling field-effect transistors (TFETs) offer great potentials in future low-power electronics application due to their steep subthreshold slope and large ``on" current. Their 3D quantum transport study using non-equilibrium Green's function method is computationally very intensive, in particular when combined with multiband approaches such as the eight-band $\bm{k}\cdot\bm{p}$ method. To reduce the numerical cost, an efficient reduced-order method is developed in this article and applied to study homojunction InAs and heterojunction GaSb-InAs nanowire TFETs. Device performances are obtained for various channel widths, channel lengths, crystal orientations, doping densities, source pocket lengths, and strain conditions.}

\section{Introduction}
\label{sec:1}
Scaling the supply voltage enables reduction of power consumption of integrated circuits. In order to continue reducing the supply voltage without degrading the performance, steep subthreshold swing (SS) transistors are highly needed. Steep tunneling field-effect transistors (TFETs) can achieve sub-60mV/dec SS at room temperature by using quantum-mechanical band-to-band tunneling (BTBT) \cite{ionescu2011tunnel,seabaugh2010low}. However, TFETs generally suffer from low ``on" current ($I_{ON}$) due to low tunneling probabilities. To enhance BTBT and increase $I_{ON}$, group III-V semiconductor based TFETs are very attractive since III-V materials can provide low band gap, small tunneling mass, and allow different band-edge alignments \cite{ionescu2011tunnel}.

In order to achieve the best III-V TFET performances, it is required to systematically optimize various design parameters, such as the channel thickness, channel length, crystal orientations, doping densities, etc. In addition, many schemes have been proposed to further boost $I_{ON}$. The first scheme is to embed a pocket doping between the source and the channel \cite{nagavarapu2008tunnel,jhaveri2011effect}. The pocket increases the electric field near the tunnel junction and thus improves the $I_{ON}$ and SS. 2-D quantum simulations has also been performed for this kind of device \cite{verreck2013quantum}. The second scheme is to replace the homojunction with a heterojunction, for instance, a GaSb/InAs broken-gap heterojunction \cite{mohata2011demonstration,dey2013high}, or an InGaAs/InAs heterojunction \cite{zhao2014InGaAs}. Due to the band offset between the two materials, the tunneling barrier height and distance are greatly reduced. 2-D and 3-D quantum transport simulations have also been performed for these heterojunction TFETs \cite{luisier2009performance,jiang2013atomistic,avci2013heterojunction}. Other schemes include strain engineering \cite{conzatti2011simulation,brocard2013design}, grading of the molar fraction in the source region \cite{brocard2013design,brocard2014large}, adding a doped underlap layer between source and channel \cite{sharma2014GaSb}, and embedding a quantum well in the source \cite{pala2015exploiting}.

To understand the device physics, predict the performance, and optimize the design parameters of these structures, an efficient quantum transport solver is highly needed. The BTBT process can be accurately accounted for by combining non-equilibrium Green's function (NEGF) approach \cite{datta2005quantum} with tight binding or eight-band $\bm{k}\cdot\bm{p}$ Hamiltonian. Unfortunately, these multi-band NEGF studies require huge computational resources due to the large Hamiltonian matrix. To improve their efficiency, equivalent but greatly reduced tight-binding models can be constructed for silicon nanowires (SiNWs) \cite{mil2012equivalent}, which greatly speed up the simulation of p-type SiNW MOSFETs even in the presence of phonon scattering. Recently, this method has been extended to simulate III-V nanowire MOSFETs and heterojunction TFETs \cite{afzalian2015mode}. Note that construction of the reduced tight binding models requires sophisticated optimization process. A mode space $\bm{k}\cdot\bm{p}$ approach is also proposed for p-type SiNW MOSFETs and InAs TFETs \cite{shin2009full}, which has been employed to simulate strain-engineered and heterojunction nanowire TFETs \cite{conzatti2011simulation,brocard2013design}. Though optimization process is not needed, this approach selects the modes only at the $\Gamma$ point, i.e., at $k=0$, which is inefficient to expand the modes that are far away from $k=0$.

In this work, we propose to construct the reduced-order $\bm{k}\cdot\bm{p}$ models with multi-point expansion \cite{huang2013model,huang2014model}. We also extend this method to be able to simulate heterojunction devices. This efficient quantum transport solver is then applied to optimize device configurations such as crystal orientation, channel width, and channel length. Various performance boosters such as source pocket, heterojunction, and strain will be explored. Homojunction InAs and heterojunction GaSb/InAs nanowire TFETs will be the focus of this study.

The device structure is described in Section 2. The $\bm{k}\cdot\bm{p}$ method is developed in Section 3, where the eight-band $\bm{k}\cdot\bm{p}$ Hamiltonian and its matrix rotations are reviewed first. The rotated Hamiltonian is then discretized in a mixed real and spectral space. The accuracy of the $\bm{k}\cdot\bm{p}$ method is benchmarked by comparing the band structures with tight binding method for several nanowire cross sections. The reduced-order NEGF method is developed in Section 4, where the reduced-order NEGF equations are summarized first. Then the problem of spurious bands, particular for the multi-point expansion, is identified. Afterwards, a simple procedure to eliminate these spurious bands is proposed. The method is finally validated by checking the band structures as well as the I-V curves. In Section 5, extensive simulations are carried out to understand and optimize the TFETs under different application requirements. Conclusions are drawn in Section 6.

\section{Device Structure}
\begin{figure}[htbp] \centering
\includegraphics[width=4.5in]{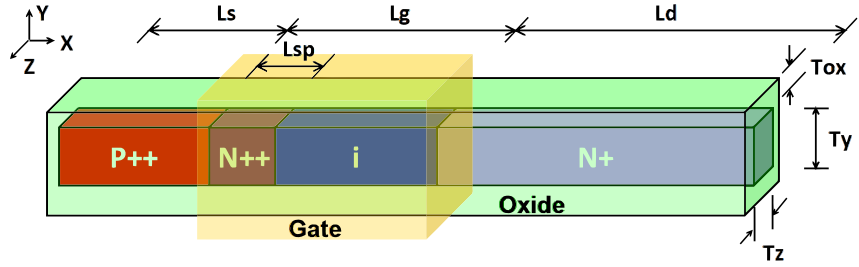}
\caption{GAA TFET with p-n-i-n doping profile. The transport direction is $x$ while the confinement directions are $y$ and $z$. The source, source pocket, gate, and drain lengths are $L_s$, $L_{sp}$, $L_g$, and $L_d$, respectively, with doping density in the source, pocket, and drain, $N_s$, $N_{sp}$, and $N_d$, respectively. Nanowire width and thickness are $T_z$ and $T_y$. Oxide layer thickness is denoted by $T_{ox}$, its dielectric constant is $\varepsilon_{ox}$.}
\label{fig_tfet_homo}
\end{figure}

The n-type gate-all-around (GAA) nanowire TFET to be simulated is illustrated in Fig. \ref{fig_tfet_homo}. The nanowire is P++ doped in the source and N+ doped in the drain, while it is intrinsic in the channel. A thin layer of N++ doping is inserted between the source and the channel to form a source pocket. The nanowire is surrounded by the oxide layer, through which the gate controls the channel (and the pocket).

For homojunction TFETs in this study, all the source, pocket, channel, and drain are made of material InAs, because high ``on" current is possible due to its small direct band gap and light effective masses \cite{luisier2009atomistic}.

For heterojunction TFETs in this study, GaSb is used for the source while InAs is used for the pocket, channel and drain. These two materials form broken-gap heterojunction at the source-channel (or source-pocket) interface, though in reality staggered-gap heterojunction is formed due to lateral confinements \cite{luisier2009performance,jiang2013atomistic}.

Different channel width, gate length, and pocket length will be studied. The channel crystal orientations will be varied from [100] (with (010) and (001) surfaces), [110] (with (-1,1,0) and (001) surfaces), to [111] (with (-1,1,0) and (-1,-1,2) surfaces). The uniaxial stress will be applied along the $x$ direction while the biaxial stress will be applied in the $y$ and $z$ directions.

\section{The $\bm{k}\cdot\bm{p}$ Method}
\label{sec:3}
\subsection{The Eight-Band $\bm{k}\cdot\bm{p}$ Hamiltonian}
To describe the band structure involving both the conduction and valence bands of III-V compound semiconductor materials, a widely used approach is the eight-band $\bm{k}\cdot\bm{p}$ model. When the eight basis functions are chosen to be spin-up and spin-down $s$ and $p$ atomic orbital-like states, the zincblende Hamiltonian can be written as \cite{gershoni1993calculating,enders1996exact,wu2006impact},
\begin{equation}\label{eq1.1}
\bm{H}^{8}=
       \left(
         \begin{array}{cc}
           \bm{H}^4 & \bm{0} \\
           \bm{0}   & \bm{H}^4\\
         \end{array}
       \right)
      +\left(
         \begin{array}{cc}
           \bm{G}_{so} & \bm{\Gamma} \\
           -\bm{\Gamma}^*   & \bm{G}_{so}^*\\
         \end{array}
       \right)
      +\left(
         \begin{array}{cc}
           \bm{H}^4_{str} & \bm{0} \\
           \bm{0}   & \bm{H}^4_{str}\\
         \end{array}
       \right),
\end{equation}
where the first part is spin-independent, the second part accounts for spin-orbit coupling, and the last part is deformation
potential contribution due to strain.

With operator ordering (for heterostructures) taken into account, the four band Hamiltonian $\bm{H}^4$ is
\begin{equation}\label{eq1.2}
\bm{H}^{4}=
\left(
\begin{array}{cc}
E_c\bm{I}_{1}+\bm{H}_{cc}    &\bm{H}_{cv}   \\
\bm{H}_{vc}    &E_v^\prime\bm{I}_{3}+\bm{H}_{vv}   \\
\end{array}
\right),
\end{equation}
where $\bm{I}_{1}$ and $\bm{I}_{3}$ are the $1\times1$ and $3\times3$ identity matrices, and
\begin{equation}\label{eq1.3}
\bm{H}_{cc}=k_xA_{c}k_x+k_yA_{c}k_y+k_zA_{c}k_z,
\end{equation}
\begin{equation}\label{eq1.4}
\bm{H}_{cv}=\left(
\begin{array}{ccc}
iP^+k_x+ik_xP^- &iP^+k_y+ik_yP^- &iP^+k_z+ik_zP^-
\end{array}
\right),
\end{equation}
\begin{equation}\label{eq1.5}
\bm{H}_{vc}=\left(
\begin{array}{ccc}
-ik_xP^+-iP^-k_x  &-ik_yP^+-iP^-k_y  &-ik_zP^+-iP^-k_z
\end{array}
\right)^T,
\end{equation}
\begin{equation}\label{eq1.6}
\bm{H}_{vv}=\left(
\begin{array}{ccc}
k_{x}Lk_{x}+k_{y}Mk_{y}+k_{z}Mk_{z} &k_{x}N^+k_{y}+k_{y}N^-k_{x} &k_{x}N^+k_{z}+k_{z}N^-k_{x} \\
k_{y}N^+k_{x}+k_{x}N^-k_{y} &k_{y}Lk_{y}+k_{z}Mk_{z}+k_{x}Mk_{x} &k_{y}N^+k_{z}+k_{z}N^-k_{y} \\
k_{z}N^+k_{x}+k_{x}N^-k_{z} &k_{z}N^+k_{y}+k_{y}N^-k_{z} &k_{z}Lk_{z}+k_{x}Mk_{x}+k_{y}Mk_{y}
\end{array}
\right).
\end{equation}
Here the parameter $E_c=E_v+E_g$ is the conduction band edge with $E_v$ being the valence band edge and $E_g$ the band gap. $E_v^\prime=E_v-\Delta/3$ is the valence band edge in the absence of spin-orbit coupling, with $\Delta$ being the spin-orbit split-off energy. $P$ is proportional to the momentum matrix element and can be evaluated by its equivalent energy $E_p=2m_0P^2/\hbar^2$. $A_c$ is determined from the conduction band effective mass $m_c^*$,
\begin{equation}\label{eq1.2}
A_c=\frac{\hbar^2}{2m_c^*}-\frac{2P^2}{3E_g}-\frac{P^2}{3\left(E_g+\Delta\right)}.
\end{equation}
The parameters $L$, $M$, and $N$ are related to the Luttinger parameters $\gamma_1$, $\gamma_2$, and $\gamma_3$,
\begin{equation}
L=-\frac{\hbar^2}{2m_0}\left(\gamma_1+4\gamma_2\right)+\frac{P^2}{E_g},
\end{equation}
\begin{equation}
M=-\frac{\hbar^2}{2m_0}\left(\gamma_1-2\gamma_2\right),
\end{equation}
\begin{equation}
N=-\frac{\hbar^2}{2m_0}\left(6\gamma_3\right)+\frac{P^2}{E_g}.
\end{equation}
While the widely used symmetrized operator ordering evenly divides the terms leading to $P^+=P^-=P/2$ and $N^+=N^-=N/2$, the correct Burt-Foreman ordering \cite{Foreman97} divides the terms according to different bands' contribution, which leads to $P^+=P$, $P^-=0$, $N^-=M-\hbar^2/2m_0$, and $N^+=N-N^-$.

The spin-orbit terms $\bm{G}_{so}$ and $\bm{\Gamma}$ are
\begin{equation}\label{eq1.7}
\bm{G}_{so}=\frac{\Delta}{3}
\left(
\begin{array}{cccc}
0    & 0    & 0     & 0  \\
0    & 0    & -i     & 0      \\
0    & i   & 0     & 0      \\
0    & 0    & 0     & 0  \\
\end{array}
\right),\indent
\bm{\Gamma}=\frac{\Delta}{3}
\left(
\begin{array}{cccc}
0    & 0    & 0     & 0  \\
0    & 0    & 0     & 1  \\
0    & 0    & 0     & -i \\
0    & -1   & i     & 0  \\
\end{array}
\right).
\end{equation}

The strain Hamiltonian is \cite{Bahder90}
\begin{equation}
\bm{H}_{str}^4=
\left(
\begin{array}{cc}
\bm{H}^{cc}_{str}    &\bm{0}   \\
\bm{0}                   &\bm{H}^{vv}_{str}  \\
\end{array}
\right),
\end{equation}
where
\begin{equation}
\bm{H}^{cc}_{str}=a_c\left(\epsilon_{xx}+\epsilon_{yy}+\epsilon_{zz}\right),
\end{equation}
\begin{equation}
\bm{H}^{vv}_{str}=\left(
\begin{array}{ccc}
l\epsilon_{xx}+m\left(\epsilon_{yy}+\epsilon_{zz}\right) & n\epsilon_{xy} &n\epsilon_{xz}\\
n\epsilon_{xy} &l\epsilon_{yy}+m\left(\epsilon_{xx}+\epsilon_{zz}\right) &n\epsilon_{yz} \\
n\epsilon_{xz} &n\epsilon_{yz} &l\epsilon_{zz}+m\left(\epsilon_{xx}+\epsilon_{yy}\right) \\
\end{array}
\right).
\end{equation}
Here $\left(\epsilon_{xx},\epsilon_{yy},\epsilon_{zz},2\epsilon_{yz},2\epsilon_{xz},2\epsilon_{xy}\right)$ is the strain vector in Voigt's notation, $a_c$ is the deformation potential constant for the conduction band, $m=a_v-b$, $l=a_v+2b$, $n=\sqrt{3}d$, and $a_v$, $b$, $d$ are the Pikus-Bir deformation potential constants for the valence bands. Note that we only keep the $k$ independent terms.

The $\bm{k}\cdot\bm{p}$ parameters for III-V compounds and their alloys can be found in \cite{Vurgaftman01}.

The $\bm{k}\cdot\bm{p}$ Hamiltonian matrix defined in the above is in terms of $\bm{k}$ in the crystal coordinate system (CCS). In practice, nanostructures can grow in different crystal directions, and thus the quantization directions and periodic directions are aligned with device coordinate system (DCS). Therefore it is more convenient to work in the DCS; and it requires coordinate transformation of the Hamiltonian matrix.

We first define a $3\times3$ unitary rotation matrix from DCS to CCS $\bm{R}_{D\rightarrow C}$, so that the $3\times1$ $\bm{k}$ vectors in the CCS and DCS, i.e., $\bm{k}_C$ and $\bm{k}_D$, are related by
\begin{equation}\label{eq_rotation}
\bm{k}_C=\bm{R}_{D\rightarrow C}\cdot\bm{k}_D.
\end{equation}
Note that the rows of $\bm{R}_{D\rightarrow C}$ are the coordinates of the CCS unit vectors in the DCS.

Then we rotate the $\bm{k}\cdot\bm{p}$ matrix element by element.
Each of the second order in $\bm{k}$ terms in the CCS is of the form $\bm{k}_C^T\bm{H}_C^{\left(2\right)}\bm{k}_C$. Substituting \eqref{eq_rotation}, we have,
\begin{equation}
\bm{k}_C^T\bm{H}_C^{\left(2\right)}\bm{k}_C
=\left(\bm{k}_D^T\bm{R}_{D\rightarrow C}^T\right)\bm{H}_C^{\left(2\right)}\left(\bm{R}_{D\rightarrow C}\bm{k}_D\right)
=\bm{k}_D^T\bm{H}_D^{\left(2\right)}\bm{k}_D.
\end{equation}
From above we can identify
\begin{equation}
\bm{H}_D^{\left(2\right)}=\bm{R}_{D\rightarrow C}^T\bm{H}_C^{\left(2\right)}\bm{R}_{D\rightarrow C}.
\end{equation}

Each of the first order in $\bm{k}$ terms in the CCS is of the form $\bm{k}_C^T\bm{H}_{C,R}^{\left(1\right)}+\bm{H}_{C,L}^{\left(1\right)}\bm{k}_C$, where $\bm{H}_{C,R}^{\left(1\right)}$ is a $3\times1$ matrix and $\bm{H}_{C,L}^{\left(1\right)}$ is an $1\times3$ matrix. Substituting \eqref{eq_rotation}, we have,
\begin{align}
\bm{k}_C^T\bm{H}_{C,R}^{\left(1\right)}+\bm{H}_{C,L}^{\left(1\right)}\bm{k}_C
&=\bm{k}_D^T\bm{R}_{D\rightarrow C}^T \bm{H}_{C,R}^{\left(1\right)}+\bm{H}_{C,L}^{\left(1\right)}\bm{R}_{D\rightarrow C}\bm{k}_D \nonumber\\
&=\bm{k}_D^T\bm{H}_{D,R}^{\left(1\right)}+\bm{H}_{D,L}^{\left(1\right)}\bm{k}_D.
\end{align}
From above we can identify
\begin{equation}
\bm{H}_{D,R}^{\left(1\right)}=\bm{R}_{D\rightarrow C}^T \bm{H}_{C,R}^{\left(1\right)},
\end{equation}
and
\begin{equation}
\bm{H}_{D,L}^{\left(1\right)}=\bm{H}_{C,L}^{\left(1\right)}\bm{R}_{D\rightarrow C}.
\end{equation}

The $\bm{k}$ independent terms, such as the band edges and spin-orbit constants, do not need to rotate.

The strain and stress components are usually set in the DCS, however the strain components in the CCS are those that enter the $\bm{k\cdot p}$ Hamiltonian \cite{Esseni11}.

The rotation for the strain is given by
\begin{equation}\label{eq_strain_rotation}
\bm{\epsilon}_{C,3\times3}=\bm{R}_{D\rightarrow C}\cdot\bm{\epsilon}_{D,3\times3}\cdot \bm{R}_{D\rightarrow C}^{-1},
\end{equation}
where $\bm{\epsilon}_{C,3\times3}$ and $\bm{\epsilon}_{D, 3\times3}$ are the $3\times3$ strain matrices in the CCS and DCS, respectively. Similar rotation holds for $\bm{\sigma}_{C,3\times3}$ and $\bm{\sigma}_{D,3\times3}$, the stress matrices in the CCS and DCS.

It is convenient to transform directly the strain vector by
\begin{equation}\label{eq_strain_transformation}
\bm{\epsilon}_{C,6}=\bm{R}_{6,D\rightarrow C}\cdot\bm{\epsilon}_{D, 6},
\end{equation}
where $\bm{\epsilon}_{C,6}$ and $\bm{\epsilon}_{D, 6}$ are the $6\times1$ strain vectors in the CCS and DCS, respectively. $\bm{R}_{6,D\rightarrow C}$ is the $6\times6$ transformation matrix, whose elements can be found by expanding equation \eqref{eq_strain_rotation}. Similar rotation holds for $\bm{\sigma}_{C,6}$ and $\bm{\sigma}_{D, 6}$, the stress vectors in the CCS and DCS.

Finally, $\bm{\sigma}_{C,6}$ can be converted to $\bm{\epsilon}_{C,6}$ via three elastic constants $C_{11}$, $C_{12}$, and $C_{44}$,
\begin{equation}
\left(
\begin{array}{c}
\epsilon_{C,xx}\\
\epsilon_{C,yy}\\
\epsilon_{C,zz}\\
\epsilon_{C,yz}\\
\epsilon_{C,xz}\\
\epsilon_{C,xy}\\
\end{array}
\right)
=
\left(
\begin{array}{cccccc}
C_{11} &C_{12} &C_{12} &0 &0 &0\\
C_{12} &C_{11} &C_{12} &0 &0 &0\\
C_{12} &C_{12} &C_{11} &0 &0 &0\\
0 &0 &0 &2C_{44} &0 &0\\
0 &0 &0 &0 &2C_{44} &0\\
0 &0 &0 &0 &0 &2C_{44}\\
\end{array}
\right)^{-1}
\left(
\begin{array}{c}
\sigma_{C,xx}\\
\sigma_{C,yy}\\
\sigma_{C,zz}\\
\sigma_{C,yz}\\
\sigma_{C,xz}\\
\sigma_{C,xy}\\
\end{array}
\right).
\end{equation}

\subsection{The Discretized Hamiltonian}
For nanostructures, the periodicity is broken by the finite sizes and the external potentials. The eigen states can be found by solving the following coupled differential equation for envelop function $\bm{F}_m$ ($m=1,2,\cdots,8$),
\begin{equation}\label{eq2.1}
\sum_{n=1}^{8}[{\bm{H}}^{8}_{mn}\left(-i\nabla\right)+V\left(\bm{r}\right)\delta_{mn}]\bm{F}_n\left(\bm{r}\right)=E\bm{F}_m\left(\bm{r}\right)
\end{equation}
where $V\left(\bm{r}\right)$ is the slowly varying perturbed potential distribution, and operator ${\bm{H}}^{8}_{mn}\left(-i\nabla\right)$ is the element of ${\bm{H}}^8\left(\bm{k}\right)$ with $\bm{k}$ replaced by the differential operator $-i\nabla$.

In order to solve \eqref{eq2.1} numerically, the operator needs to be discretized first. To have a discretized form that is compact and valid for arbitrary nanowire orientation, we rewrite the eight-band $\bm{k}\cdot\bm{p}$ operator (considering operator ordering) in \eqref{eq2.1} as,
\begin{equation}\label{eq2.2}
\bm{H}=\sum_{\alpha,\beta=x,y,z}\partial_\alpha\bm{H}_{\alpha,\beta}\partial_\beta+
           \sum_{\alpha=x,y,z}\left(\bm{H}_{\alpha,L}\partial_\alpha+\partial_\alpha\bm{H}_{\alpha,R}\right)+
           \bm{H}_0
\end{equation}
where the matrices $\bm{H}_{\alpha,\beta}$, $\bm{H}_{\alpha,L}$, $\bm{H}_{\alpha,R}$, and $\bm{H}_0$ are the material- and orientation- dependent coefficients containing contributions from L\"{o}wdin's renormalization, spin-orbit interaction, and strain.

In Ref. \cite{huang2013model}, finite difference method (FDM) is adopted and it results in extremely sparse matrices. Therefore, the Bloch modes can be obtained efficiently with sparse matrix solvers. In fact, with shift-and-invert strategy implemented, the Krylov subspace based eigenvalue solver converges very quickly, as the eigenvalues of interest (close to the valence band top) distribute in a very small area. However, it is found that the Krylov subspace method is less efficient in the eight-band case. The reason is that the eigenvalues of interest distribute over a larger area, as both conduction and valence bands are to be sought and between them there is a band gap.

Therefore, the method used in Ref. \cite{shin2009full} is employed, which is also generalized to arbitrary crystal orientations and to heterojunctions here. In this method, the transport direction is still discretized by FDM while the transverse directions are discretized by spectral method. Spectral method has high spectral accuracy (i.e., the error decreases exponentially with the increase of discretization points $N$) if the potential distribution is smooth \cite{paussa2010pseudospectral}. This is true for devices that do not have any explicit impurities or surface roughness. So, the Hamiltonian matrix size of a layer, i.e., $N_t$, can be kept very small (although it is less sparse or even dense), making direct solution of the eigenvalue problem possible.

To discretize the operator \eqref{eq2.2}, the longitudinal component of the unknown envelope function is discretized with second-order central FDM,
\begin{equation}\label{eq2.3}
\partial_x \left(\bm{H}_{x,R}\psi\right)|_{x=x_i}\approx\frac{\bm{H}_{x,R}\left(x_{i+1}\right)\psi\left(x_{i+1}\right)-\bm{H}_{x,R}\left(x_{i-1}\right)\psi\left(x_{i-1}\right)}{2\Delta x},
\end{equation}
\begin{equation}\label{eq2.4}
\bm{H}_{x,L}\partial_x\psi|_{x=x_i}\approx\bm{H}_{x,L}\left(x_{i}\right)\frac{\psi\left(x_{i+1}\right)-\psi\left(x_{i-1}\right)}{2\Delta x},
\end{equation}
\begin{align}\label{eq2.5}
\partial_x\left(\bm{H}_{x,x}\partial_x\psi\right)|_{x=x_i}
&\approx\frac{\bm{H}_{x,x}\left(x_{i+1}\right)+\bm{H}_{x,x}\left(x_{i}\right)}{2\left(\Delta x\right)^2}\psi\left(x_{i+1}\right)\nonumber\\
&-\frac{\bm{H}_{x,x}\left(x_{i+1}\right)+2\bm{H}_{x,x}\left(x_{i}\right)+\bm{H}_{x,x}\left(x_{i-1}\right)}{2\left(\Delta x\right)^2}\psi\left(x_{i}\right)\nonumber\\
&+\frac{\bm{H}_{x,x}\left(x_{i-1}\right)+\bm{H}_{x,x}\left(x_{i}\right)}{2\left(\Delta x\right)^2}\psi\left(x_{i-1}\right),
\end{align}
where $\Delta x$ is the grid spacing.

The transversal components are expanded using Fourier series \cite{shin2009full}, i.e.,
\begin{equation}\label{eq2.6}
\phi_{p,q}\left(y_m,z_n\right)=\frac{2}{\sqrt{N_yN_z}}\sin\left(k_py_m\right)\sin\left(k_qz_n\right),
\end{equation}
where $N_y$ and $N_z$ are the number of real space grid points in the $y$ and $z$ directions respectively,
$m$ and $n$ $\left(1\leq m \leq N_y, 1\leq n \leq N_z\right)$ are the coordinates of the $R$th grid point in real space,
$p$ and $q$ $\left(1\leq p \leq N_y, 1\leq q \leq N_z\right)$ are the coordinates of the $S$th grid point in the Fourier space,
\begin{equation}\label{eq2.7}
k_p=\frac{p\pi}{T_y},\quad k_q=\frac{q\pi}{T_z},
\end{equation}
where $T_y$ ($T_z$) is the nanowire thickness in the $y$ ($z$) direction. Note that hard wall boundary condition is enforced at the interfaces between the oxide layer and the semiconductor nanowire when the basis function \eqref{eq2.6} is used.

Operating \eqref{eq2.2} on \eqref{eq2.6}, multiplying the result with \eqref{eq2.6} and performing integrations, we get the discretized form. It is block tridiagonal,
\begin{equation}\label{eq2.8}
\bm{H}=\left(
                \begin{array}{ccccc}
                       \bm{D}_{1}   & \bm{T}_{1,2}      & \bm{0}            &                   &   \\
                       \bm{T}_{1,2}^\dag  & \bm{D}_{2}  & \bm{T}_{2,3}      &                   &   \\
                                        & \ddots          & \ddots          & \ddots            &   \\
                                        &                 & \bm{T}_{N_x-2,N_x-1}^\dag & \bm{D}_{N_x-1}  & \bm{T}_{N_x-1,N_x}  \\
                                        &                 &  \bm{0}          & \bm{T}_{N_x-1,N_x}^\dag   & \bm{D}_{N_x}  \\
                     \end{array}
                \right),
\end{equation}
where $\bm{D}_{i}$ is the on-site Hamiltonian for layer $i$ ($1\leq i\leq N_x$), $\bm{T}_{i,i+1}$ ($1\leq i\leq N_x-1$) is the coupling Hamiltonian between adjacent layers, and $N_x$ is the number of grids in the longitudinal direction $x$.

The $\left(S,S^\prime\right)$ block of $\bm{D}_{i}$ can be written down using very simple prescription,
\begin{align}\label{eq2.9}
\bm{D}_{i}^{S,S^\prime}&=\left[\left(\bm{H}^{i}_{y,L}+\bm{H}^{i}_{y,R}\right)\frac{4k_p^\prime}{\pi}\frac{p}{p^2-p^{\prime 2}}\right]\delta_{p+p^\prime,odd}\delta_{q,q^\prime}\\
&+\left[\left(\bm{H}^{i}_{z,L}+\bm{H}^{i}_{z,R}\right)\frac{4k_q^\prime}{\pi}\frac{q}{q^2-q^{\prime 2}}\right]\delta_{q+q^\prime,odd}\delta_{p,p^\prime}\nonumber\\
&-\left[\left(\bm{H}^{i}_{y,z}+\bm{H}^{i}_{z,y}\right)\frac{4k_p^\prime}{\pi}\frac{p}{p^2-p^{\prime 2}}\frac{4k_q^\prime}{\pi}\frac{q}{q^2-q^{\prime 2}}\right]\delta_{p+p^\prime,odd}\delta_{q+q^\prime,odd}\nonumber\\
&+\left[\bm{H}^{i}_0+\left(\bm{H}^{i+1}_{x,x}+2\bm{H}^{i}_{x,x}+\bm{H}^{i-1}_{x,x}\right)\frac{1}{2\left(\Delta x\right)^2}
+\bm{H}^{i}_{y,y}k_p^2+\bm{H}^{i}_{z,z}k_q^2\right]\delta_{p,p^\prime}\delta_{q,q^\prime},\nonumber
\end{align}
where $\left(p,q\right)$ and $\left(p^\prime,q^\prime\right)$
are the coordinates of the $S$th and $S^\prime$th grid points, respectively. $\delta$ is Kronecker delta function, for instance, $\delta_{q+q^\prime,odd}$ is equal to 1 (0) if $q+q^\prime$ is an odd (even) number.

Similarly, the $\left(S,S^\prime\right)$ block of $\bm{T}_{i,i+1}$ can be written as,
\begin{align}\label{eq2.10}
\bm{T}_{i,i+1}^{S,S^\prime}&=\left[-\left(\bm{H}^{i+1}_{x,x}+\bm{H}^{i}_{x,x}\right)\frac{1}{2\left(\Delta x\right)^2}
+\left(\bm{H}^{i+1}_{x,R}+\bm{H}^{i}_{x,L}\right)\frac{1}{2\Delta x}\right]\delta_{p,p^\prime}\delta_{q,q^\prime}\nonumber\\
&-\left[\left(\bm{H}^{i+1}_{x,y}+\bm{H}^{i}_{y,x}\right)\frac{1}{2\Delta x}\frac{4k_p^\prime}{\pi}\frac{p}{p^2-p^{\prime 2}}\right]\delta_{p+p^\prime,odd}\delta_{q,q^\prime}\nonumber\\
&-\left[\left(\bm{H}^{i+1}_{x,z}+\bm{H}^{i}_{z,x}\right)\frac{1}{2\Delta x}\frac{4k_q^\prime}{\pi}\frac{q}{q^2-q^{\prime 2}}\right]\delta_{q+q^\prime,odd}\delta_{p,p^\prime}.
\end{align}

In this work, we use $\Delta x=0.2\rm{nm}$ and have limited $S$ to be $1\leq S\leq183$ by employing the index scheme in \cite{shin2009full}. This means 183 Fourier series are used to expand each wave function component, which is found to be sufficient. The dimension of $\bm{D}_{i}$ is thus $N_t=183\times8=1464$.

\begin{figure}[h]
\centering
\includegraphics[width=2.2in]{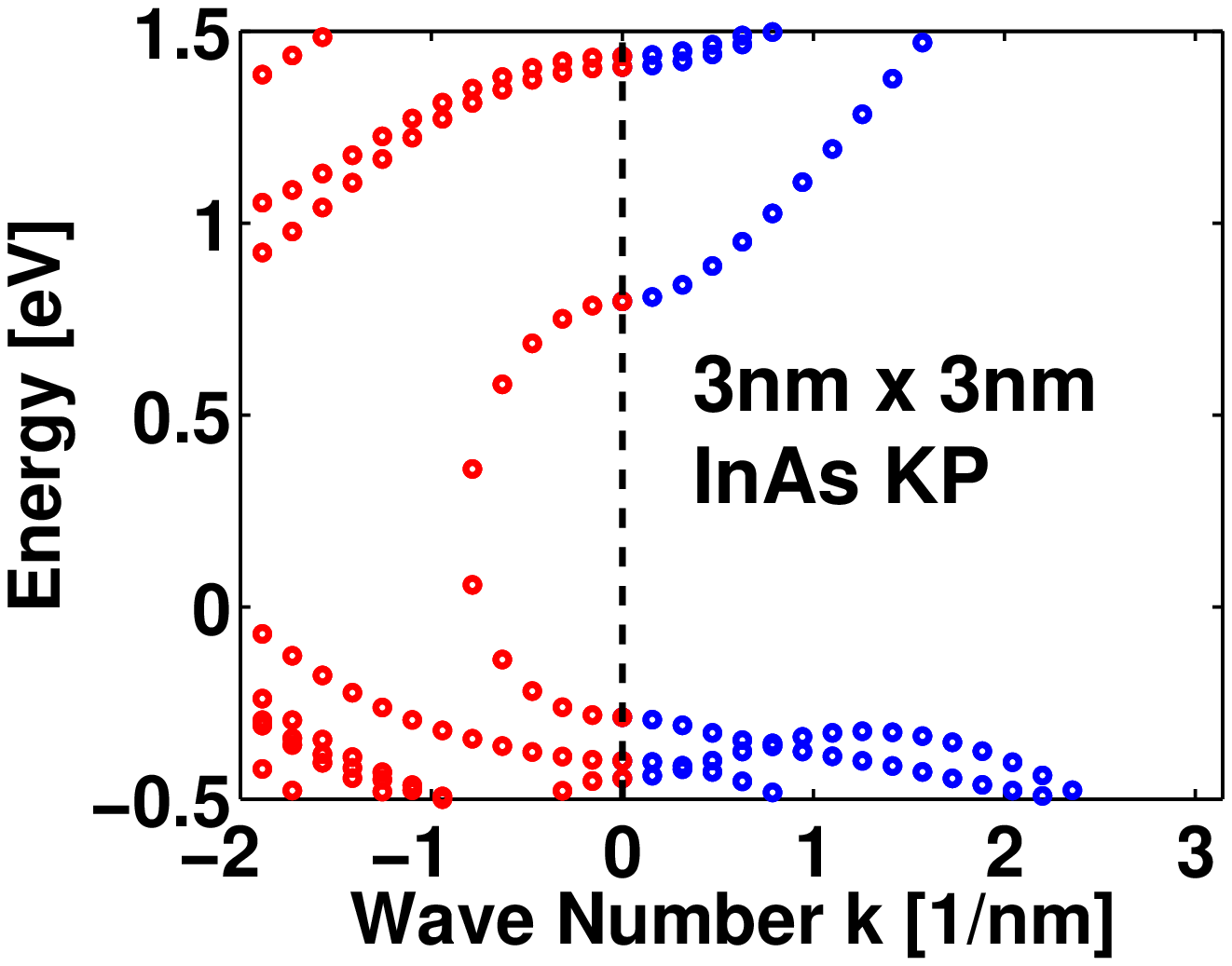}
\includegraphics[width=2.2in]{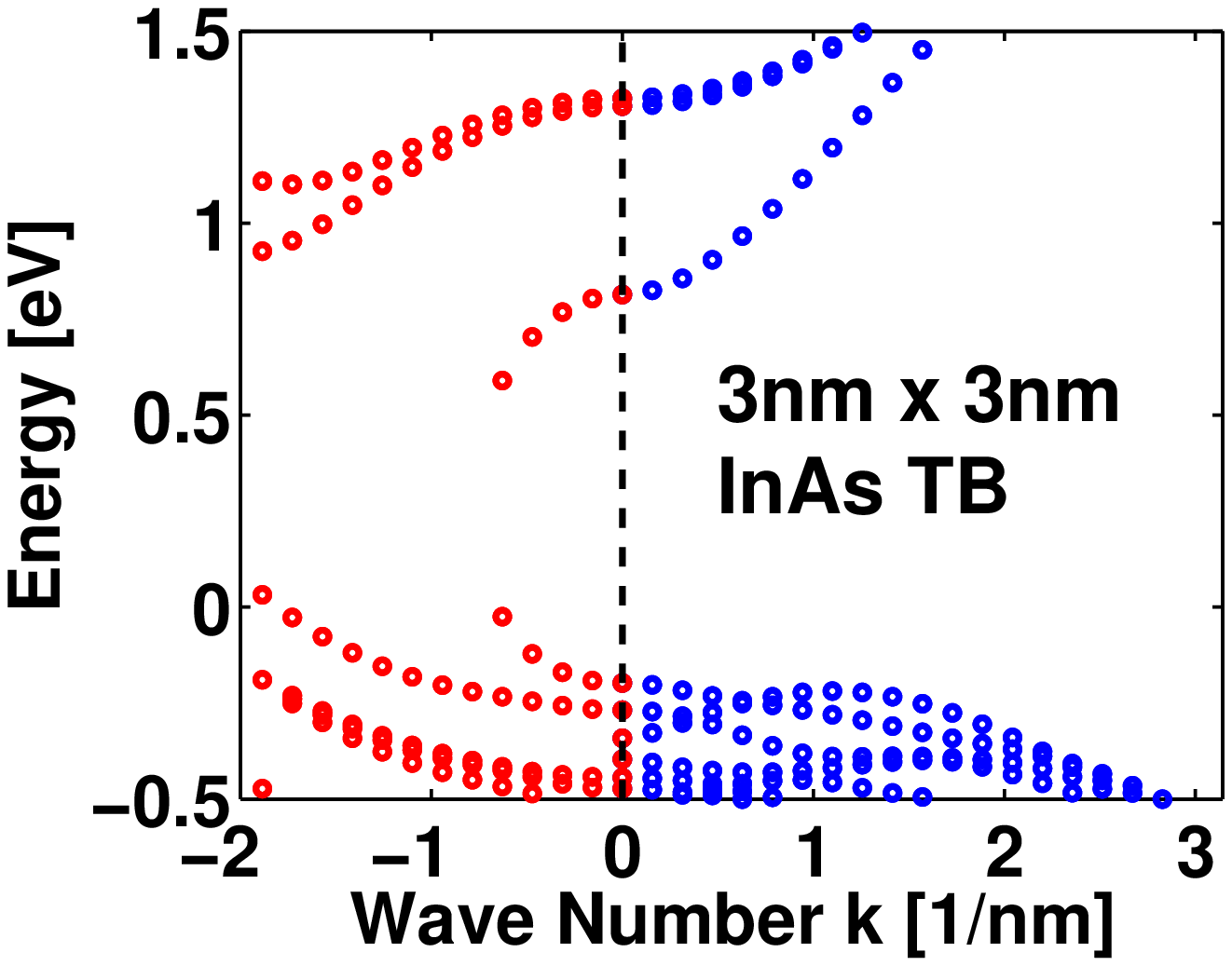}
\includegraphics[width=2.2in]{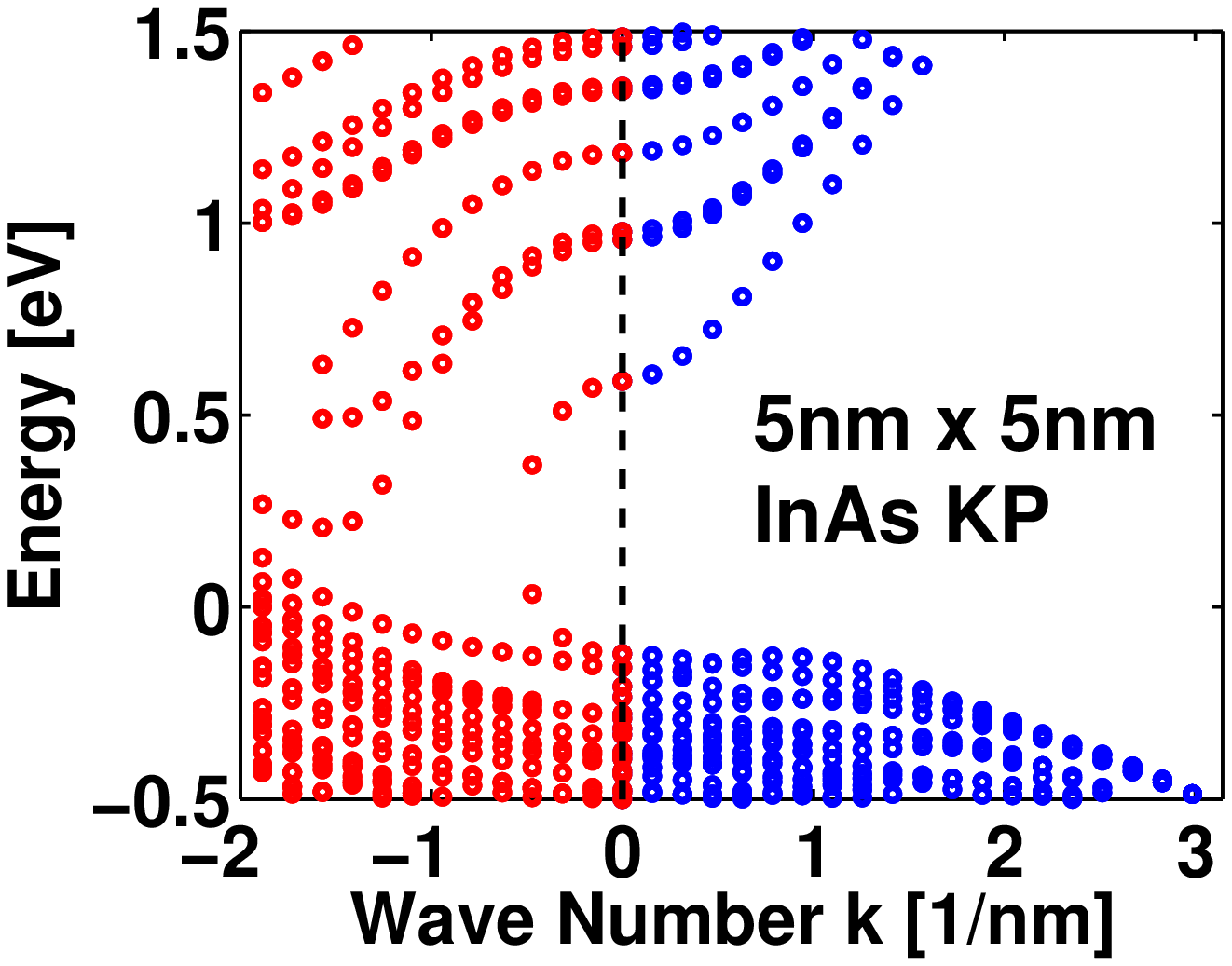}
\includegraphics[width=2.2in]{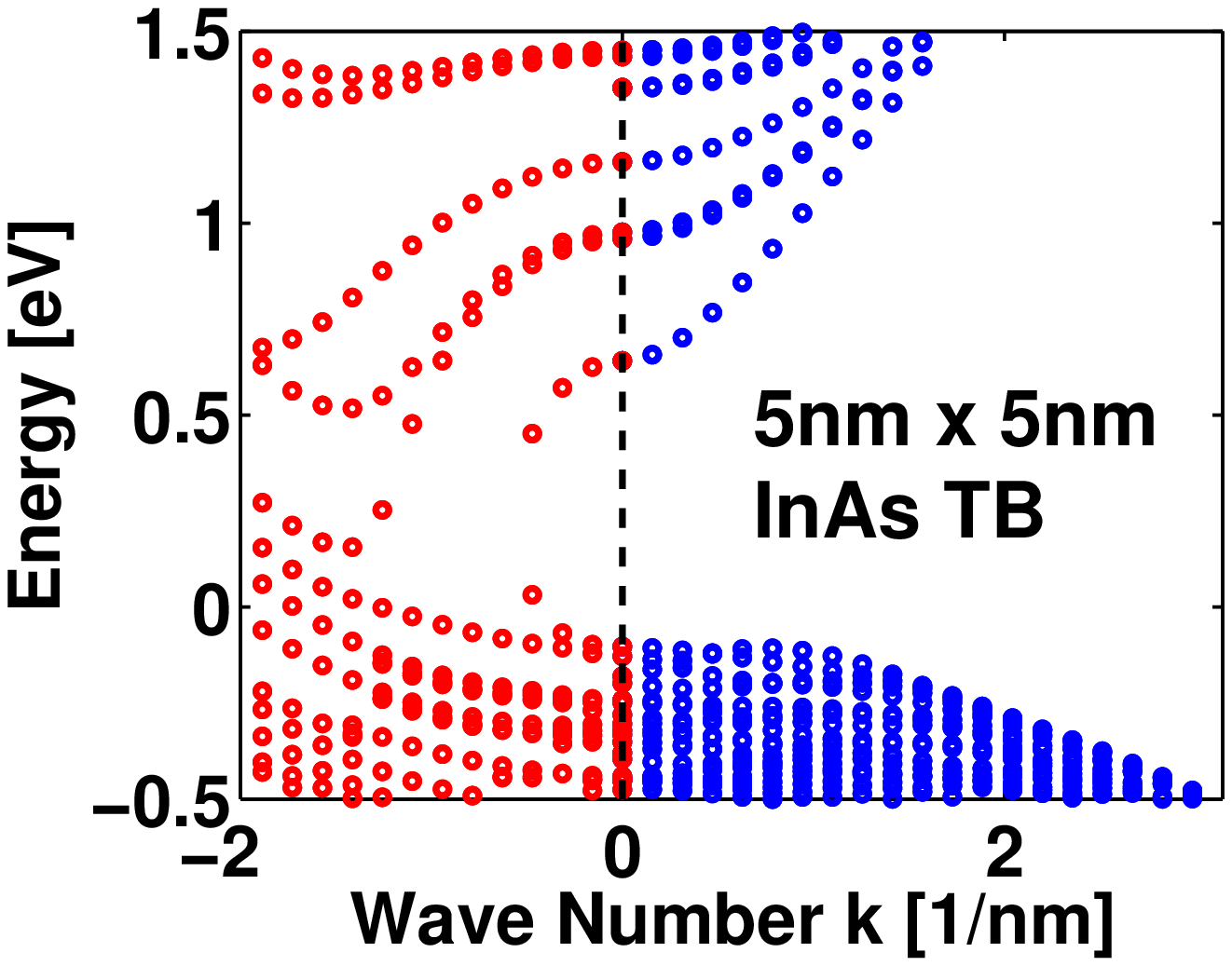}
\includegraphics[width=2.2in]{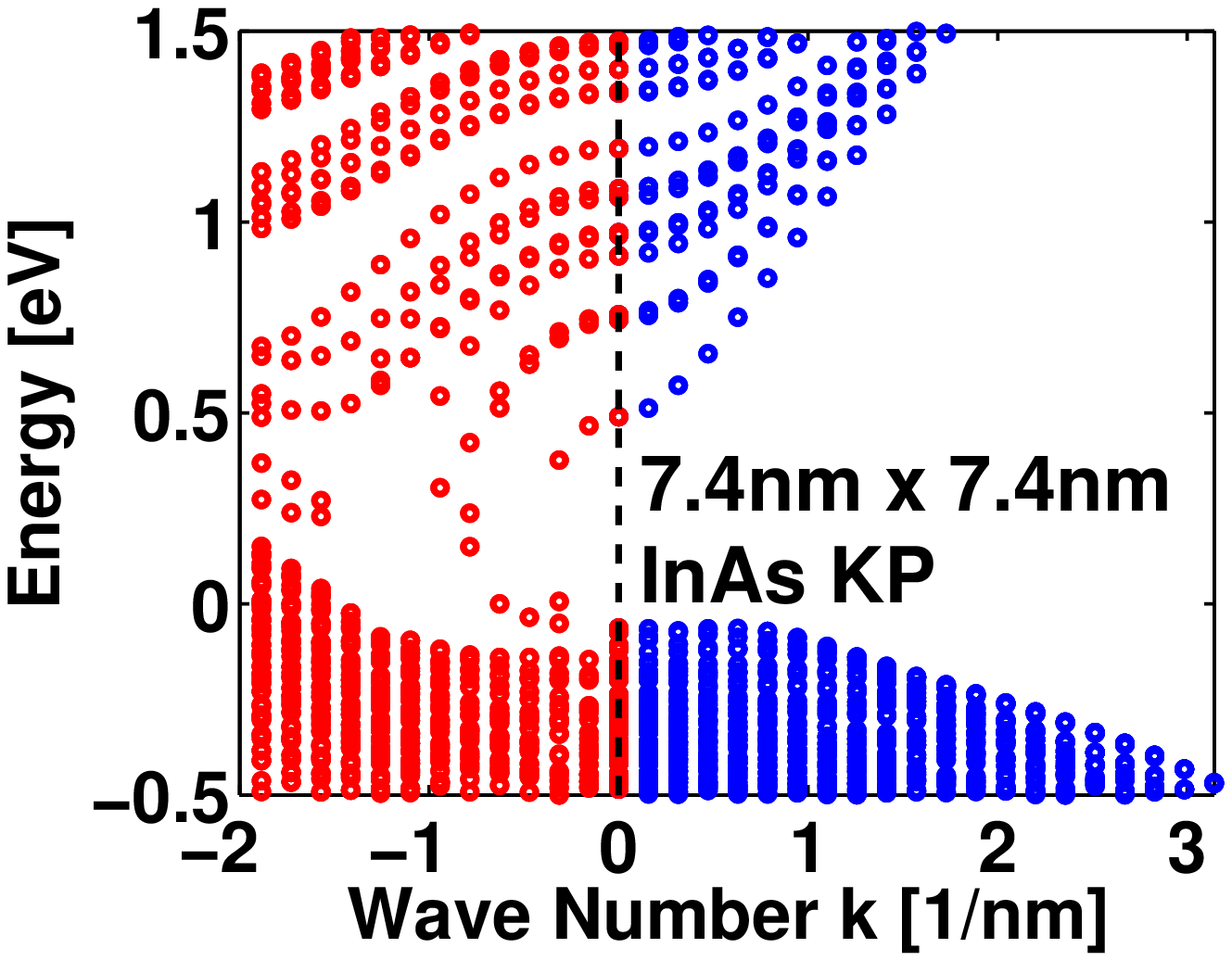}
\includegraphics[width=2.2in]{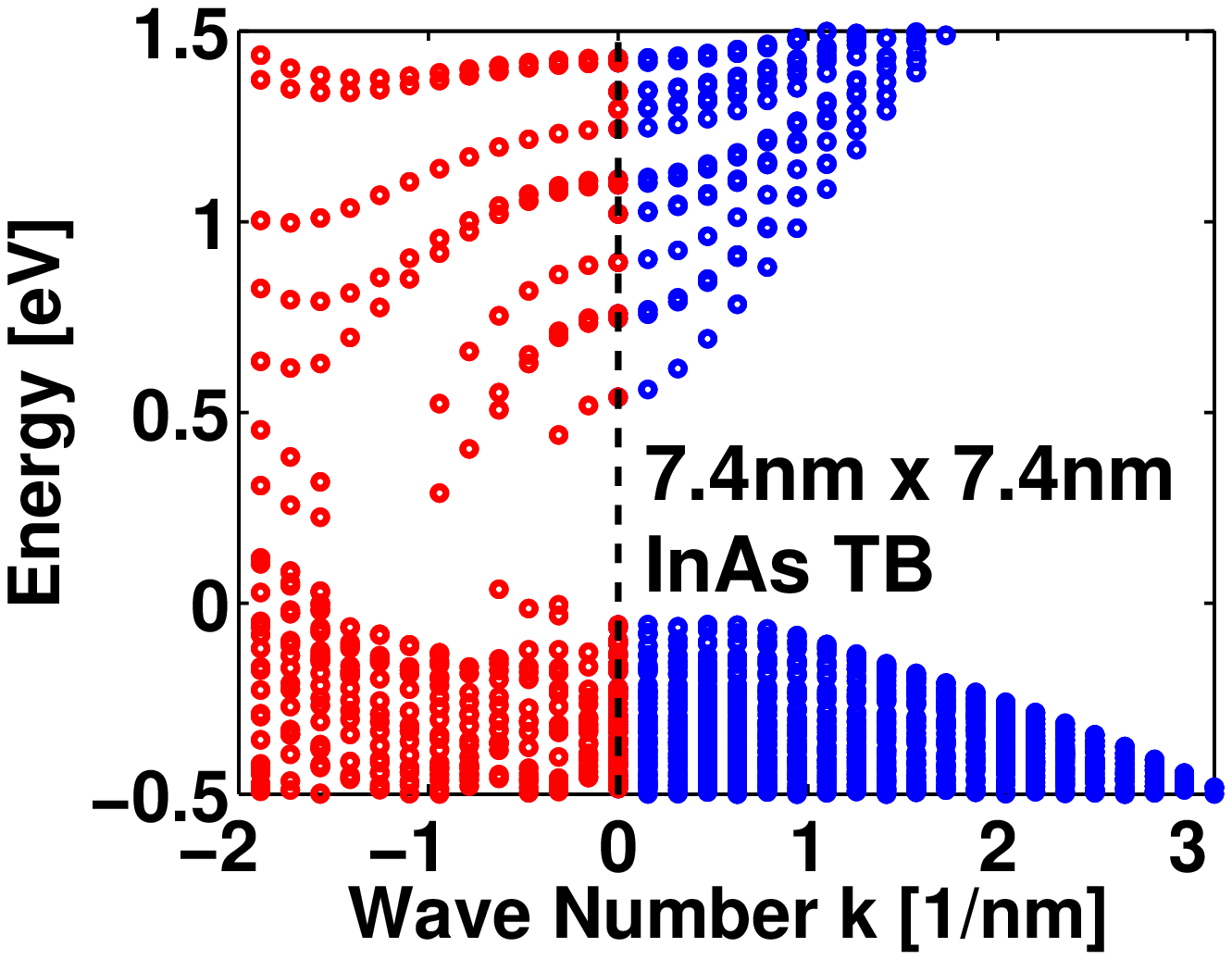}
\caption{Complex and real band structures of a $3\rm{nm}\times3\rm{nm}$ (top), a $5\rm{nm}\times5\rm{nm}$ (middle), and a $7.4\rm{nm}\times7.4\rm{nm}$ (bottom) cross-section InAs nanowire in the [100] orientation (with (010) and (001) surfaces). Left: eight-band $\bm{k}\cdot\bm{p}$ results; right: $sp^3s^*$ spin-orbit TB results.}
\label{fig_kp_vs_tb_InAs}
\end{figure}

\subsection{Comparison with TB Results}
Since $\bm{k}\cdot\bm{p}$ method is only valid in a small region around the $\Gamma$ point, there is a concern whether the $\bm{k}\cdot\bm{p}$ method is accurate for small nanostructures. A comparison with full-band tight binding (TB) results will help answer this question. In order to have a fair comparison for confined structures, the $\bm{k}\cdot\bm{p}$ parameters are fit to the bulk TB calculations. At first the bulk band structure is computed using the $sp^3s^*$ spin-orbit TB model, from which we have the band gap, spit-off energy, electron effective mass, heavy hole and light hole effective masses in both the [100] and [111] directions. These immediately determine $\bm{k}\cdot\bm{p}$ parameters $E_g$, $\Delta$, $m_c^*/m_0$, $\gamma_1$, $\gamma_2$, and $\gamma_3$ \cite{gershoni1993calculating},
\begin{equation}
\frac{m_0}{m^*_{hh}(100)}=\gamma_1-2\gamma_2,
\end{equation}
\begin{equation}
\frac{m_0}{m^*_{lh}(100)}=\gamma_1+2\gamma_2,
\end{equation}
\begin{equation}
\frac{m_0}{m^*_{hh}(111)}=\gamma_1-2\gamma_3,
\end{equation}
\begin{equation}
\frac{m_0}{m^*_{lh}(111)}=\gamma_1+2\gamma_3.
\end{equation}
The remaining parameter $E_p$ is then slightly reduced from experiment value so as to avoid spurious solution \cite{veprek2007ellipticity}.
The fitted parameters for materials InAs and GaSb used in this work are list in Table 1, which slightly differ from those in \cite{Vurgaftman01}. The valence band offset (VBO) of GaSb relative to InAs is taken from \cite{Vurgaftman01}.

\begin{table}
\caption{Material parameters for InAs and GaSb at T=300K.}
\label{tab:1}       
\begin{tabular}{p{1.6cm}p{1.1cm}p{1.1cm}p{1.1cm}p{1.1cm}p{1.1cm}p{1.1cm}p{1.1cm}p{1.4cm}}
\hline\noalign{\smallskip}
Parameters & $E_g$ (eV) & $\Delta$ (eV) & $m_c^*/m_0$ & $\gamma_1$ & $\gamma_2$ &  $\gamma_3$ & $E_p$ (eV) & VBO (eV)\\
\noalign{\smallskip}\svhline\noalign{\smallskip}
InAs       & 0.368  & 0.381    & 0.024    & 19.20     & 8.226    &  9.033   & 18.1  & 0\\
GaSb       & 0.751  & 0.748    & 0.042    & 13.27     & 4.97     &  5.978   & 21.2  & 0.56\\
\hline\noalign{\smallskip}
\end{tabular}
\end{table}

Fig. \ref{fig_kp_vs_tb_InAs} compares the eight-band $\bm{k}\cdot\bm{p}$ and $sp^3s^*$ spin-orbit TB band structures of three InAs nanowires with $3\rm{nm}\times3\rm{nm}$, $5\rm{nm}\times5\rm{nm}$, and $7.4\rm{nm}\times7.4\rm{nm}$ cross sections, respectively. For $\bm{k}\cdot\bm{p}$, hard wall boundaries are imposed at the four surfaces while for TB the surface atoms are passivated with hydrogen atoms. Good matches are observed except for the $3\rm{nm}\times3\rm{nm}$ case, where the $\bm{k}\cdot\bm{p}$ model has larger separation of subbands though the band gaps are close.

\begin{figure}[h]
\centering
\includegraphics[width=2.2in]{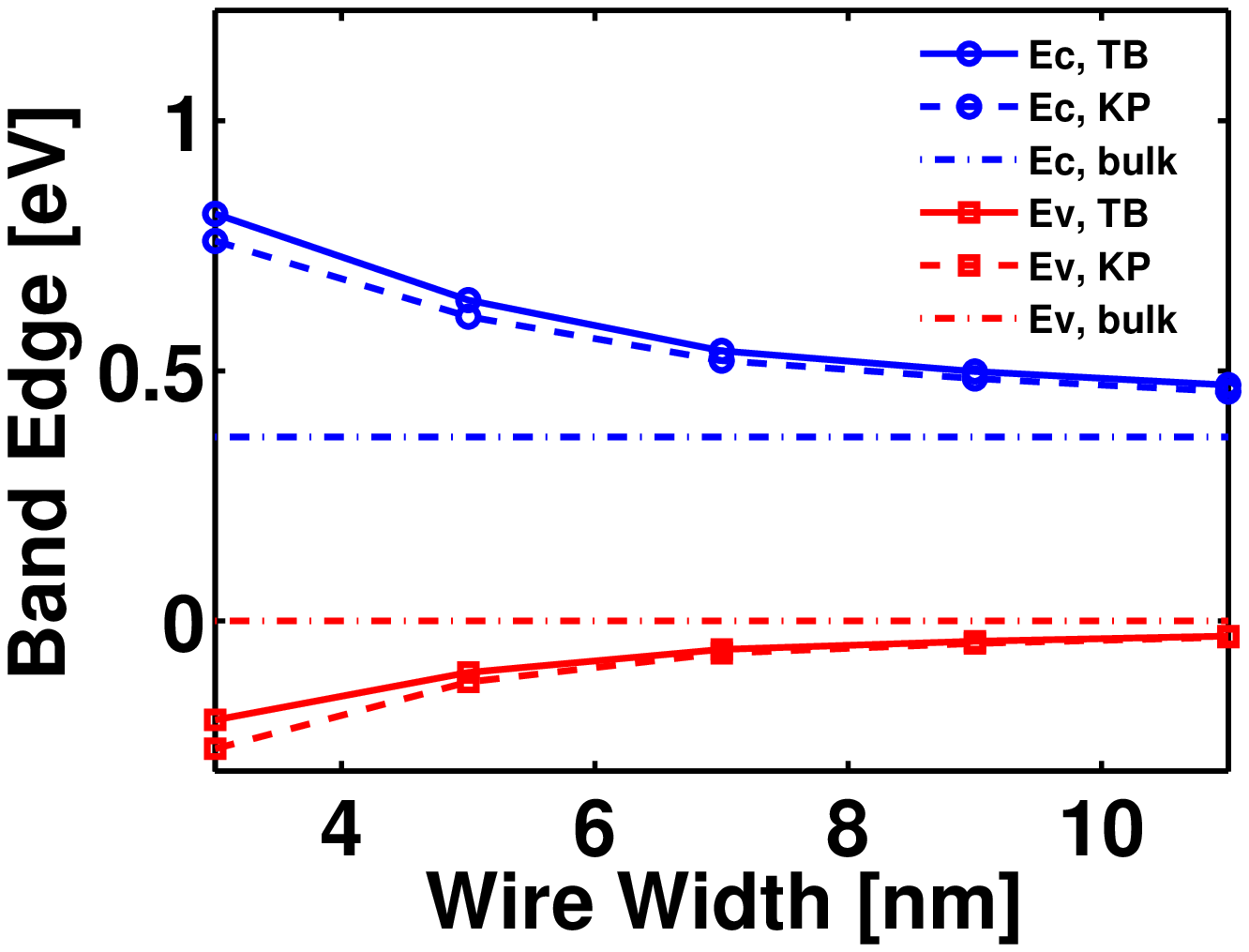}
\includegraphics[width=2.2in]{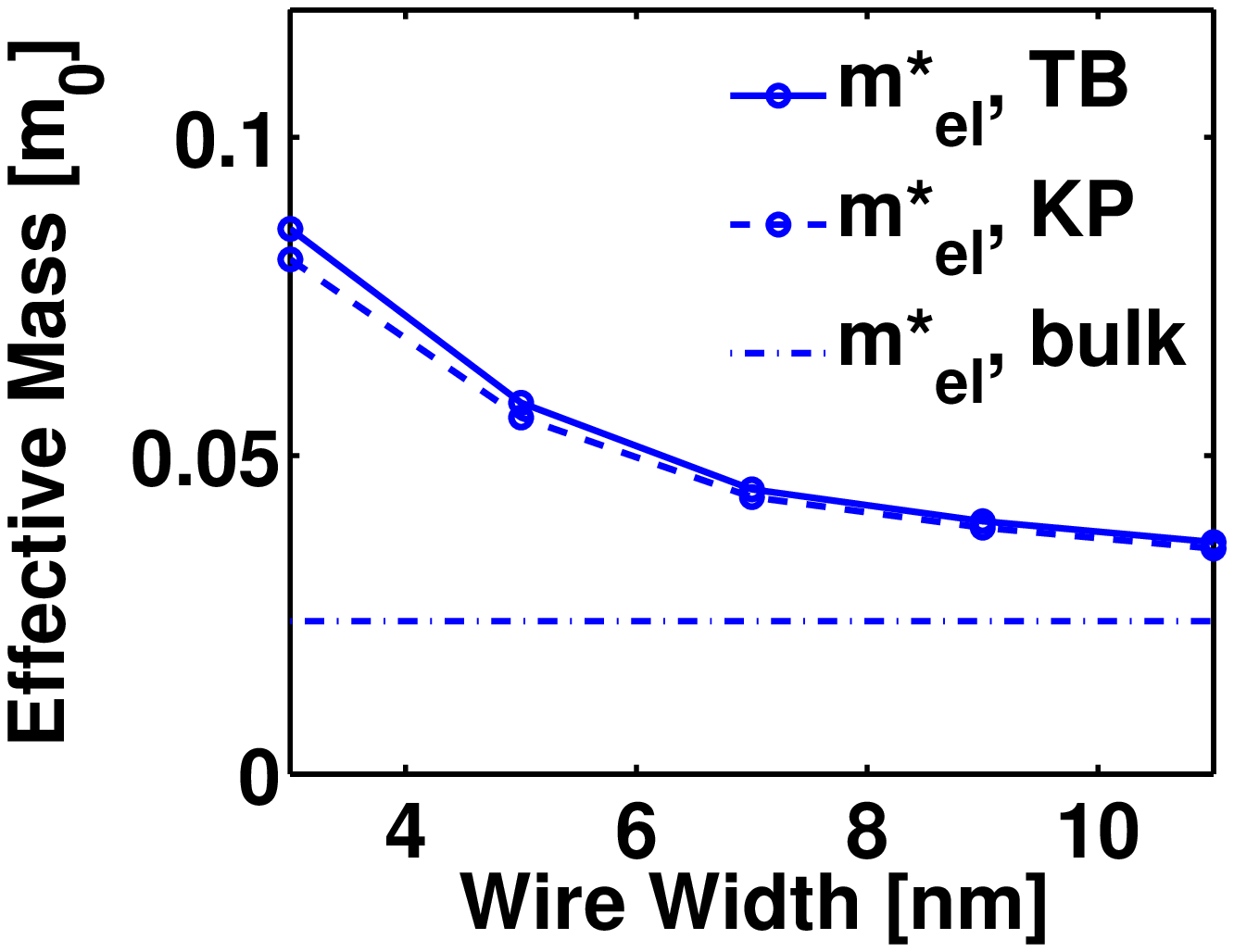}
\caption{Band edges (left) and electron effective mass (right) as functions of InAs wire cross section.}
\label{fig_band_edge_mass_wire}
\end{figure}

The band edges and effective masses as functions of nanowire cross-section size are plotted in Fig. \ref{fig_band_edge_mass_wire}. The $\bm{k}\cdot\bm{p}$ and TB results match quite well, except that $\bm{k}\cdot\bm{p}$ band edges are slightly shifted downwards for small nanowires. These two models predict the same trends, i.e., as the nanowire size decreases both the band gap and the electron effective masses increase.

\begin{figure}[h]
\centering
\includegraphics[width=2.2in]{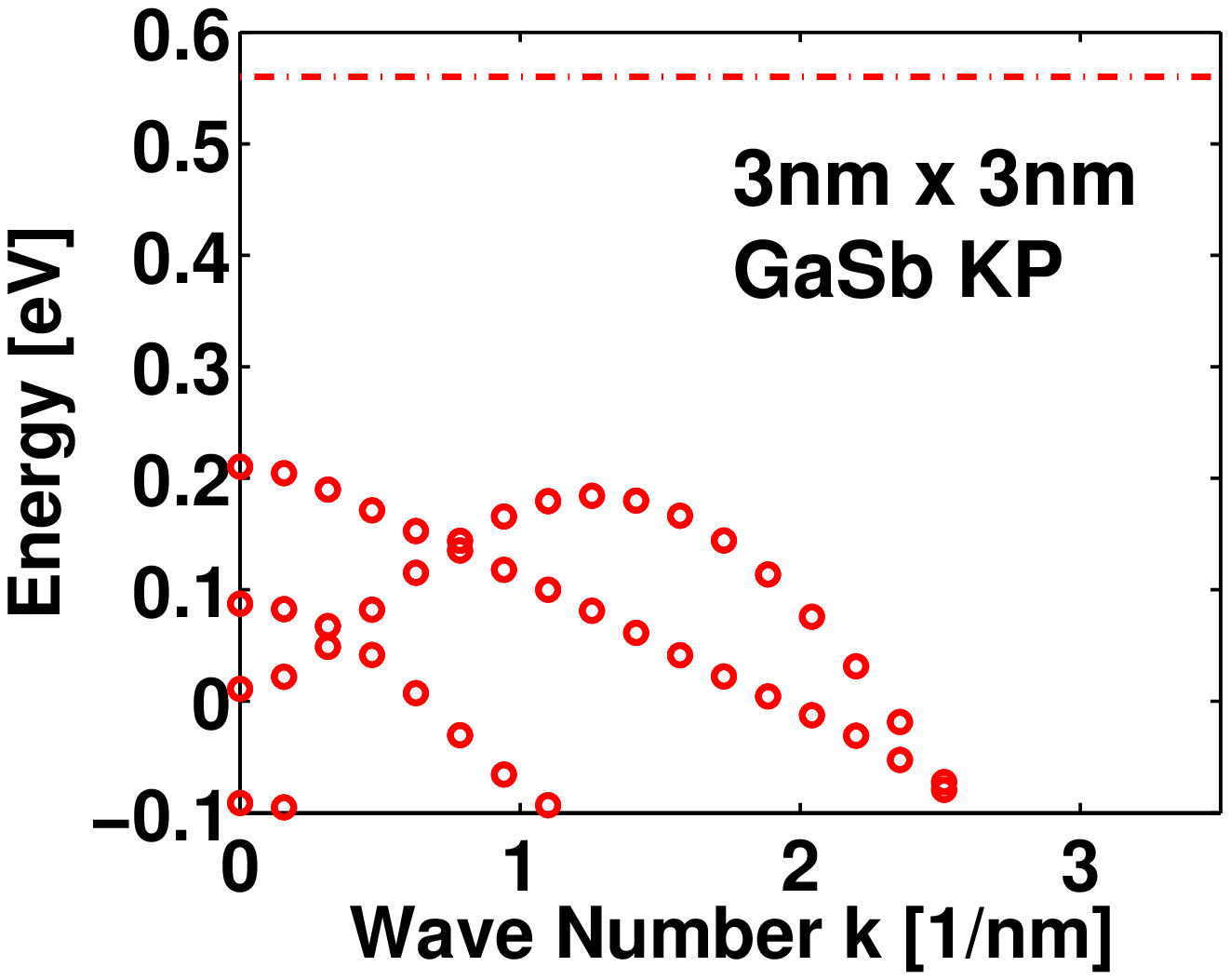}
\includegraphics[width=2.2in]{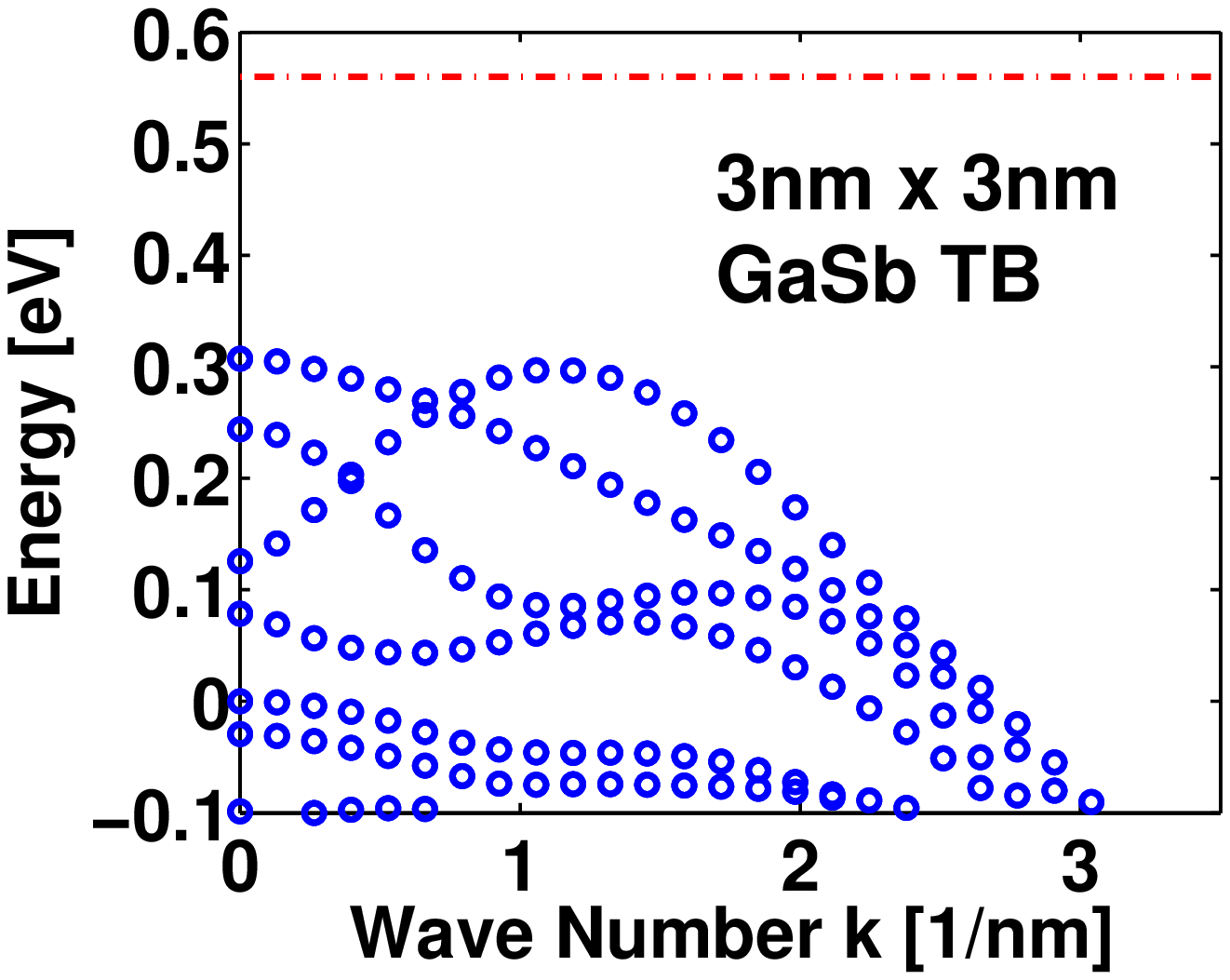}
\includegraphics[width=2.2in]{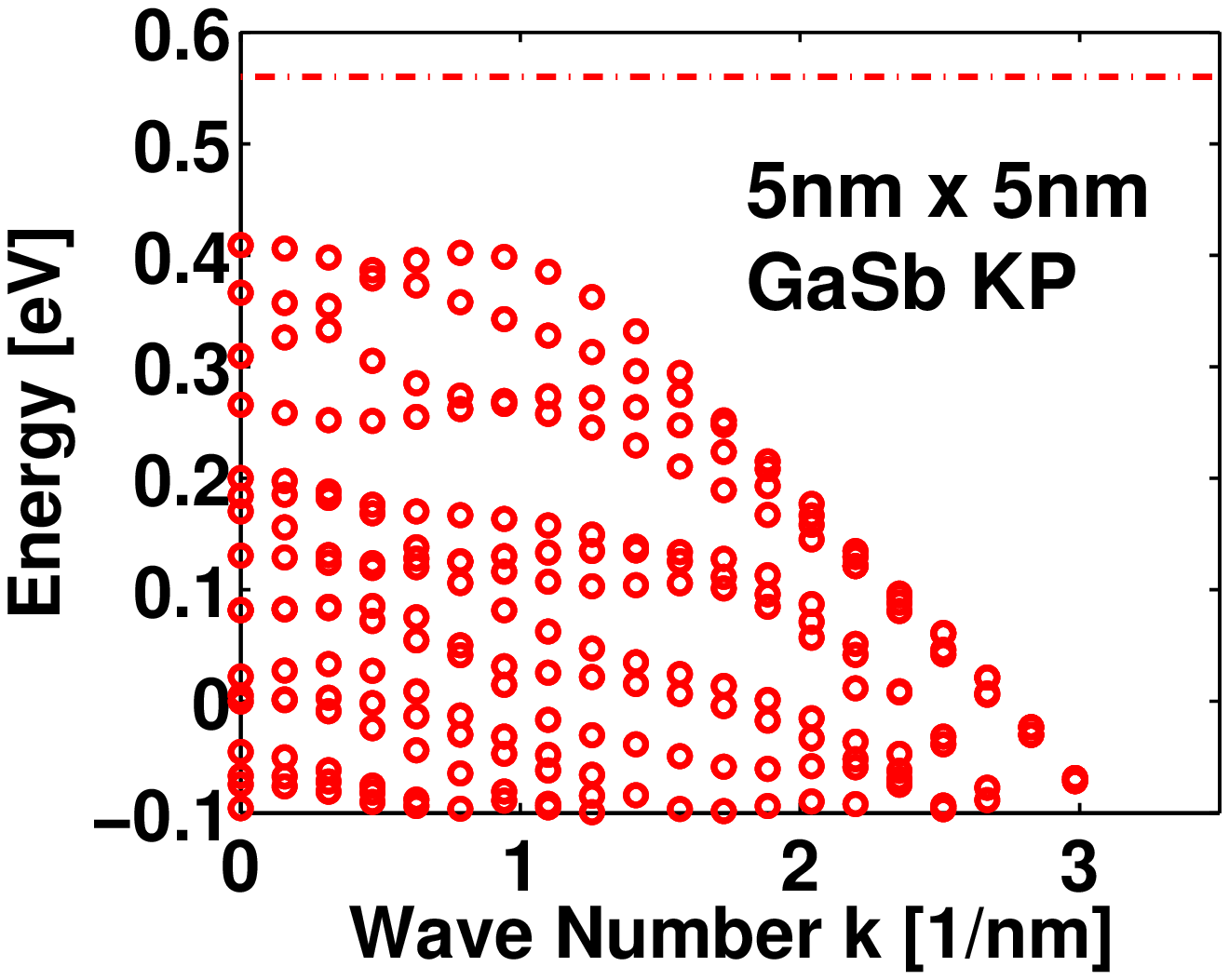}
\includegraphics[width=2.2in]{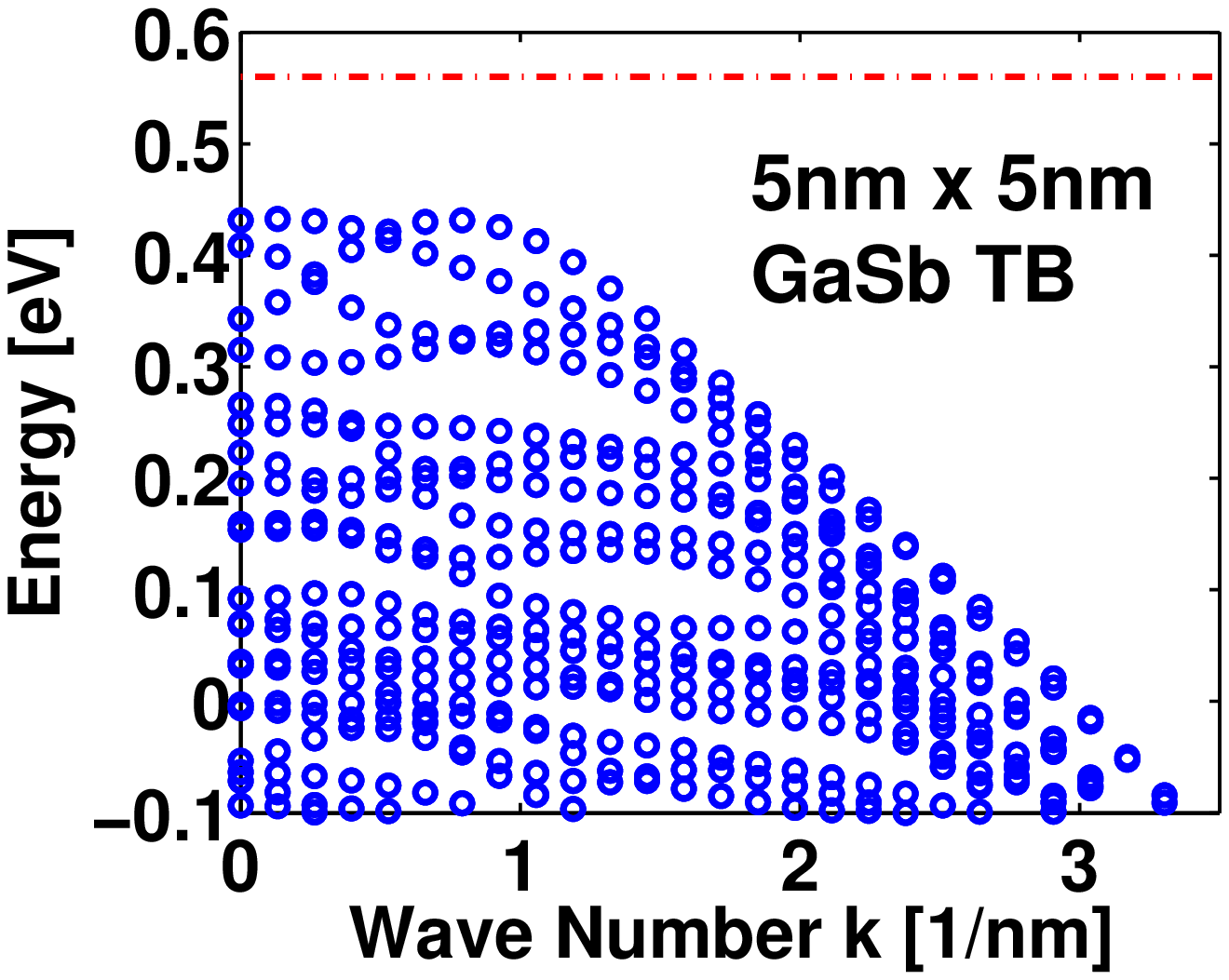}
\includegraphics[width=2.2in]{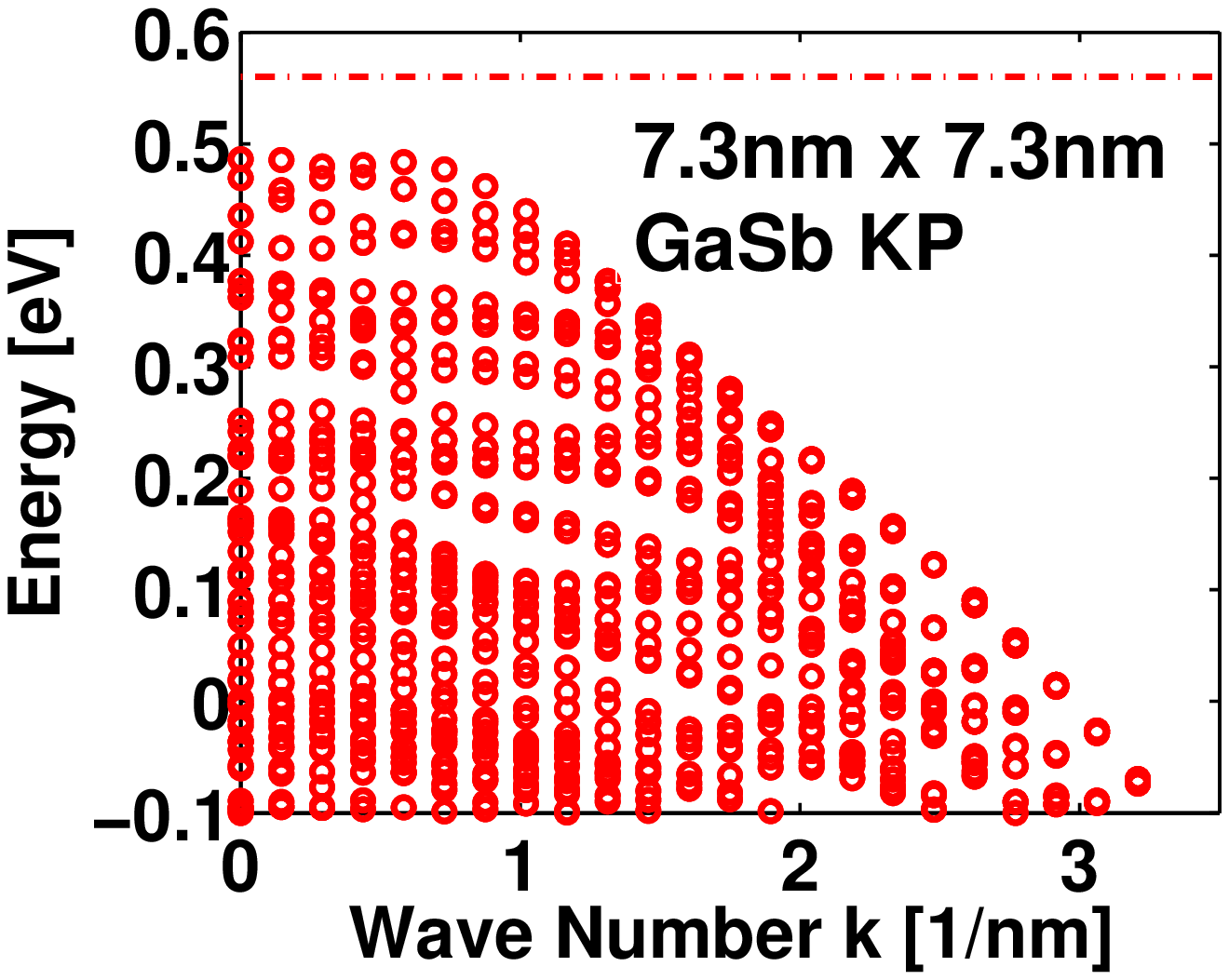}
\includegraphics[width=2.2in]{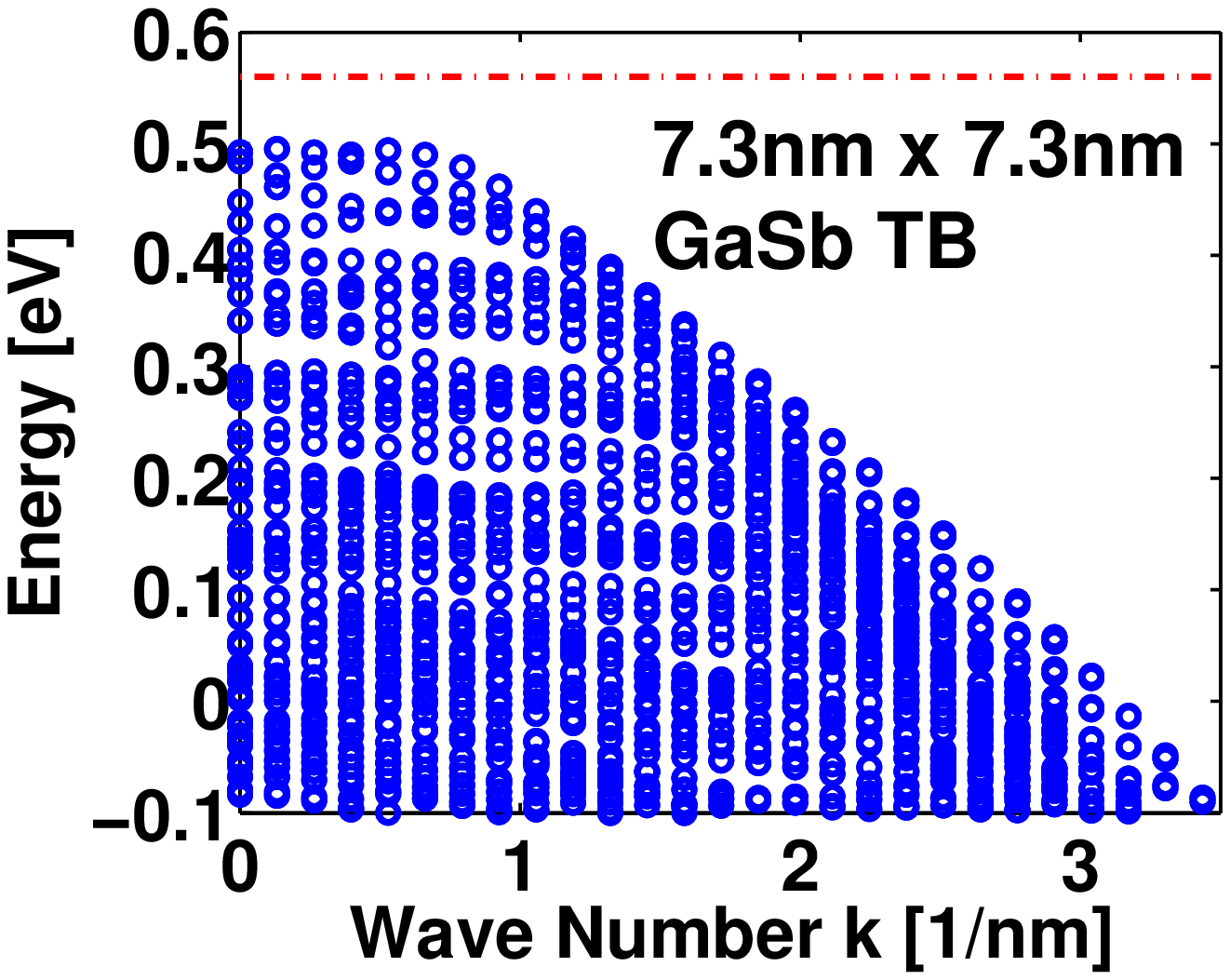}
\caption{Valence band structures of a $3\rm{nm}\times3\rm{nm}$ (top), a $5\rm{nm}\times5\rm{nm}$ (middle), and a $7.3\rm{nm}\times7.3\rm{nm}$ (bottom) cross-section GaSb nanowire in the [100] orientation (with (010) and (001) surfaces). Left: eight-band $\bm{k}\cdot\bm{p}$ results; right: $sp^3s^*$ spin-orbit TB results. The bulk valence band edge is shown in dash-dot line.}
\label{fig_kp_vs_tb_GaSb}
\end{figure}

Fig. \ref{fig_kp_vs_tb_GaSb} compares $\bm{k}\cdot\bm{p}$ and TB band structures of three GaSb nanowires with $3\rm{nm}\times3\rm{nm}$, $5\rm{nm}\times5\rm{nm}$, and $7.3\rm{nm}\times7.3\rm{nm}$ cross sections, respectively. Only valence band is shown since it is the most relevant for transport in heterojunction GaSb/InAs TFET application here. Similar to InAs case, quantitative matches are observed except for the $3\rm{nm}\times3\rm{nm}$ case where $\bm{k}\cdot\bm{p}$ model predicts lower valence band edge and larger subband energies.

\section{Reduced-Order NEGF Method}
\subsection{Reduced-Order NEGF Equations}
The NEGF equations for the retarded and lesser Green's function, $\bm{G}^R$ and $\bm{G}^{<}$, in the mixed real and Fourier space can be written as,
\begin{equation}\label{eq2}
\left[E\bm{I}-\bm{H}-\bm{V}-\bm{\Sigma}^{R}\left(E\right)\right]\bm{G}^{R}\left(E\right)=\bm{I},
\end{equation}
\begin{equation}\label{eq2}
\bm{G}^{<}\left(E\right)=\bm{G}^{R}\left(E\right)\bm{\Sigma}^{<}\left(E\right)\bm{G}^{R\dagger}\left(E\right),
\end{equation}
where $\bm{H}$ is the block three-diagonal $\bm{k}\cdot\bm{p}$ Hamiltonian of the isolated device, $\bm{V}$ is potential term that is block diagonal, and $\bm{\Sigma}^R$ ($\bm{\Sigma}^{<}$) is the retarded (lesser) self-energy matrix due to the semi-infinite leads, which is non-zero only in the first and last blocks. Phonon scattering has a very modest effect on the I-V curve \cite{conzatti2011simulation} and coherent transport is sufficient for III-V homojunction and heterojunction TFETs with direct band gap \cite{luisier2010simulation,Koswatta2010on}, thus it is excluded in this work.

As the matrices involved are very large, to solve $\bm{G}^R$ and $\bm{G}^{<}$ efficiently for many different energy $E$, the reduced-order matrix equations can be constructed,
\begin{equation}\label{eq3}
\left[E\bm{\widetilde{I}}-\bm{\widetilde{H}}-\bm{\widetilde{V}}-\bm{\widetilde{\Sigma}}^R\left(E\right)\right]\bm{\widetilde{G}}^R\left(E\right)=\bm{\widetilde{I}},
\end{equation}
\begin{equation}\label{eq2}
\bm{\widetilde{G}}^{<}\left(E\right)=\bm{\widetilde{G}}^{R}\left(E\right)\bm{\widetilde{\Sigma}}^{<}\left(E\right)\bm{\widetilde{G}}^{R\dagger}\left(E\right),
\end{equation}
and the reduced-order Green's functions $\bm{\widetilde{G}}^R\left(E\right)$ and $\bm{\widetilde{G}}^{<}\left(E\right)$ are to be solved. Here, the reduced Hamiltonian, potential, self energy, and Green's function are
\begin{align}\label{eq4}
&\bm{\widetilde{H}}=\bm{U}^{\dag}\bm{H}\bm{U}, \quad \bm{\widetilde{V}}=\bm{U}^{\dag}\bm{V}\bm{U},\nonumber\\
&\bm{\widetilde{\Sigma}}^{R,<}\left(E\right)=\bm{U}^{\dag}\bm{\Sigma}^{R,<}\left(E\right)\bm{U}, \quad \bm{\widetilde{G}}^{R,<}\left(E\right)=\bm{U}^{\dag}\bm{G}^{R,<}\left(E\right)\bm{U},
\end{align}
where $\bm{U}$ is a block-diagonal transformation matrix containing the reduced basis $\bm{U}_i$ of each layer $i$ (with dimension $N_t\times N_m$, where $N_m$ is the number of reduced basis).

The $-i\bm{\widetilde{G}}^{<}\left(E\right)$ gives the electron density. The hole density $i\bm{\widetilde{G}}^{>}\left(E\right)$ is obtained by subtracting electron density from the spectral function
\begin{equation}
i\bm{\widetilde{G}}^>\left(E,x_i\right)=-2Im\{\bm{\widetilde{G}}^R\left(E,x_i\right)\}+i\bm{\widetilde{G}}^<\left(E,x_i\right).
\end{equation}
In TFET, electrons can tunnel from valence band into conduction band and leave holes in the valence band. The charge density involving both electrons and holes is calculated by the method similar to Ref. \cite{guo2004toward}.
\begin{eqnarray}
\bm{\widetilde{Q}}\left(x_i\right)&=\left(ie\right)\int dE\frac{1}{2} \left[sgn\left(E-E_N\left(x_i\right)\right)+1\right]\cdot\bm{\widetilde{G}}^<\left(E,x_i\right)\nonumber\\
&+\frac{1}{2}\left[-sgn\left(E-E_N\left(x_i\right)\right)+1\right]\cdot\bm{\widetilde{G}}^>\left(E,x_i\right),
\end{eqnarray}
where $E_N\left(x_i\right)$ is the layer dependent threshold (charge neutral level), which is taken as the mid band gap $E_N\left(x_i\right)=0.5\left[E_v\left(x_i\right)+E_c\left(x_i\right)\right]+\bar{V}\left(x_i\right)$ where $\bar{V}\left(x_i\right)$ is the average potential of layer $x_i$. $sgn$ is the sign function. This model basically says that if a carrier is above (below) the threshold it is considered as an electron (hole). The required diagonal blocks of $\bm{\widetilde{G}}^{R,<}\left(E\right)$, i.e., $\bm{\widetilde{G}}^{R,<}\left(E,x_i\right)$, can be calculated with efficient recursive Green's function (RGF) algorithm \cite{lake1997single}, since the matrices are still block three-diagonal after the transformation.

The integrated $\bm{\widetilde{Q}}\left(x_i\right)$ is then transformed back into real space,
\begin{equation}
Q\left(\bm{r}\right)=diag\left(\bm{U}^\prime\bm{U}\bm{\widetilde{Q}}\bm{U}^{\dag}\bm{U}^{\prime\dagger}\right),
\end{equation}
where $\bm{U}^\prime$ is the transformation matrix from Fourier space to real space. Note that only diagonal terms are needed, which can be utilized to relieve the computational cost of back transformation. The transmission coefficient (and then ballistic current) can be calculated directly in the reduced space.

The problem now is how to construct this transformation matrix $\bm{U}$ so that the reduced system is as small as possible, and yet it still accurately describes the original system.
To construct the reduced basis $\bm{U}_i$ for layer $i$, the Hamiltonian of layer $i$ is repeated to form an infinite periodic nanowire. The reduction comes from the fact that only the electrons near the conduction band bottom and valence band top are important in the transport process. To approximate the band structure over that small region, $\bm{U}_i$ then consists of the sampled Bloch modes with energy lying in that region. Multiple-point $k$ space sampling and (or) $E$ space sampling can be employed as has been demonstrated for the three- and six-band cases \cite{huang2013model}. Here $k$ space sampling is adopted since $E$ space sampling is more costly and that the eight-band matrix is larger than the six- or three-band case.

\subsection{Spurious Band Elimination}
For three- and six-band $\bm{k}\cdot\bm{p}$ models, as is shown in \cite{huang2013model}, by sampling the Bloch modes at multiple points in the $k$ space and (or) $E$ space, a significantly reduced Hamiltonian can be constructed that describes very well the valence band top, based on which p-type SiNW FETs are simulated with good accuracy and efficiency. However, direct extension of this method to eight-band $\bm{k}\cdot\bm{p}$ model fails. The problem is that the reduced model constructed by multi-point expansion generally leads to some spurious bands, a situation similar to constructing the equivalent tight binding models \cite{mil2012equivalent}, rendering the reduced model useless.

\begin{figure}[htbp] \centering
\includegraphics[width=4.5in]{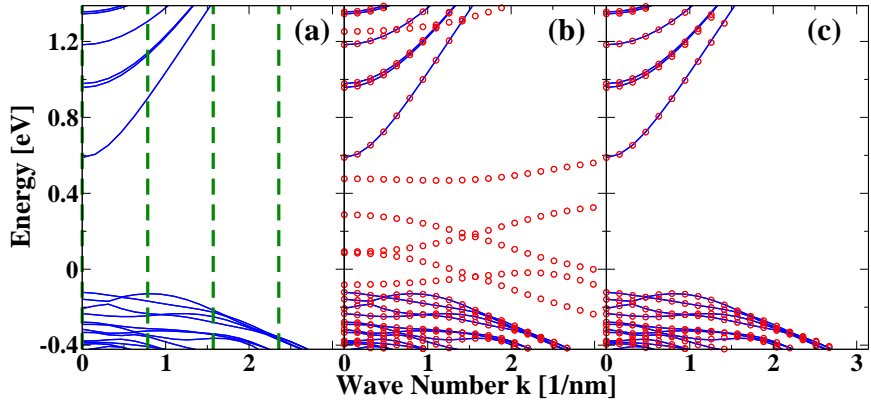}
\caption{$E$-$k$ diagrams of a $5\rm{nm}\times5\rm{nm}$ InAs nanowire in the [100] direction. (a) Exact solution (blue lines) and the sampling lines (dashed green lines). (b) Comparison between exact solution (blue lines) and reduced-order model solution (red circles) with spurious bands showing up. (c) Comparison between exact solution (blue lines) and reduced-order model solution (red circles) with spurious bands removed.}
\label{fig_5nm_wire_100}
\end{figure}

As an example, Fig. \ref{fig_5nm_wire_100}(a) plots the $E$-$k$ dispersion for an ideal InAs nanowire orientated in the [100] direction. Fig. \ref{fig_5nm_wire_100}(b) is the result using the reduced Hamiltonian $\bm{\widetilde{H}}$. The reduced basis $\bm{U}_i$ ($i$ is arbitrary here) is constructed by sampling the Bloch modes evenly in the Brillouin zone (at $k=0$, $\pm\pi/4$, $\pm2\pi/4$, and $\pm3\pi/4$ $[1/\rm{nm}]$, as denoted in green lines in Fig. \ref{fig_5nm_wire_100}(a)), with the energy $E \in [E_v-0.3\rm{eV}, E_c+0.8\rm{eV}]$ ($E_v$ and $E_c$ are the confined valence and conduction band edges), which results in $N_m=134$ modes. Note that the modes at negative $k$ can be obtained by a transformation of those at positive $k$ \cite{huang2013model}. Clearly, the reduced Hamiltonian reproduces quite well the exact dispersion in that energy window, demonstrating the number of sampling points is sufficient. However, there are also some spurious bands appearing in the conduction band and even in the band gap, making the reduced model useless. Moreover, different sampling points or sampling windows would change the number and position of the spurious bands. This situation is not encountered in the three- or six-band model involving only the valence bands, or in the one-band effective mass model involving the conduction band only. It should be caused by the coupling between the conduction and valence bands which makes the eight-band model indefinite. The coupling is important for materials with narrow band gaps.

The spurious bands must be suppressed. To this end, a singular value decomposition (SVD) is applied to the matrix $\bm{U}_i$. It is found that the singular values spread from a large value down to zero, indicating there are some linearly dependent modes. These linearly dependent modes give rise to null space of the reduced model and therefore must be removed. It is further found that the normal bands are mainly contributed by singular vectors having large singular values, in contrast to the spurious bands where singular vectors with small singular values also have large contribution. By removing the vectors with small singular values, i.e., vectors with $v\leq v_{th}$ where $v_{th}=0.20$ is the threshold, a new reduced basis $\bm{\widetilde{U}}_i$ is generated with $\widetilde{N}_m=76$. Using this new reduced basis, a new reduced Hamiltonian is constructed with its $E$-$k$ diagram given in Fig. \ref{fig_5nm_wire_100}(c). It is observed that all the spurious bands have been eliminated at a cost of slightly compromised accuracy. The reduction ratio is $\widetilde{N}_m/N_t=76/1464=5.19\%$, which is quite significant.

The value of $v_{th}$ is found to be crucial. A small $v_{th}$ might be insufficient to remove all the spurious bands while a large $v_{th}$ may degrade the accuracy severely. Moreover, adjustment of $v_{th}$ may be required when different sampling points or sampling energy windows are used. To determine $v_{th}$ automatically, we propose a search process as follows:

1. Sample enough Bloch modes and store them in matrix $\bm{B}$.
Suppose $I$ points are sampled in the $k$ space, and $m_{i}$ modes with energy $E \in [E_1, E_2]$ are obtained at the $i$th point $k_i$ ($1\leq i\leq I$), then the size of matrix $\bm{B}$ is $N_t\times N_m$, where $N_m=\sum_{i=1}^I m_i$.

2. Do SVD of $\bm{B}$, i.e., $\bm{B}=\bm{U}\bm{\Sigma}\bm{V}^\dagger$.

3. Set an initial value for $v_{th}$. Let us use $v_{th}=0$ here.

4. Use $v_{th}$ to construct a reduced basis $\widetilde{\bm{U}}$ by removing the singular vectors with $v<v_{th}$ in $\bm{U}$.
The size of $\widetilde{\bm{U}}$ will be $N_t\times \widetilde{N}_m$.

5. Use $\widetilde{\bm{U}}$ to build a reduced Hamiltonian $\widetilde{\bm{H}}$.
For each layer of $\widetilde{\bm{H}}$, the size will be $\widetilde{N}_m \times \widetilde{N}_m$.

6. Solve the $E$-$k$ relation of $\widetilde{\bm{H}}$ for certain $k_i$, obtaining $\widetilde{m}_{i}$ modes with $E \in [E_1, E_2]$.
It is found that $k_i=0$ is a good choice.

7. If $\widetilde{m}_{i}>m_{i}$ (which means that there are still some spurious bands), increase $v_{th}$ appropriately and go back to step 4. Otherwise, stop.

The above search process is fast, since step 5 and step 6 are much cheaper than step 1 although they have to be repeated many times.
In fact, the complexity of step 1 is $I\times O\left(N_t^3\right)$, step 2 is $O\left(N_tN_m^2\right)$, step 5 is $O\left(\widetilde{N}_m N_t^2\right)$, and step 6 is $O\left(\widetilde{N}_m^3\right)$. Note that $\widetilde{N}_m<N_m<N_t$.

\begin{figure}[h]
\centering
\includegraphics[width=4.5in]{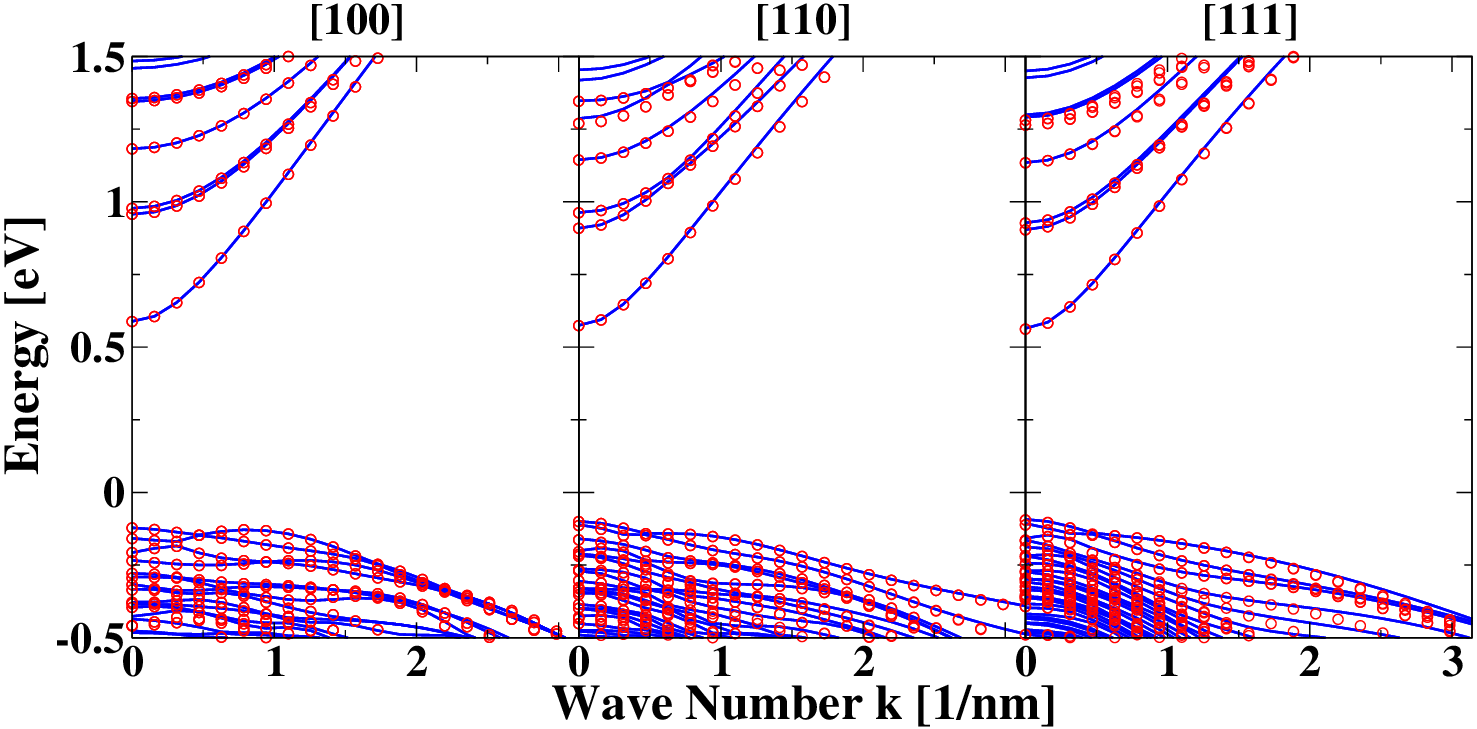}
\caption{$E$-$k$ diagrams of $5\rm{nm}\times5\rm{nm}$ InAs nanowires in the [100], [110], and [111] orientations. The blue lines are the exact solutions, while the red circles are the solutions of the reduced-order models. The valid energy window of the reduced-order models is $[E_v-0.3\rm{eV}, E_c+0.8\rm{eV}]$. The orders of the reduced models are 76, 66, and 62, respectively.}
\label{fig_5nm_wire}
\end{figure}

\begin{figure}[h]
\centering
\includegraphics[width=4.5in]{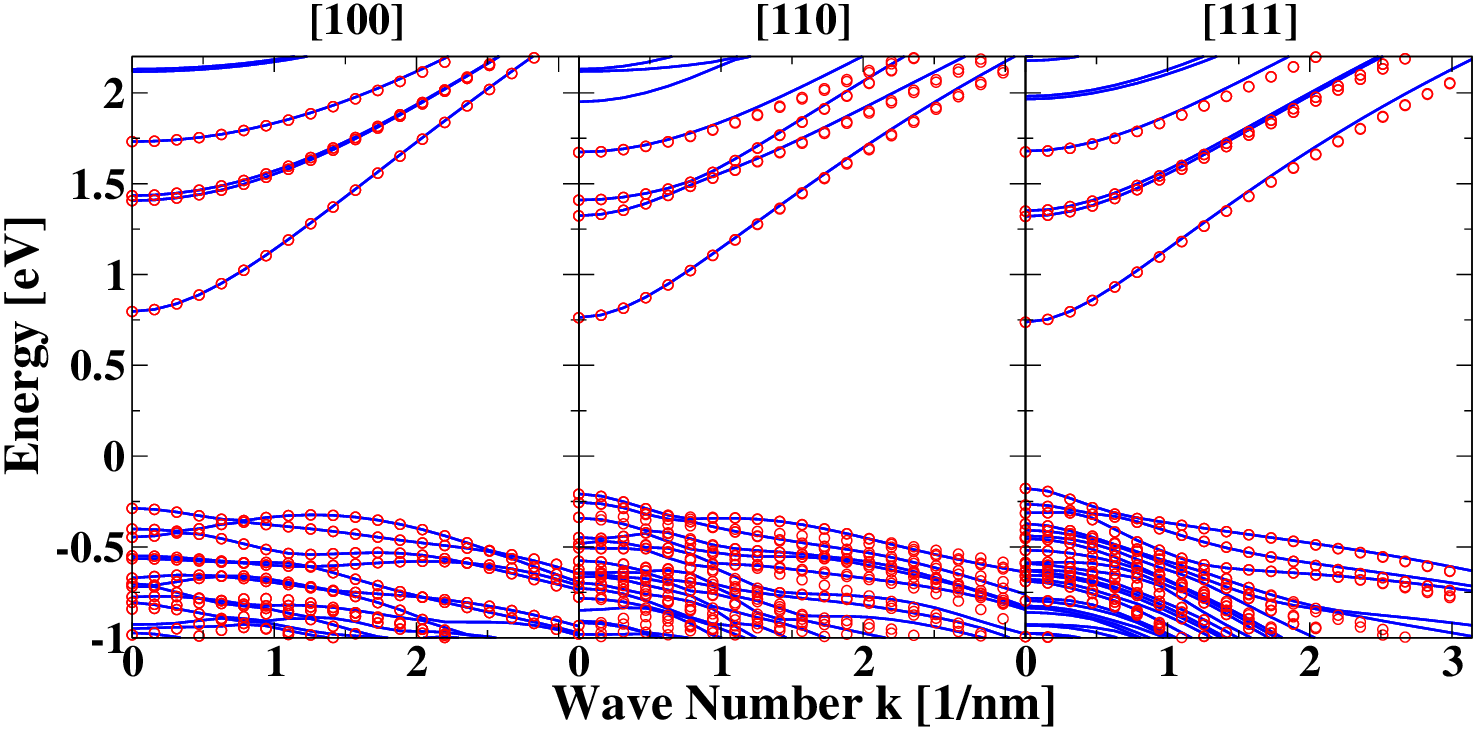}
\caption{$E$-$k$ diagrams of $3\rm{nm}\times3\rm{nm}$ InAs nanowires in the [100], [110], and [111] orientations. The blue lines are the exact solutions, while the red circles are the solutions of the reduced-order models. The valid energy window of the reduced-order models is $[E_v-0.55\rm{eV}, E_c+1.0\rm{eV}]$. The orders of the reduced models are 66, 52, and 48, respectively.}
\label{fig_3nm_wire}
\end{figure}

\begin{figure}[h]
\centering
\includegraphics[width=4.5in]{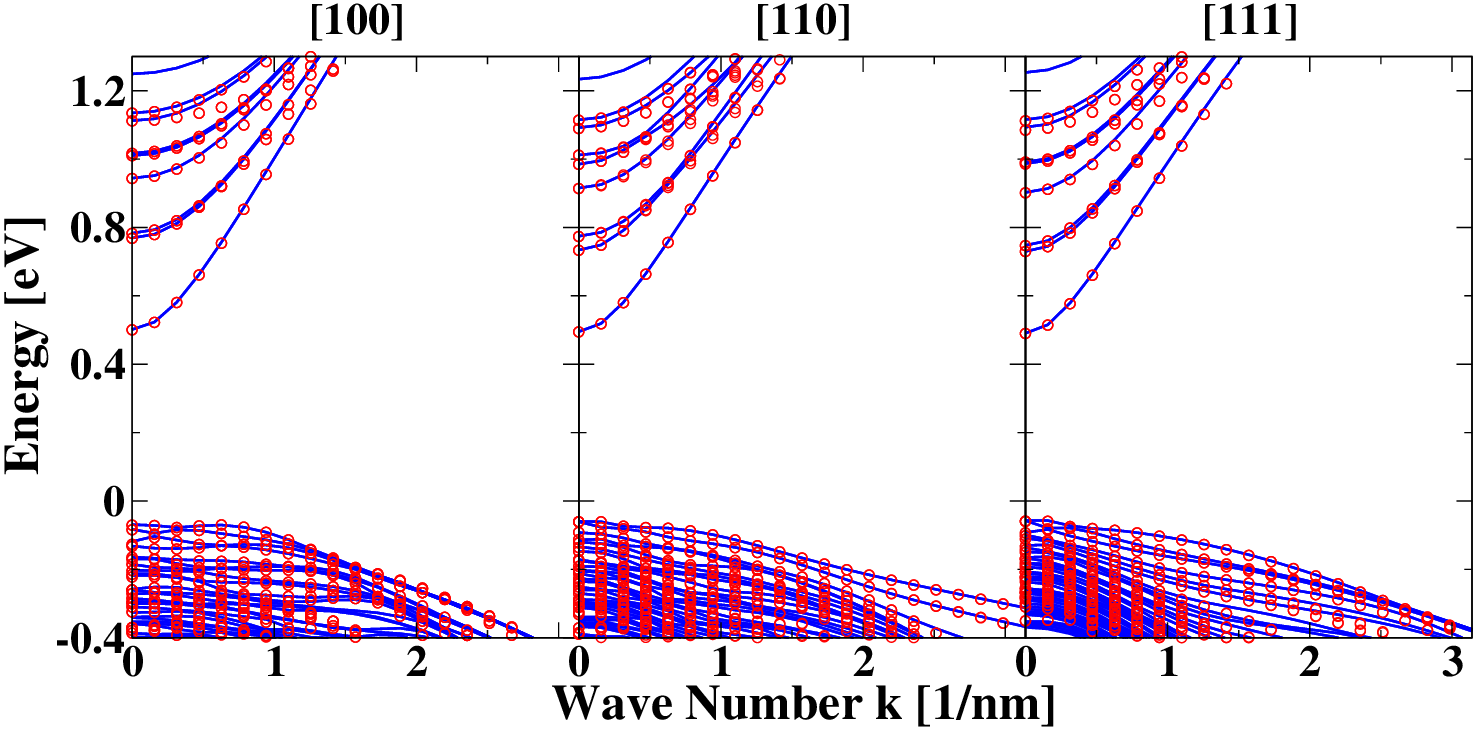}
\caption{$E$-$k$ diagrams of $7\rm{nm}\times7\rm{nm}$ InAs nanowires in the [100], [110], and [111] orientations. The blue lines are the exact solutions, while the red circles are the solutions of the reduced-order models. The valid energy window of the reduced-order models is $[E_v-0.25\rm{eV}, E_c+0.7\rm{eV}]$. The orders of the reduced models are 104, 104, and 96, respectively.}
\label{fig_7nm_wire}
\end{figure}

The $v_{th}=0.20$ used earlier is the result of the above search process. The above process also gives good results for nanowires in the [110] and [111] directions, as shown in Fig. \ref{fig_5nm_wire}. Different energy windows $E \in [E_v-0.2\rm{eV}, E_c+0.6\rm{eV}]$ and $E \in [E_v-0.4\rm{eV}, E_c+1.0\rm{eV}]$ are tested, again faithful results are obtained (not shown here). For other cross-section InAs nanowires such as the $3\rm{nm}\times3\rm{nm}$ and $7\rm{nm}\times7\rm{nm}$ in the [100], [110], and [111] orientations, this method gives reliable results, as shown in Fig. \ref{fig_3nm_wire}, and Fig. \ref{fig_7nm_wire}. It should be mentioned that this process results in a smaller basis set, which is different from the method for tight binding models \cite{mil2012equivalent} where the basis is enlarged by putting in more modes to eliminate the spurious bands.

\subsection{Error and Cost of the Reduced Models}
Now this reduced model can be applied to simulate the TFET as shown in Fig. \ref{fig_tfet_homo}. Reduced NEGF equations and Poisson equation are solved self-consistently. To improve the efficiency, the reduced basis is constructed for an ideal nanowire with its potential term set to zero, so the reduced basis just needs to be solved only once for each material and it remains unchanged during the self-consistent iterations. The potential term in real devices then merely causes transitions between these scattering states. This assumption has been adopted in Ref. \cite{mil2012equivalent} with good accuracy demonstrated. As will be shown below, it is also a fairly good approximation for the GAA nanowire TFET here.

\begin{figure}[htbp] \centering
\subfigure[]{\includegraphics[width=5.8cm]{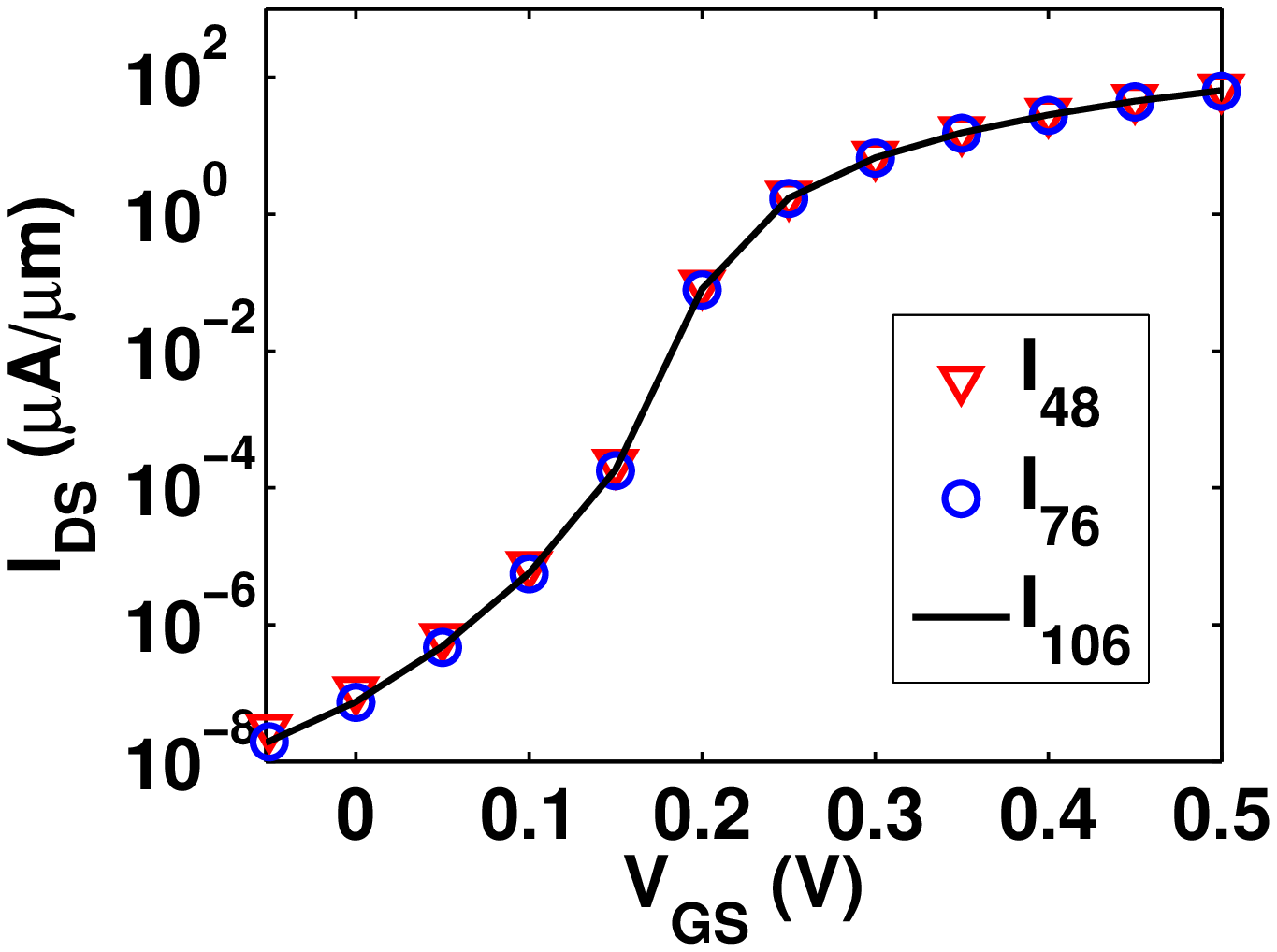}}
\subfigure[]{\includegraphics[width=5.8cm]{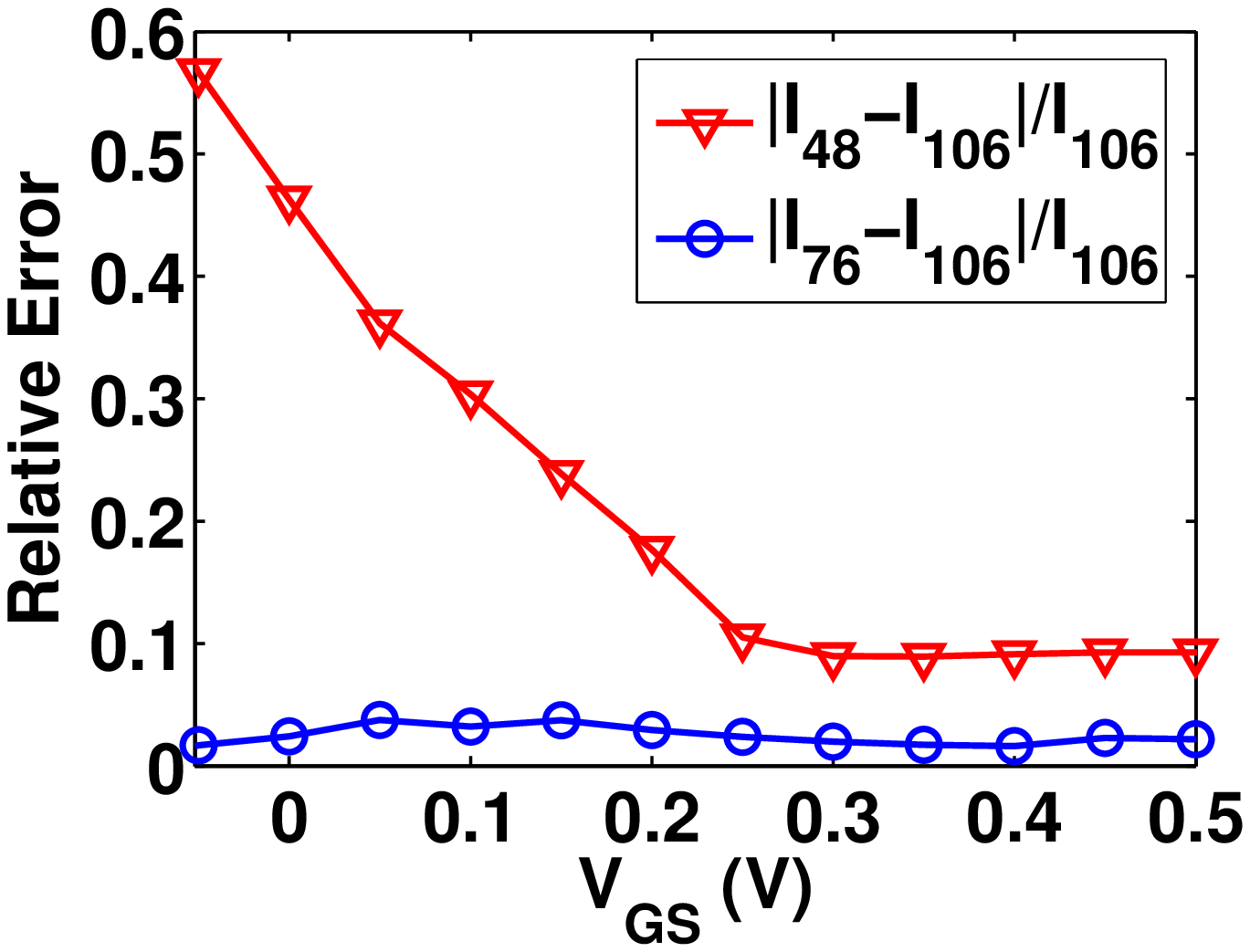}}
\caption{(a) $I_{DS}$-$V_{GS}$ transfer characteristics of a p-i-n InAs homojunction TFET as shown in Fig. 1. The nanowire is oriented in the [100] direction. $T_{ox}=1\rm{nm}$, $\varepsilon_{ox}=12.7$, $T_y=T_z=5\rm{nm}$, $L_s=15\rm{nm}$, $L_g=20\rm{nm}$, $L_d=30\rm{nm}$. The doping density is equal to $5\times10^{19}\rm{cm}^{-3}$ at the source and $5\times10^{18}\rm{cm}^{-3}$ at the drain. The drain bias is fixed to $V_{DS}=0.3\rm{V}$. (b) Relative errors of $I_{48}$ and $I_{76}$ with respect to $I_{106}$.}
\label{fig_homoIV_benchmark}
\end{figure}

The $I_{DS}$-$V_{GS}$ transfer characteristics of a $5\rm{nm}\times5\rm{nm}$ cross section InAs homojunction TFET is plotted in Fig. \ref{fig_homoIV_benchmark}(a). Three curves are compared. In the first, second, and third $I$-$V$ curve, the valid energy window is $[E_v-0.2\rm{eV}, E_c+0.6\rm{eV}]$, $[E_v-0.3\rm{eV}, E_c+0.8\rm{eV}]$, and $[E_v-0.4\rm{eV}, E_c+1.0\rm{eV}]$, respectively. The sampling k points are all at $k=0$, $\pm\pi/4$, $\pm2\pi/4$, and $\pm3\pi/4$ $[1/\rm{nm}]$. This leads to $\widetilde{N}_m=48$, $\widetilde{N}_m=76$, and $\widetilde{N}_m=106$, with corresponding $I$-$V$ curves denoted as $I_{48}$, $I_{76}$, and $I_{106}$. Here $I_{106}$ can be considered as the reference, since with larger energy window and more modes the result is expected to have better accuracy. The relative errors of $I_{48}$ and $I_{76}$ relative to $I_{106}$ are calculated and plotted in Fig. \ref{fig_homoIV_benchmark}(b). It is observed that $I_{48}$ has large deviations with respect to $I_{106}$, especially when current is small. Instead $I_{76}$ is very close to $I_{106}$ and the relative errors are less than 4\% for all bias points, indicating that the results have converged.

\begin{figure}[htbp] \centering
\subfigure[]{\includegraphics[width=5.8cm]{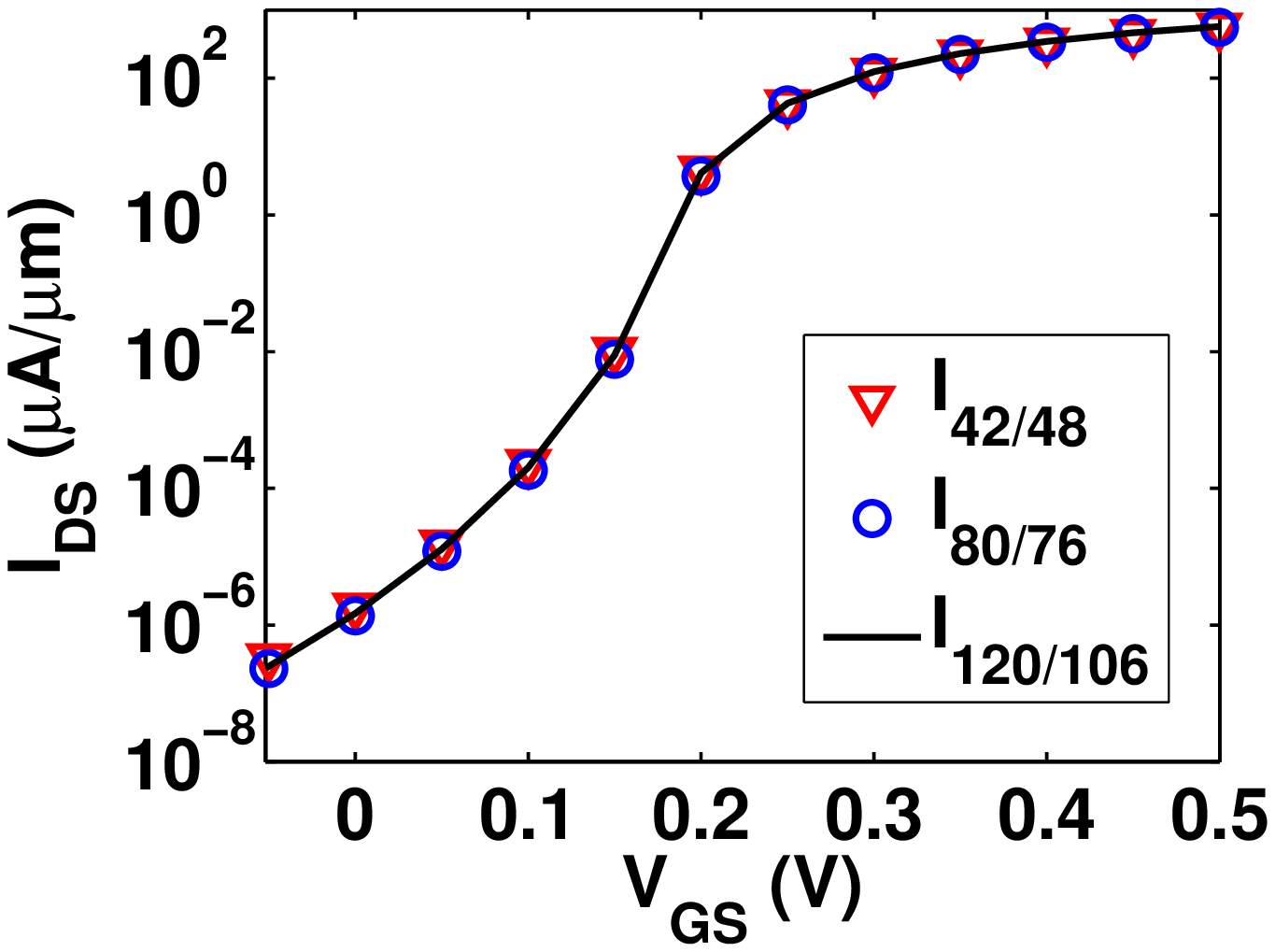}}
\subfigure[]{\includegraphics[width=5.8cm]{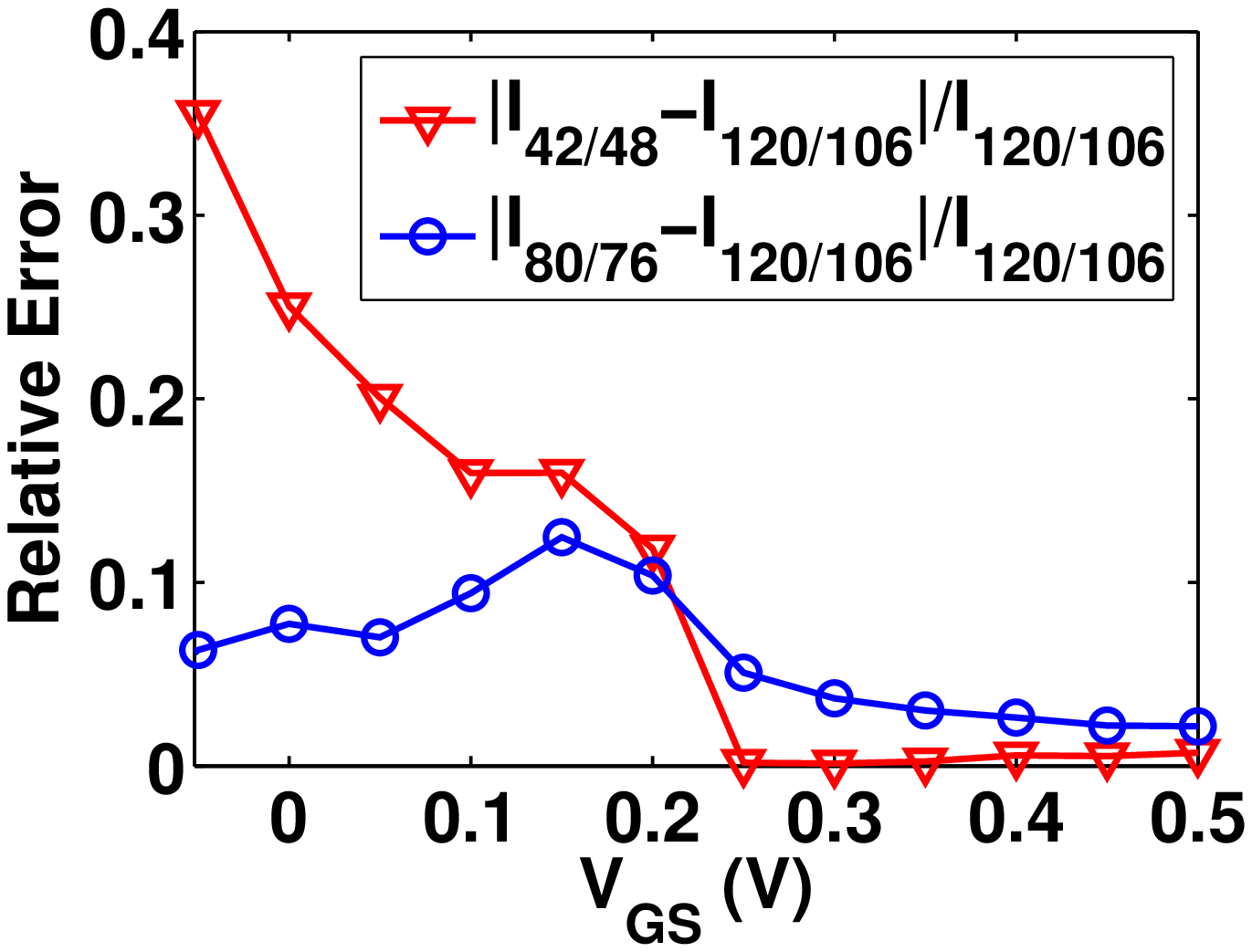}}
\caption{(a) $I_{DS}$-$V_{GS}$ transfer characteristics of a p-i-n GaSb/InAs heterojunction TFET as shown in Fig. 1. The device settings are the same as Fig. \ref{fig_homoIV_benchmark}. (b) Relative errors of $I_{42/48}$ and $I_{80/76}$ with respect to $I_{120/106}$.}
\label{fig_heteroIV_benchmark}
\end{figure}

The $I_{DS}$-$V_{GS}$ transfer characteristics of a $5\rm{nm}\times5\rm{nm}$ cross section GaSb/InAs heterojunction TFET is plotted in Fig. \ref{fig_heteroIV_benchmark}(a). Burt-Foreman operator ordering is used at the material interface though symmetrized ordering gives similar results in this case. Due to the small lattice mismatch between GaSb and InAs, strain is small and is neglected here. Again, three I-V curves are compared. In the first, second, and third $I$-$V$ curve, the energy window is $[E_v-0.3\rm{eV}, E_c+0.4\rm{eV}]$, $[E_v-0.4\rm{eV}, E_c+0.6\rm{eV}]$, and $[E_v-0.5\rm{eV}, E_c+0.8\rm{eV}]$ for GaSb, $[E_v-0.2\rm{eV}, E_c+0.6\rm{eV}]$, $[E_v-0.3\rm{eV}, E_c+0.8\rm{eV}]$, and $[E_v-0.4\rm{eV}, E_c+1.0\rm{eV}]$ for InAs. The sampling k points are all at $k=0$, $\pm\pi/4$, $\pm2\pi/4$, and $\pm3\pi/4$ $[1/\rm{nm}]$. This leads to $\widetilde{N}_m=42$, $\widetilde{N}_m=80$, and $\widetilde{N}_m=120$ for GaSb, $\widetilde{N}_m=48$, $\widetilde{N}_m=76$, and $\widetilde{N}_m=106$ for InAs, with corresponding $I$-$V$ curves denoted as $I_{42/48}$, $I_{80/76}$, and $I_{120/106}$.
The relative errors of $I_{42/48}$ and $I_{80/76}$ relative to $I_{120/106}$ are calculated and plotted in Fig. \ref{fig_heteroIV_benchmark}(b).
It is observed that $I_{42/48}$ has very small deviations with respect to $I_{106}$ when above threshold current, but large errors when below threshold current. In contrast, $I_{80/76}$ has much better accuracy below threshold but larger error above threshold. Overall, the error of $I_{80/76}$ is still acceptable for predictive device modeling.

Table 2 lists the run time details for generating the above $I$-$V$ curves. Note that homogeneous energy mesh with grid size $\Delta E=3meV$ is used which results in 359 energy points in total. Different energy points are calculated in parallel with 12 cores. All the simulations are performed on dual 8-core Intel Xeon-E5 CPUs. It is observed that the simulation time of one NEGF-Poisson iteration increases sub-linearly with $\widetilde{N}_m$; different $\widetilde{N}_m$ leads to small fluctuation of convergence (in terms of number of NEGF-Poisson iterations). In addition, the heterojunction TFET is harder to converge compared with homojunction case. Overall, the simulation time for one $I$-$V$ curve took just a few hours, suitable for device design and optimization.

\begin{table}
\caption{List of run time for the TFET simulations.}
\label{tab:2}       
\begin{tabular}{p{3.5cm}p{1.2cm}p{1.2cm}p{1.2cm}p{1.2cm}p{1.2cm}p{1.2cm}}
\hline\noalign{\smallskip}
$I$-$V$ curves     & $I_{48}$ & $I_{76}$ & $I_{106}$ & $I_{42/48}$ & $I_{80/76}$ & $I_{120/106}$\\
\noalign{\smallskip}\svhline\noalign{\smallskip}
One Iteration (minutes) & 2.38  & 2.96    & 4.09    & 2.47     & 3.02     &  4.21 \\
No. of Iterations       & 41    & 47    & 43    &  87    & 82    &  98 \\
Total (minutes)         & 97.6  & 139.1    & 175.9    &  214.9    & 247.7    &  412.6 \\
\hline\noalign{\smallskip}
\end{tabular}
\end{table}

\section{Simulation Results}
The above benchmarked quantum transport solver is used to study various device configurations as described in Section 2. Homojunction TFETs are simulated first with both n-type and p-type devices considered. Then various performance boosters are applied to the n-type devices, though the same ideas can be applied to p-type devices with qualitatively similar results expected. The device parameters are the same as those in Fig. \ref{fig_homoIV_benchmark} and Fig. \ref{fig_heteroIV_benchmark} if not stated otherwise. In the following discussions, the current is normalized to the width of the nanowire to get unit of $\mu A/\mu m$. We consider high performance (HP), low operating power (LOP), and low standby power (LSTP) applications, where the OFF currents are fixed to $10^{-1}\mu A/\mu m$, $5\times10^{-3}\mu A/\mu m$, and $10^{-5}\mu A/\mu m$, respectively.

\subsection{Homojunction TFETs}
\begin{figure}[htbp] \centering
\subfigure[]{\includegraphics[width=5.8cm]{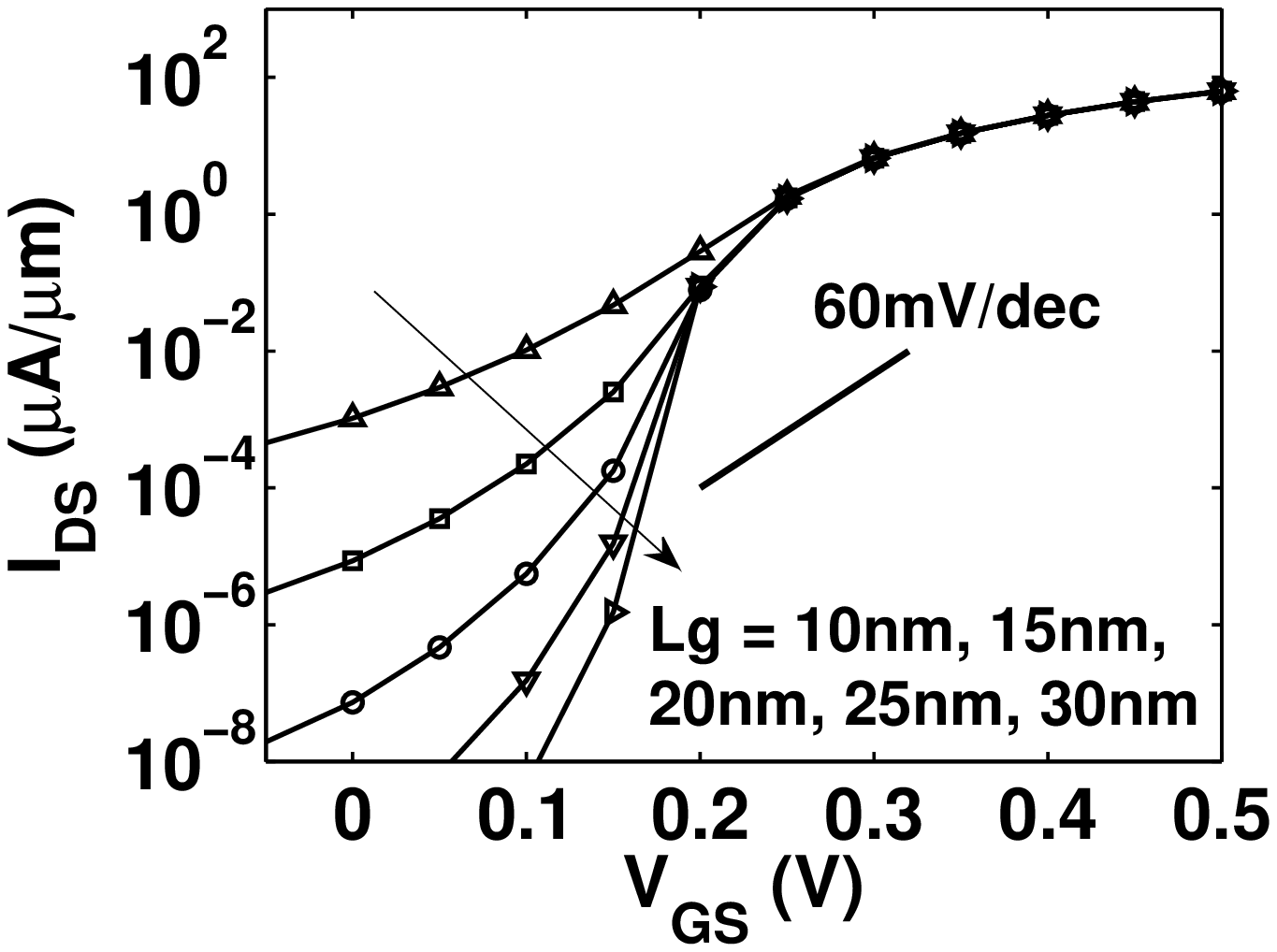}}
\subfigure[]{\includegraphics[width=5.8cm]{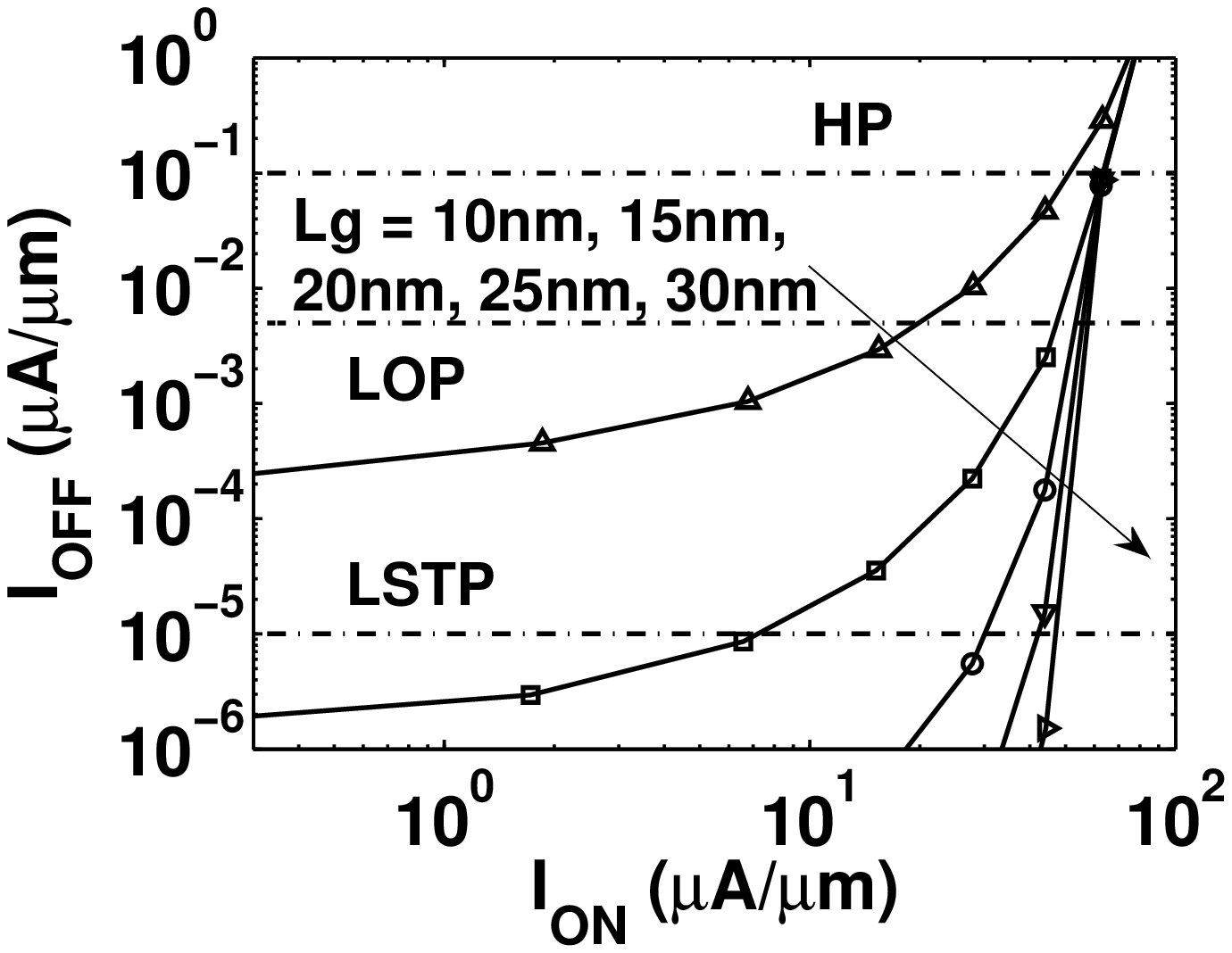}}
\caption{(a) $I_{DS}$-$V_{GS}$ curves (at $V_{DS}=0.3V$) and (b) $I_{ON}$-$I_{OFF}$ (at $V_{DD}=0.3V$) of the $5\rm{nm}\times5\rm{nm}$ cross section InAs nanowire homojunction n-type TFETs in the [100] orientation. Gate lengths of 10nm to 30nm are compared.}
\label{fig_homo_100}
\end{figure}

Fig. \ref{fig_homo_100} (a) compares the $I_{DS}$-$V_{GS}$ curves with different gate lengths. It is found that the SS improves as gate length increases, while the turn-on characteristics remain unchanged. This is understandable since longer gate length has less source-to-drain tunneling leakage. As a result, $I_{ON}$ improves when $I_{OFF}$ is fixed, as seen from Fig. \ref{fig_homo_100} (b). The $I_{ON}$ improvement is the largest for LSTP application and the smallest for the HP application. It is also found that $I_{ON}$ will saturate when gate length becomes very long; the gate length at which $I_{ON}$ saturates is shorter for HP application than for LSTP application. In the following simulations we fix the gate length to be 20nm.

\begin{figure}[htbp] \centering
\subfigure[]{\includegraphics[width=5.8cm]{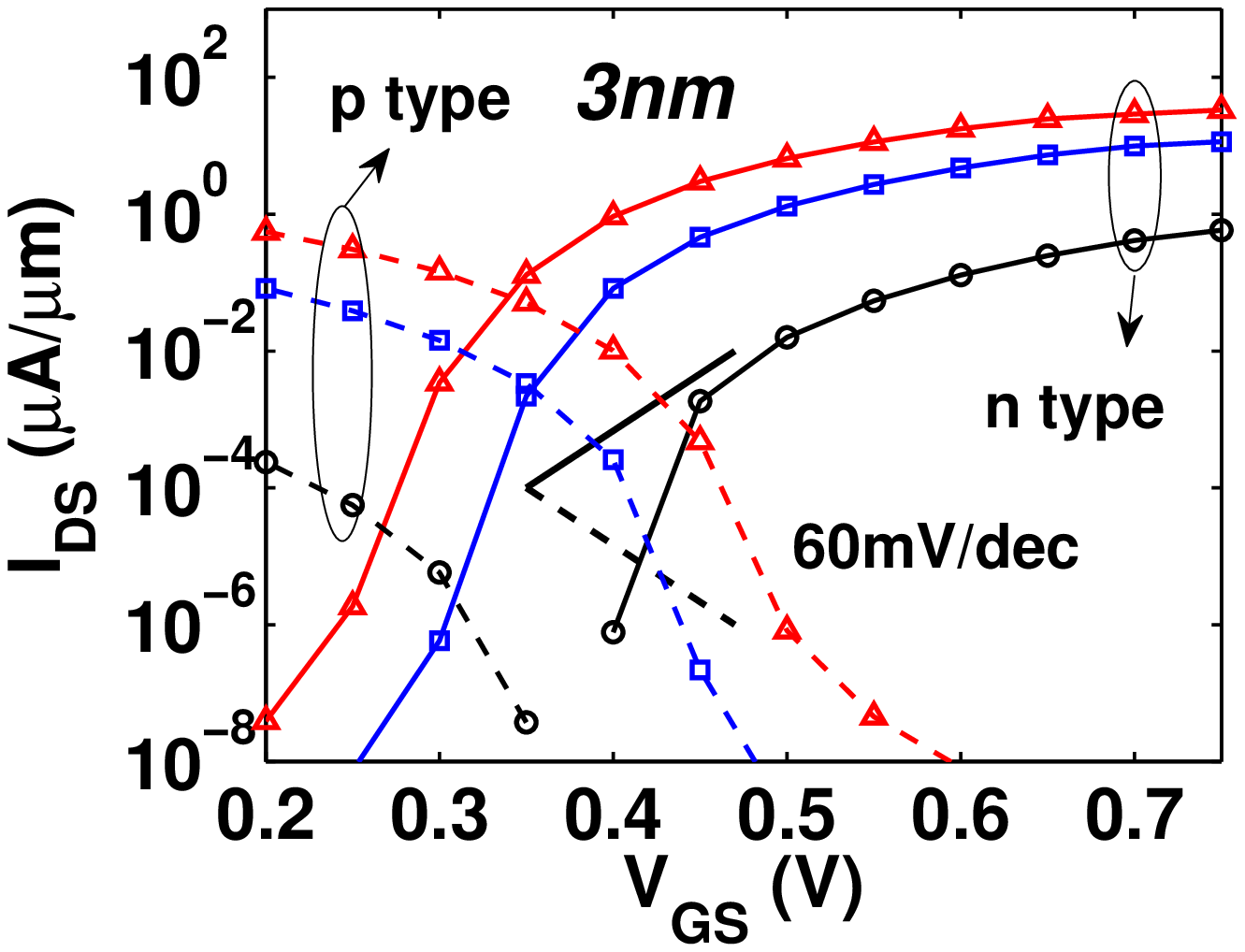}}
\subfigure[]{\includegraphics[width=5.8cm]{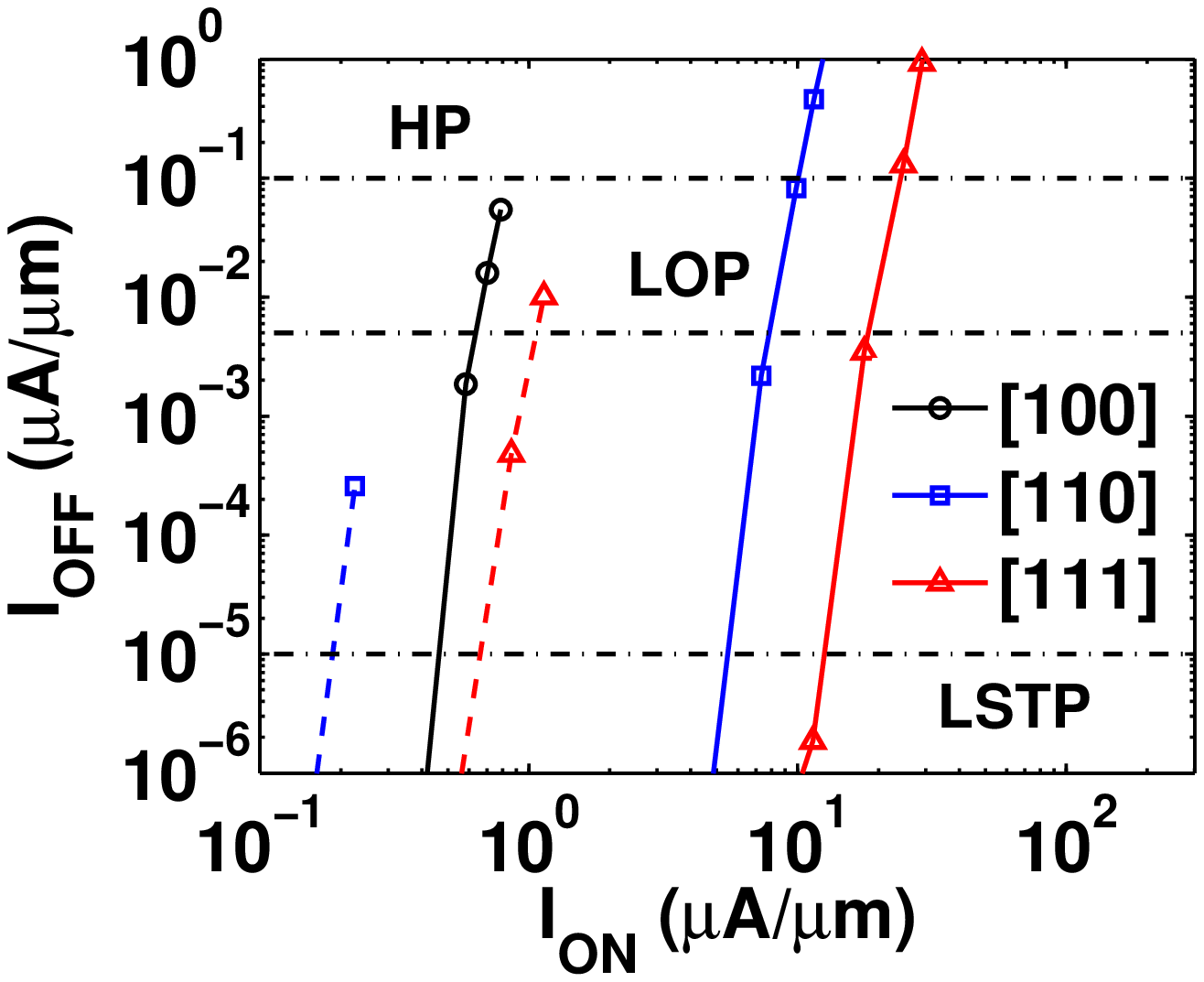}}
\caption{(a) $I_{DS}$-$V_{GS}$ curves (at $V_{DS}=0.3V$) and (b) $I_{ON}$-$I_{OFF}$ (at $V_{DD}=0.3V$) of the $3\rm{nm}\times3\rm{nm}$ cross section n-type and p-type InAs nanowire homojunction TFETs. Three transport orientations [100], [110], and [111] are considered.}
\label{fig_homo_3nm}
\end{figure}
\begin{figure}[htbp] \centering
\subfigure[]{\includegraphics[width=5.8cm]{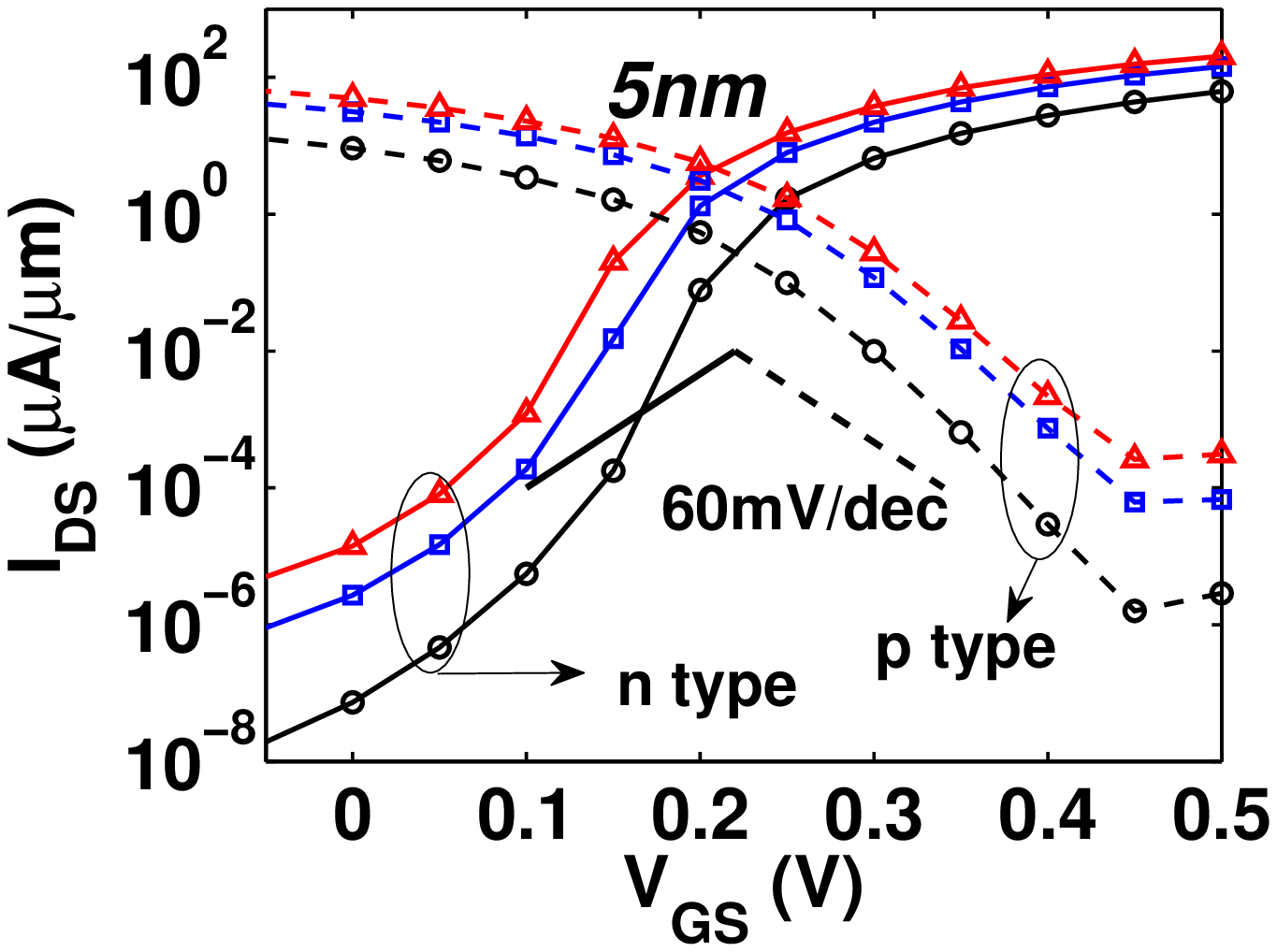}}
\subfigure[]{\includegraphics[width=5.8cm]{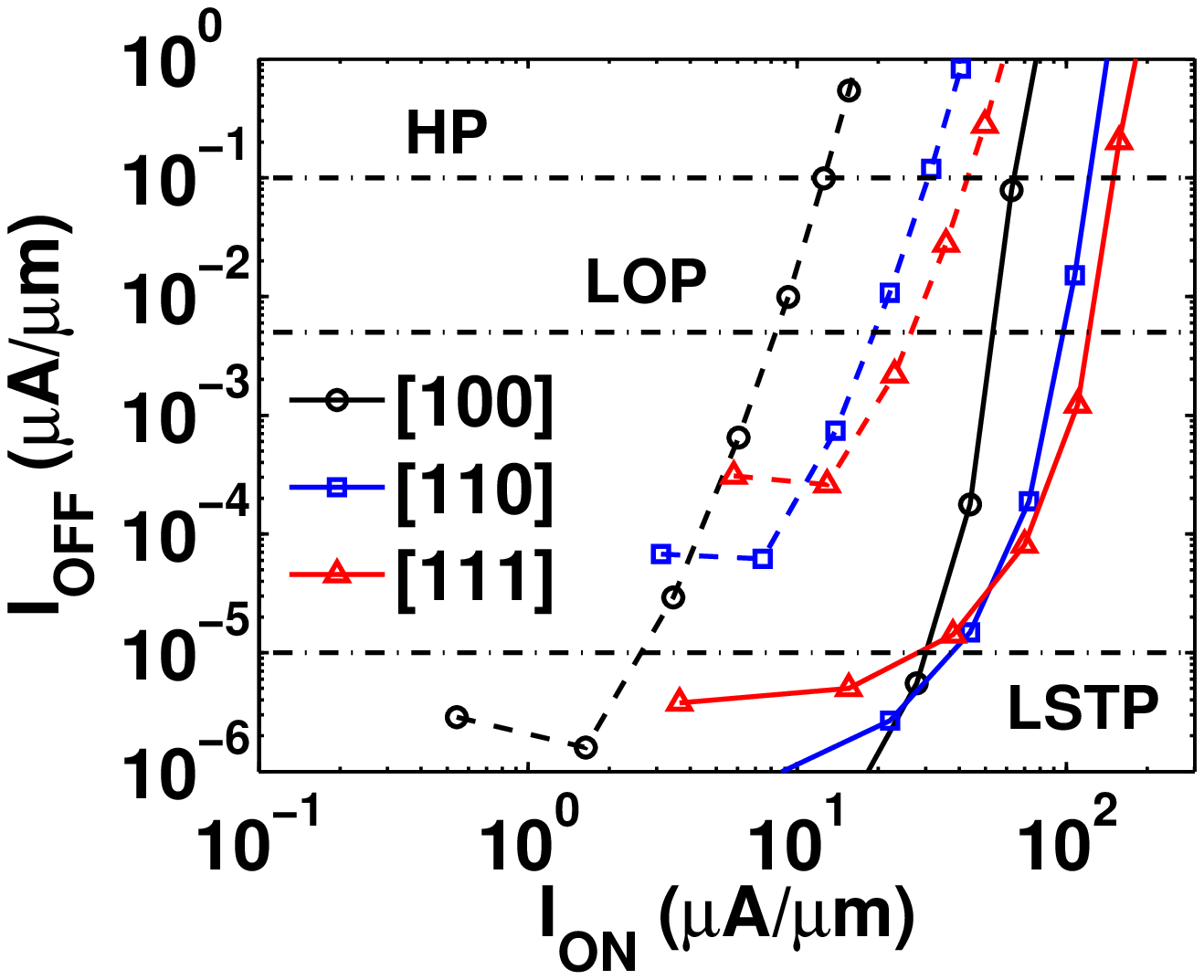}}
\caption{The same as Fig. \ref{fig_homo_3nm} but for $5\rm{nm}\times5\rm{nm}$ cross section.}
\label{fig_homo_5nm}
\end{figure}
\begin{figure}[htbp] \centering
\subfigure[]{\includegraphics[width=5.8cm]{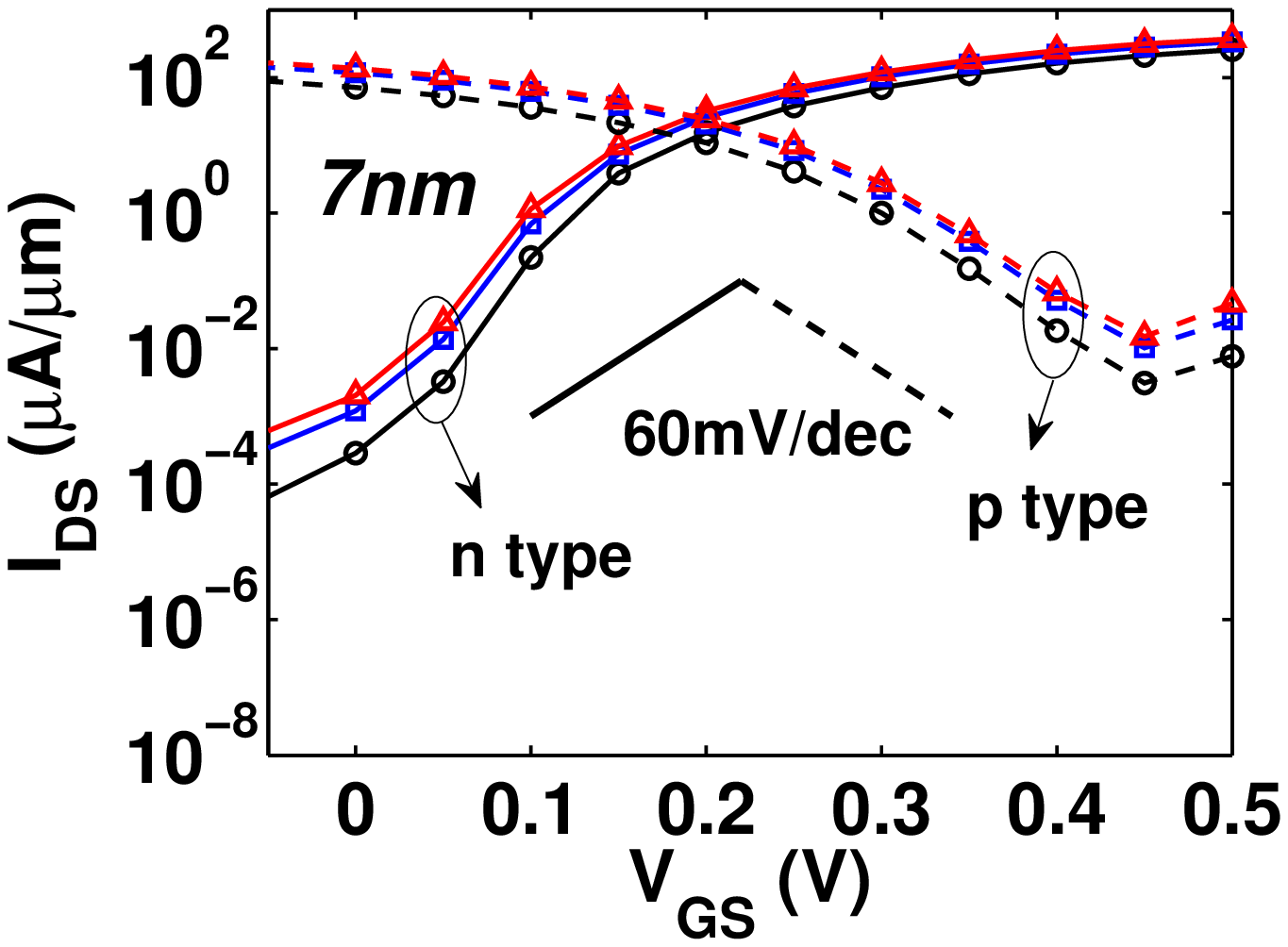}}
\subfigure[]{\includegraphics[width=5.8cm]{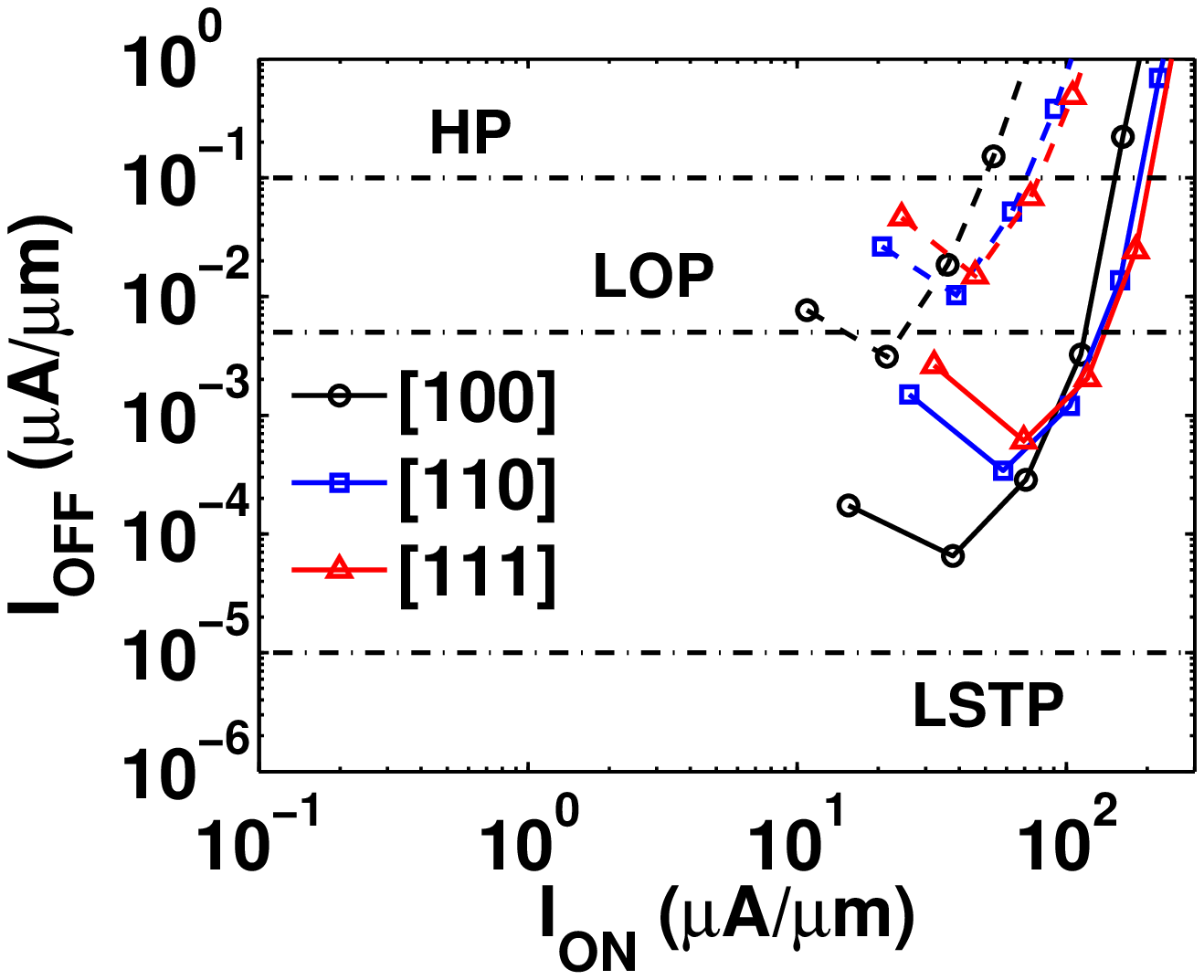}}
\caption{The same as Fig. \ref{fig_homo_3nm} but for $7\rm{nm}\times7\rm{nm}$ cross section.}
\label{fig_homo_7nm}
\end{figure}

Fig. \ref{fig_homo_3nm}, \ref{fig_homo_5nm}, and \ref{fig_homo_7nm} compare the $I_{DS}$-$V_{GS}$ and $I_{ON}$-$I_{OFF}$ characteristics of InAs nanowire TFETs for three cross section sizes and for three transport directions. For p-type devices considered here, the doping density is set to $2\times10^{19}\rm{cm}^{-3}$ at the source and $5\times10^{19}\rm{cm}^{-3}$ at the drain, $L_s=25\rm{nm}$, $L_d=15\rm{nm}$. It is observed that, for small cross sections such as the 3nm case, although the SS are very small the $I_{ON}$ are very limited, for all three orientations. This is due to their large electron effective masses and large band gaps. While for large cross sections such as the 7nm case, the SS degrades due to the smaller electron effective masses, smaller band gaps, as well as the weaker electrostatic control; the $I_{ON}$ however are very good for both HP and LOP applications. It should be noted that $I_{OFF}$ are not sufficiently small which makes them unsuitable for LSTP application. The large $I_{OFF}$ are due to the direct source-to-drain tunneling leakage and ambipolar tunneling leakage at the channel-drain junction, these two components become more pronounced when band gap becomes smaller. Therefore, for LSTP application, medium sized cross section such as the 5nm case should be a better choice; otherwise, the channel length needs to be increased and/or the drain doping density needs to be decreased to suppress the leakage.

For small wire cross section, the performances of the three orientations differ a lot. In particular, for the 3nm case, [111] orientation gives the best $I_{ON}$ for all three applications while [100] is the worst. When the cross section size increases, the three orientations tend to deliver similar performances. It also means that [111] orientation has the best cross-section scaling ability. In fact, when the confinement becomes stronger (as the nanowire size decreases) the band structure starts to differ from each other for the three orientations, as can be observed in Fig. \ref{fig_5nm_wire}, Fig. \ref{fig_3nm_wire}, and Fig. \ref{fig_7nm_wire}. In particular, the [100] orientation shows the fastest increase of band gap, while the [111] case shows the slowest increase of band gap meanwhile the hole effective mass decreases (as a result of strongly anisotropic heavy hole band).

\begin{figure}[htbp] \centering
\subfigure[]{\includegraphics[width=5.8cm]{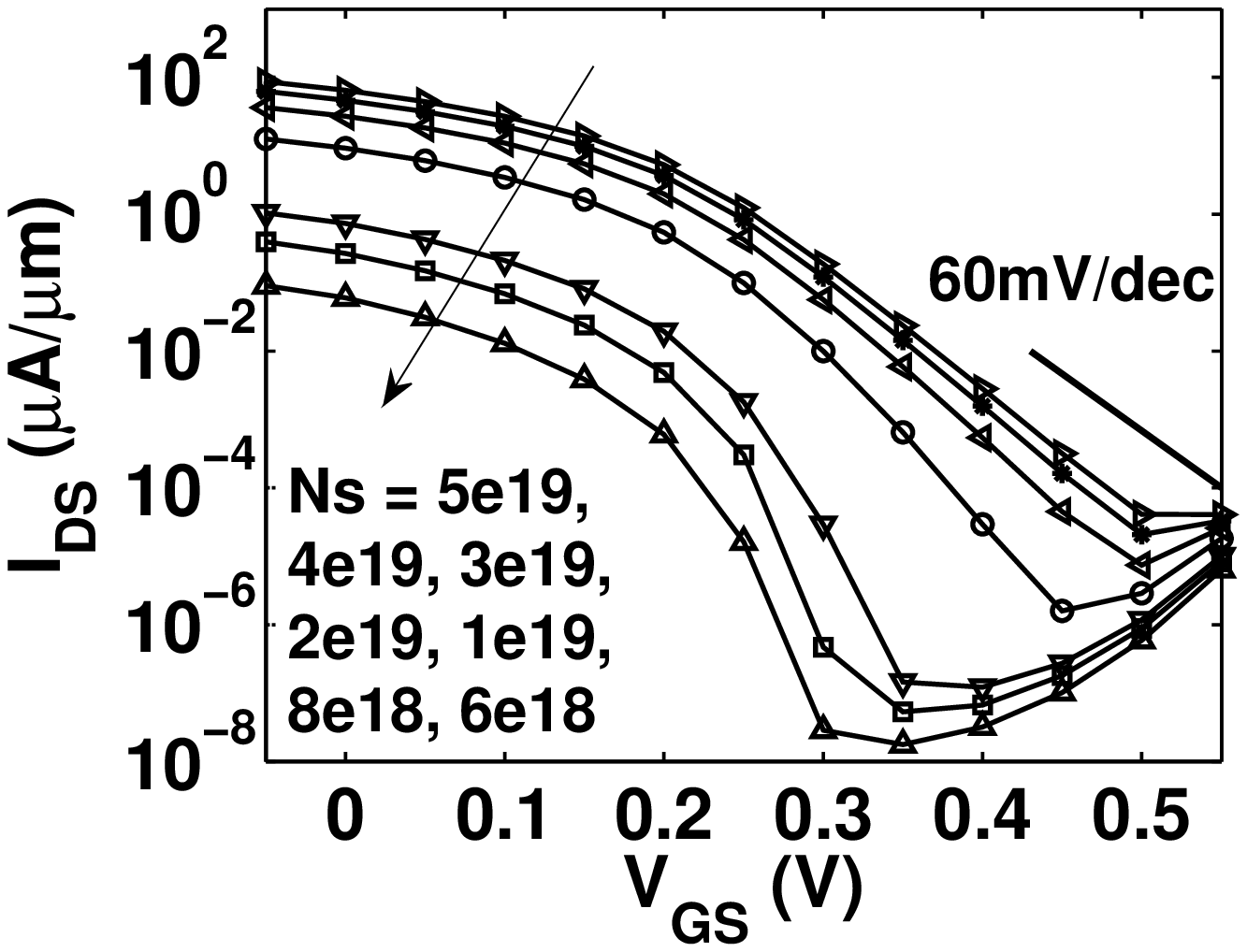}}
\subfigure[]{\includegraphics[width=5.8cm]{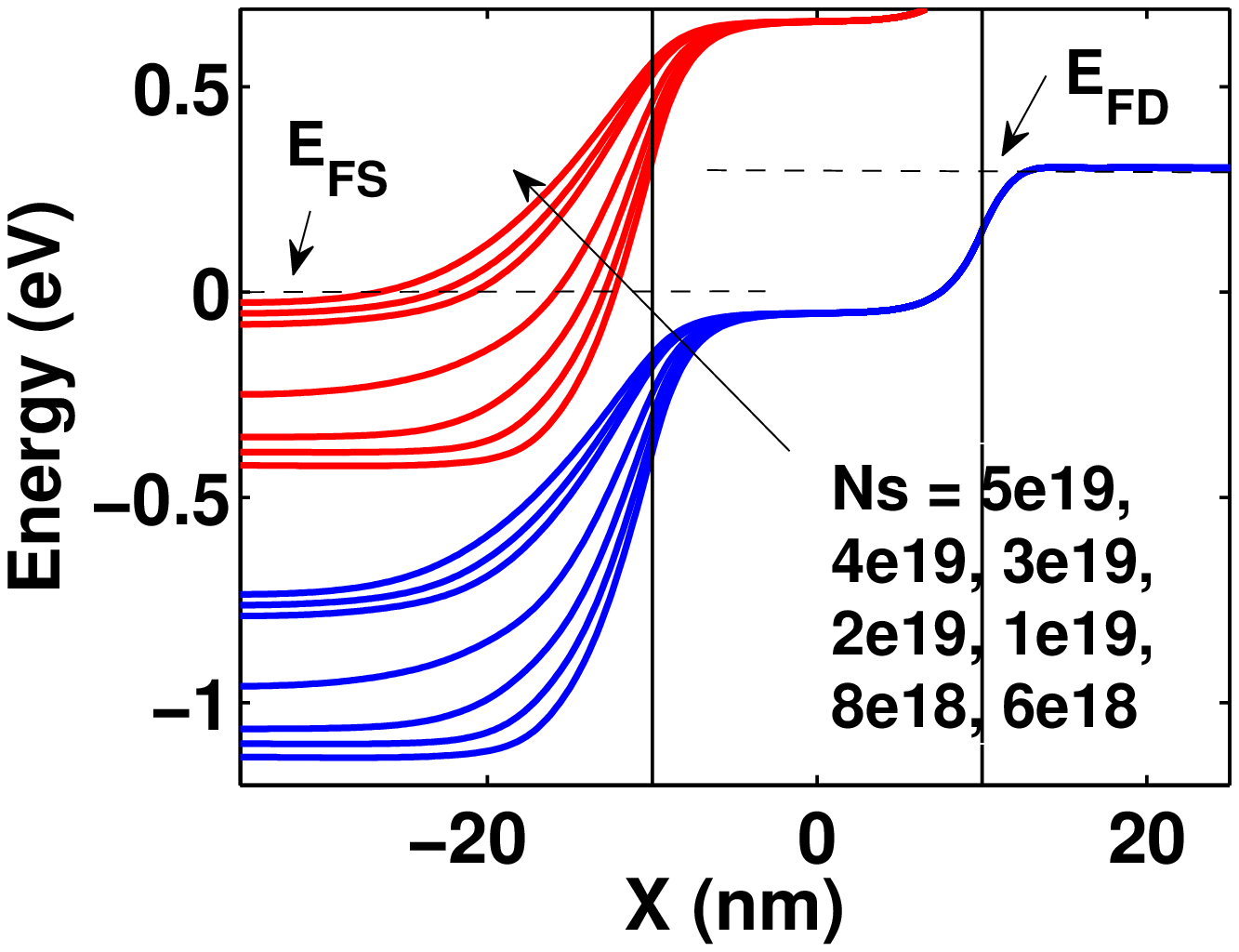}}
\caption{(a) $I_{DS}$-$V_{GS}$ curves (at $V_{DS}=0.3V$) and (b) potential profiles (at $V_{DS}=0.3V$ and $V_{GS}=0.3V$) of the $5\rm{nm}\times5\rm{nm}$ cross section InAs nanowire homojunction p-type TFETs in the [100] orientation. Source doping densities of $5\times10^{19}\rm{cm}^{-3}$ to $6\times10^{18}\rm{cm}^{-3}$ are compared.}
\label{fig_homo_doping}
\end{figure}

Comparing n-type and p-type devices, we found that p-type devices have worse SS and smaller $I_{ON}$. This is because the doping density has been set to $2\times10^{19}\rm{cm}^{-3}$ at the source side, lower than that of n-type ones (which is $5\times10^{19}\rm{cm}^{-3}$). This doping density is a compromise of SS and $I_{ON}$. In fact, as shown in Fig. \ref{fig_homo_doping}, lower doping leads to smaller $I_{ON}$ as a result of less abrupt tunneling junction. While higher doping leads to worse SS (approaching 60mV/dec) since larger Fermi degeneracy is created in the conduction band. The Fermi degeneracy creates thermal tail which counteracts the energy filtering functionality of TFETs. For 3nm (7nm) p-type TFETs here, smaller (larger) Fermi degeneracy in the source is observed because the electron mass and density of states increases (decreases) as cross section decreases (increases) (Fig. \ref{fig_band_edge_mass_wire}).

\begin{figure}[htbp] \centering
\subfigure[]{\includegraphics[width=5.8cm]{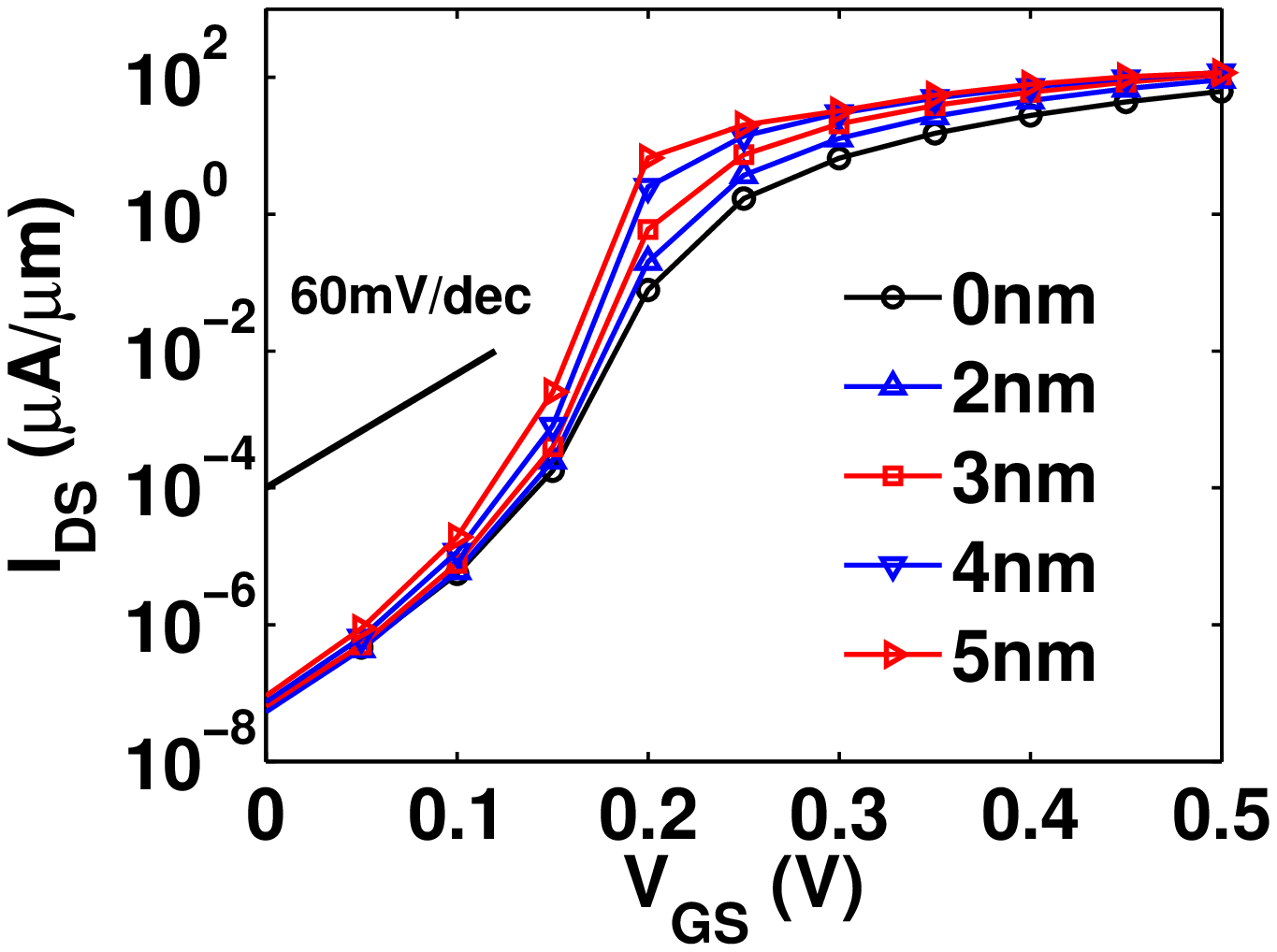}}
\subfigure[]{\includegraphics[width=5.8cm]{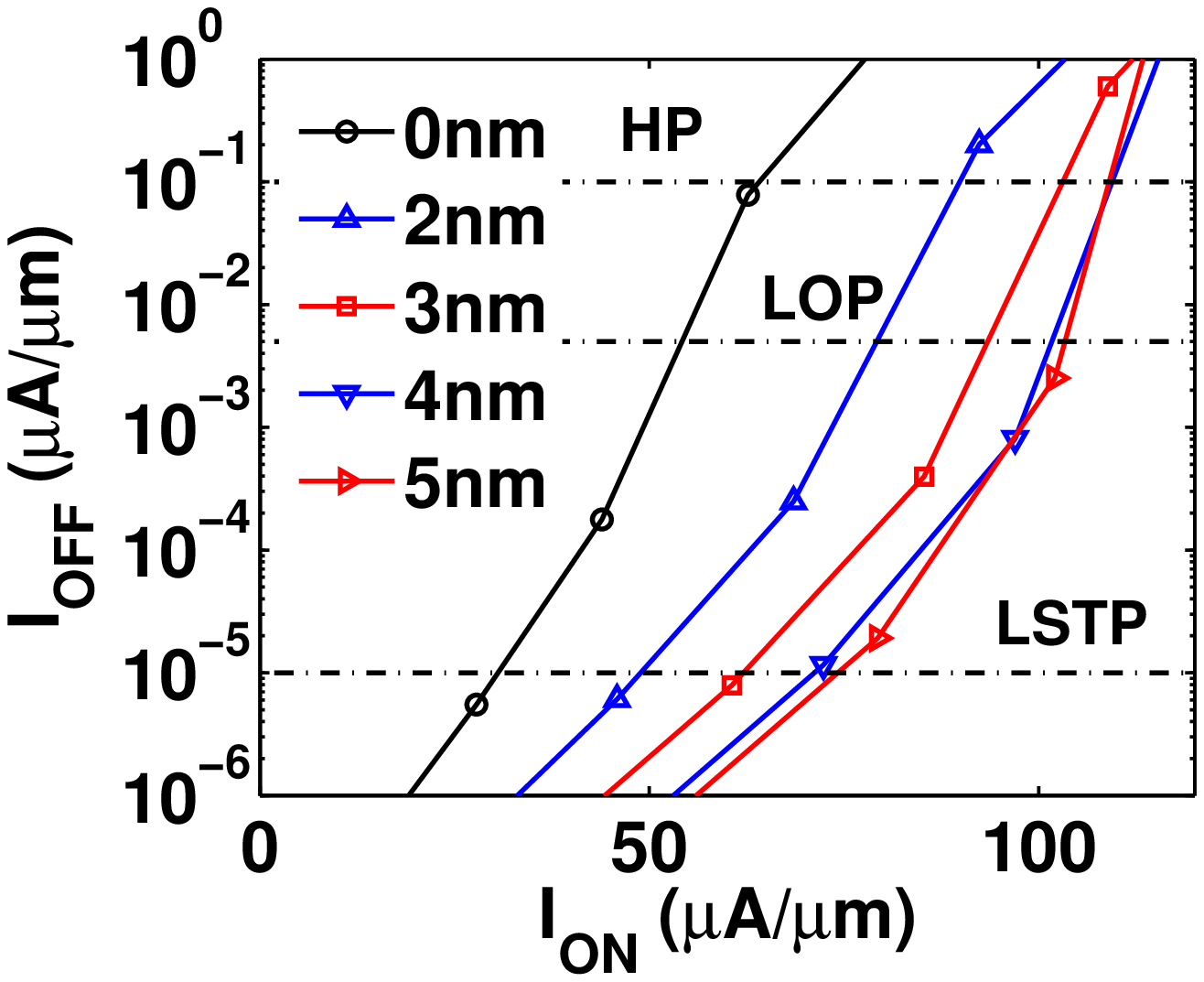}}
\subfigure[]{\includegraphics[width=5.8cm]{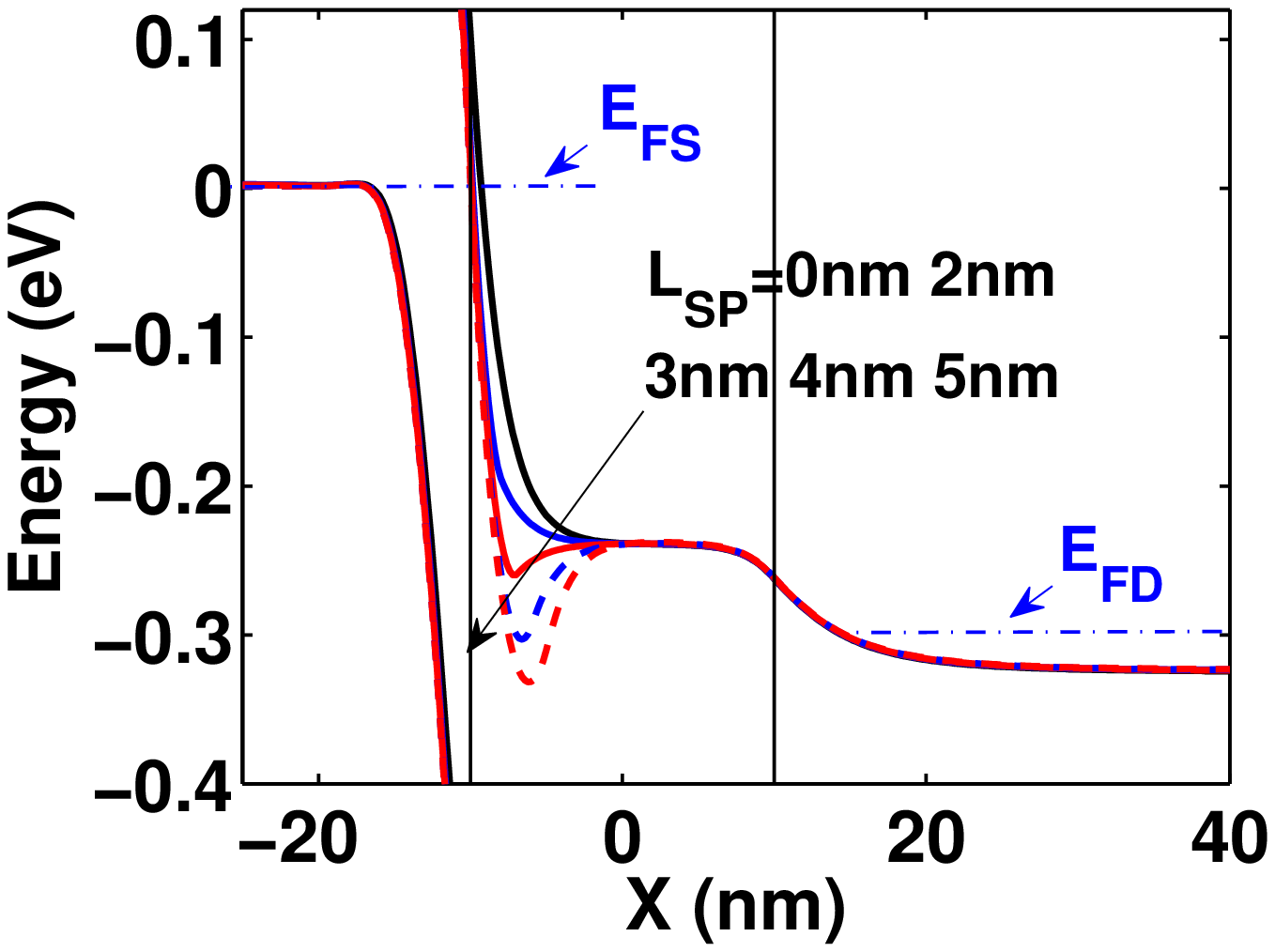}}
\subfigure[]{\includegraphics[width=5.8cm]{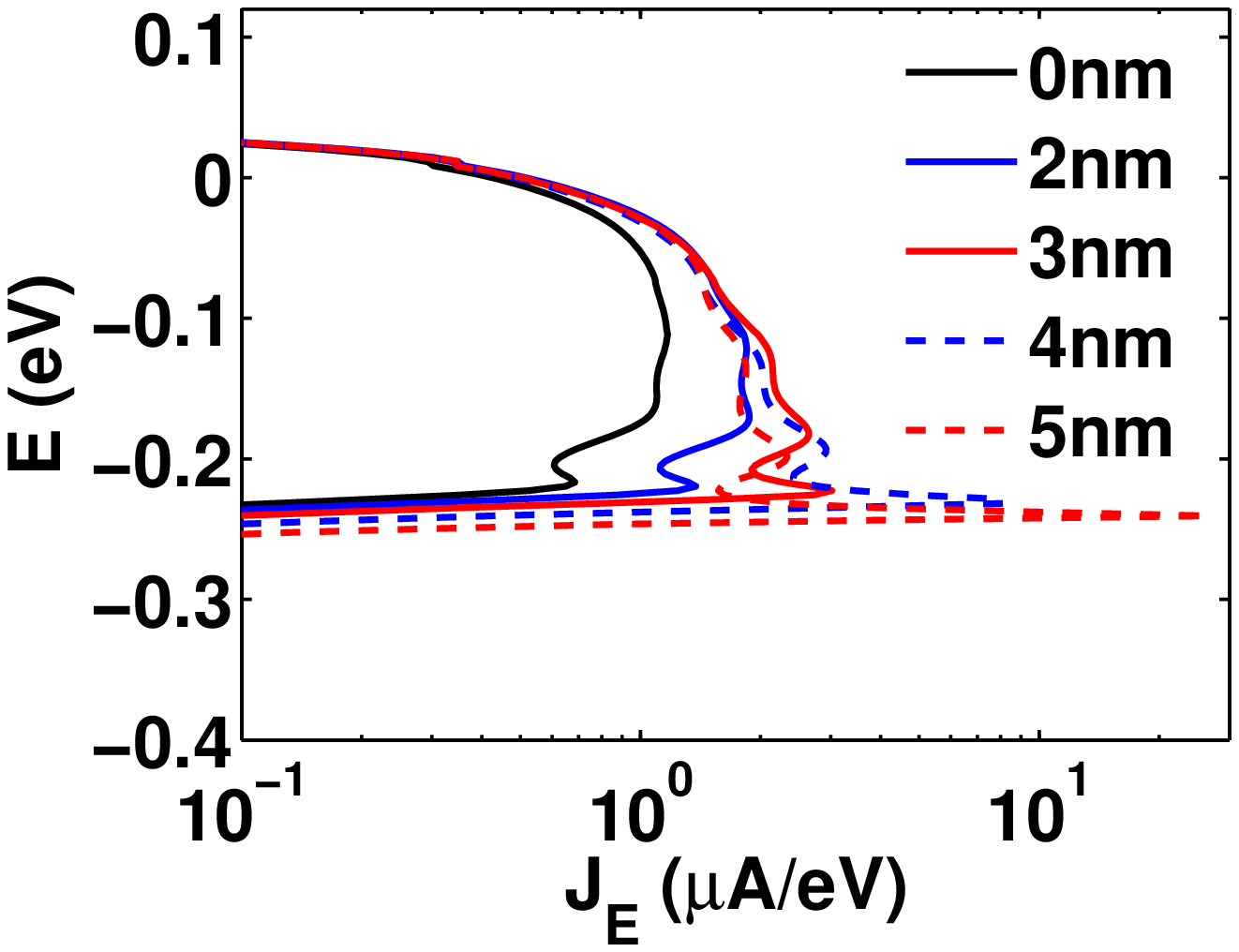}}
\caption{(a) $I_{DS}$-$V_{GS}$ curves (at $V_{DS}=0.3V$), (b) $I_{ON}$-$I_{OFF}$ (at $V_{DD}=0.3V$), (c) Potential profiles (at $V_{DS}=0.3V$ and $V_{GS}=0.45V$), and (d) current spectra (at $V_{DS}=0.3V$ and $V_{GS}=0.45V$), of the $5\rm{nm}\times5\rm{nm}$ cross section source-pocket InAs homojunction TFETs in the [100] orientation, compared with no pocket case. The doping density of the pocket is $N_{sp}=5\times10^{19}\rm{cm}^{-3}$. Pocket lengths of 2nm to 5nm are considered.}
\label{fig_homo_100_sp}
\end{figure}

\begin{figure}[htbp] \centering
\subfigure[]{\includegraphics[width=5.8cm]{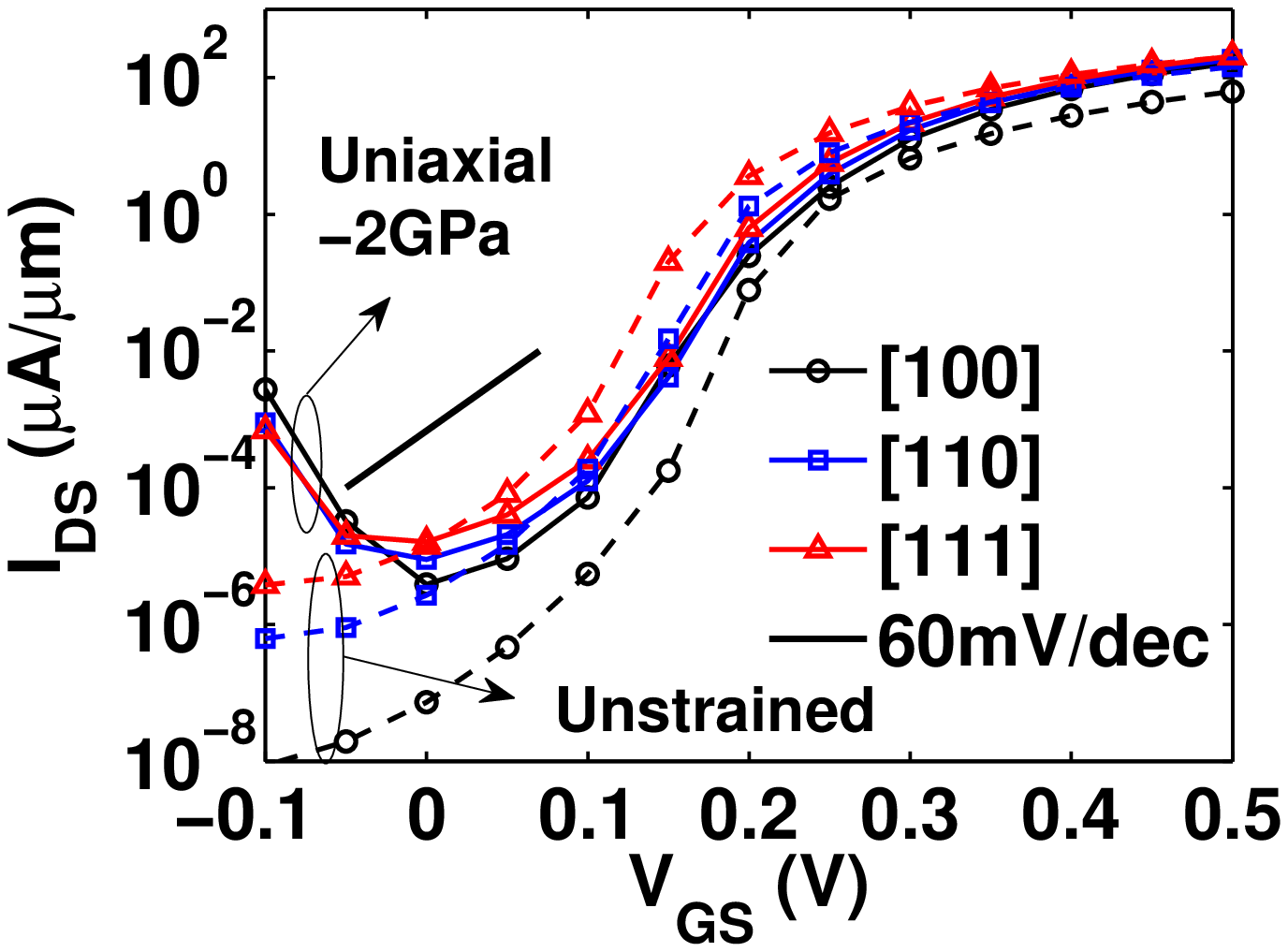}}
\subfigure[]{\includegraphics[width=5.8cm]{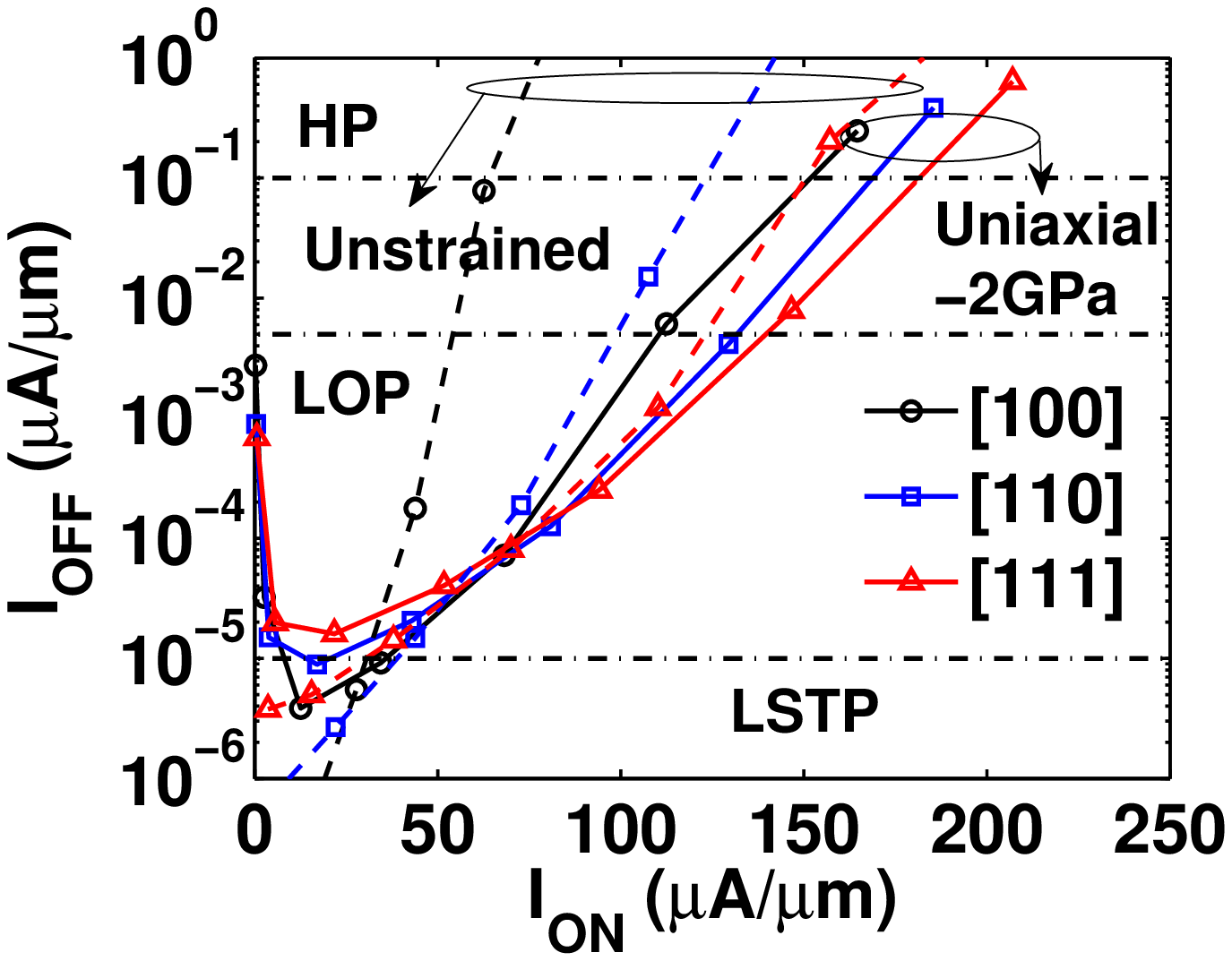}}
\subfigure[]{\includegraphics[width=5.8cm]{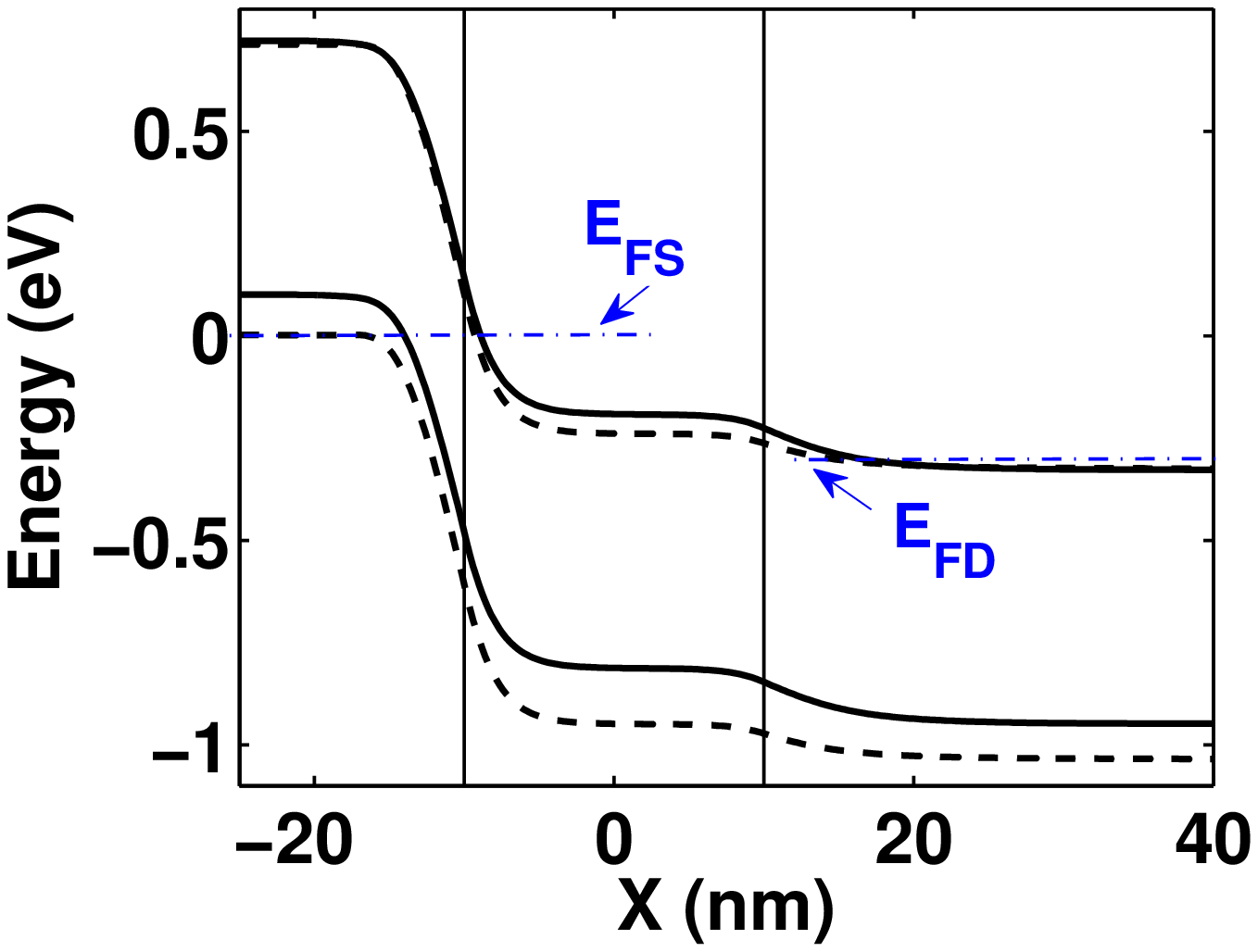}}
\subfigure[]{\includegraphics[width=5.8cm]{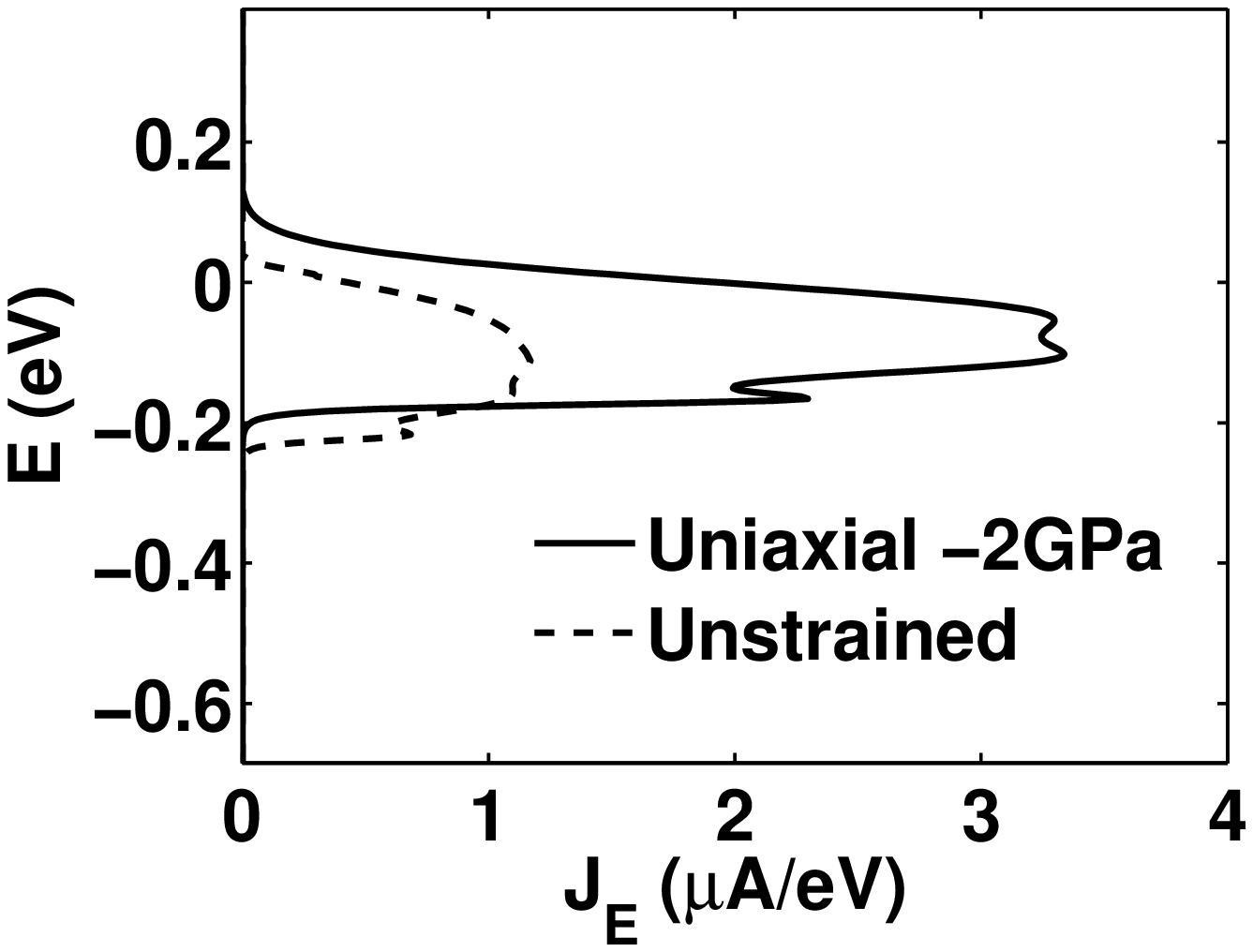}}
\caption{(a) $I_{DS}$-$V_{GS}$ curves (at $V_{DS}=0.3V$) and (b) $I_{ON}$-$I_{OFF}$ (at $V_{DD}=0.3V$) of the $5\rm{nm}\times5\rm{nm}$ cross section homojunction TFETs in the [100], [110], and [111] orientations, with uniaxial compressive stress along the transport direction, in comparison with unstrained cases. (c) Potential profiles (at $V_{DS}=0.3V$ and $V_{GS}=0.45V$) and (d) current spectra (at $V_{DS}=0.3V$ and $V_{GS}=0.45V$), of the 5nm x 5nm cross section homojunction TFET in the [100] orientation, with uniaxial compressive stress along the transport direction, in comparison with unstrained case.}
\label{fig_homo_strained}
\end{figure}

\subsection{Improvements of Homojunction TFETs}
As shown in Fig. \ref{fig_homo_100_sp} (a) and (b), the source-pocket TFETs can improve $I_{ON}$ for all HP, LOP, and LSTP applications, by up to $50\mu A/\mu m$. With 2nm to 5nm pocket lengths, $I_{ON}$ first increases and then saturates. Further increasing the pocket length will decrease $I_{ON}$ (not shown here). This can be explained by plotting the band diagram and current spectra, as shown in Fig. \ref{fig_homo_100_sp} (c) and (d). With 2nm to 5nm pocket lengths, the source pocket first increases the electric field across the source-channel tunneling junction and then starts to form a quantum well. This quantum well creates a resonant state leading to a very sharp tunneling peak. However, this peak is too narrow in energy to help the total current.

It has been shown in \cite{conzatti2011simulation} that uniaxial compressive stress and biaxial tensile stress reduce InAs nanowire band gap and effective masses, which can be used to improve TFET performances. As shown in Fig. \ref{fig_homo_strained}, the uniaxial compressive stress degrades the SS, but still improved $I_{ON}$ is observed for both HP and LOP applications, consistent with \cite{conzatti2011simulation}. The degraded SS can be explained by the large Fermi degeneracy of the source (due to lighter hole effective mass) creating thermal tail. It is found here that the strain induced $I_{ON}$ improvement is more significant in the [100] orientation than in the [110] and [111] orientations, since the strain induced band gap and effective mass reductions are more pronounced in the [100] orientation. On the other hand, uniaxial tensile stress leads to increased band gap and effective masses and thus degraded $I_{ON}$ (no shown here). Biaxial strain in the cross-section plane is hard to realize in experiments and therefore is not considered here.

\subsection{Heterojunction TFETs}
\begin{figure}[htbp] \centering
\subfigure[]{\includegraphics[width=5.8cm]{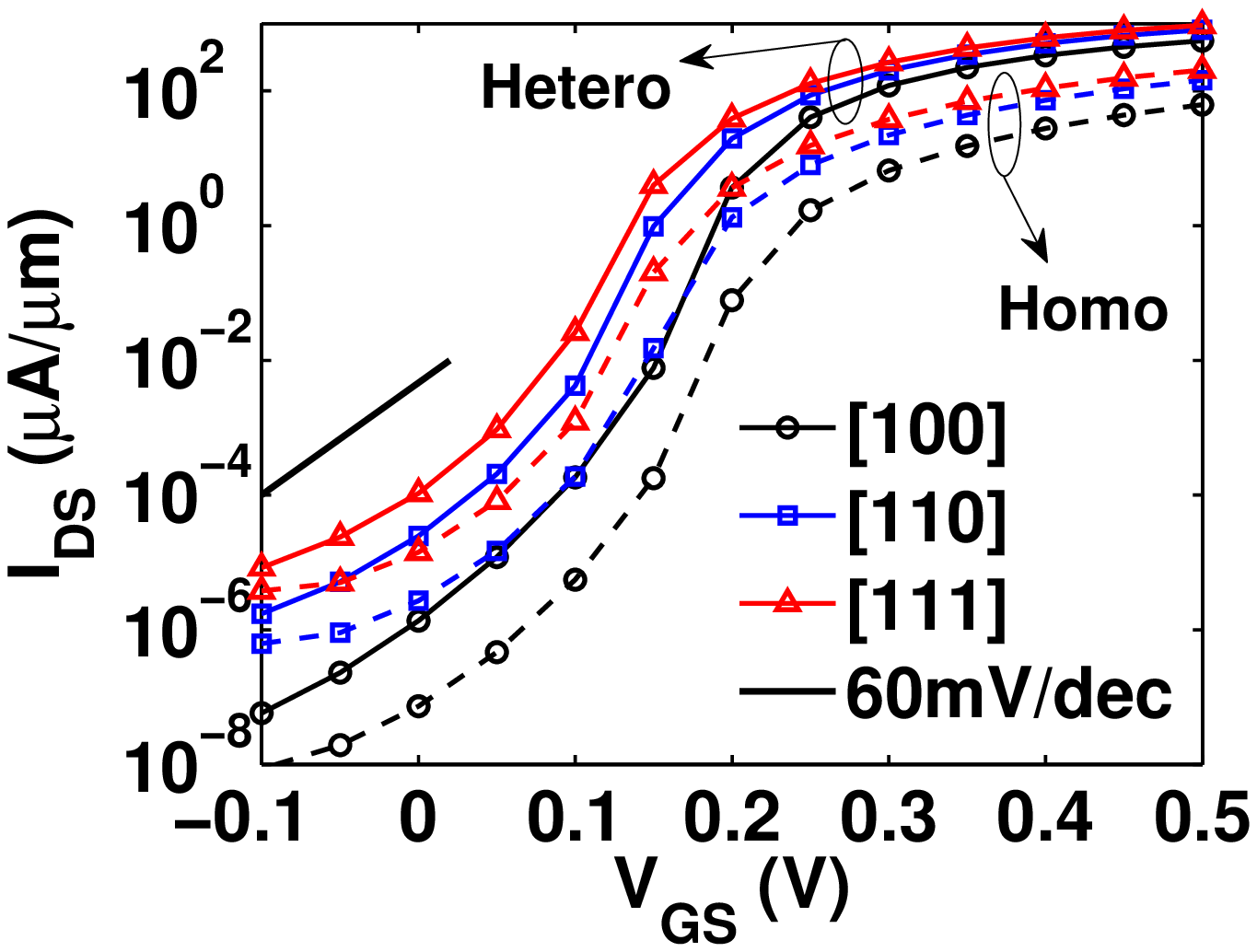}}
\subfigure[]{\includegraphics[width=5.8cm]{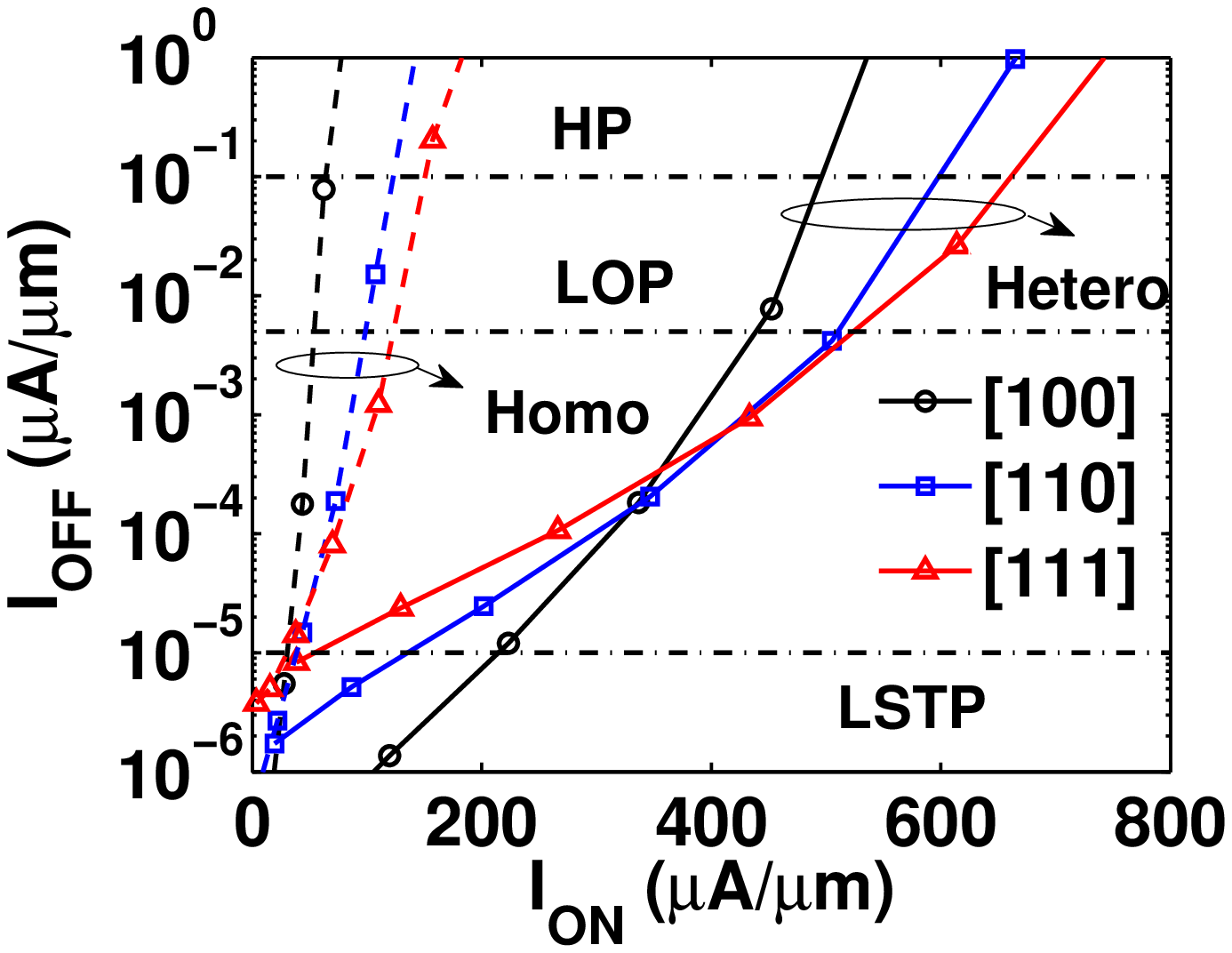}}
\subfigure[]{\includegraphics[width=5.8cm]{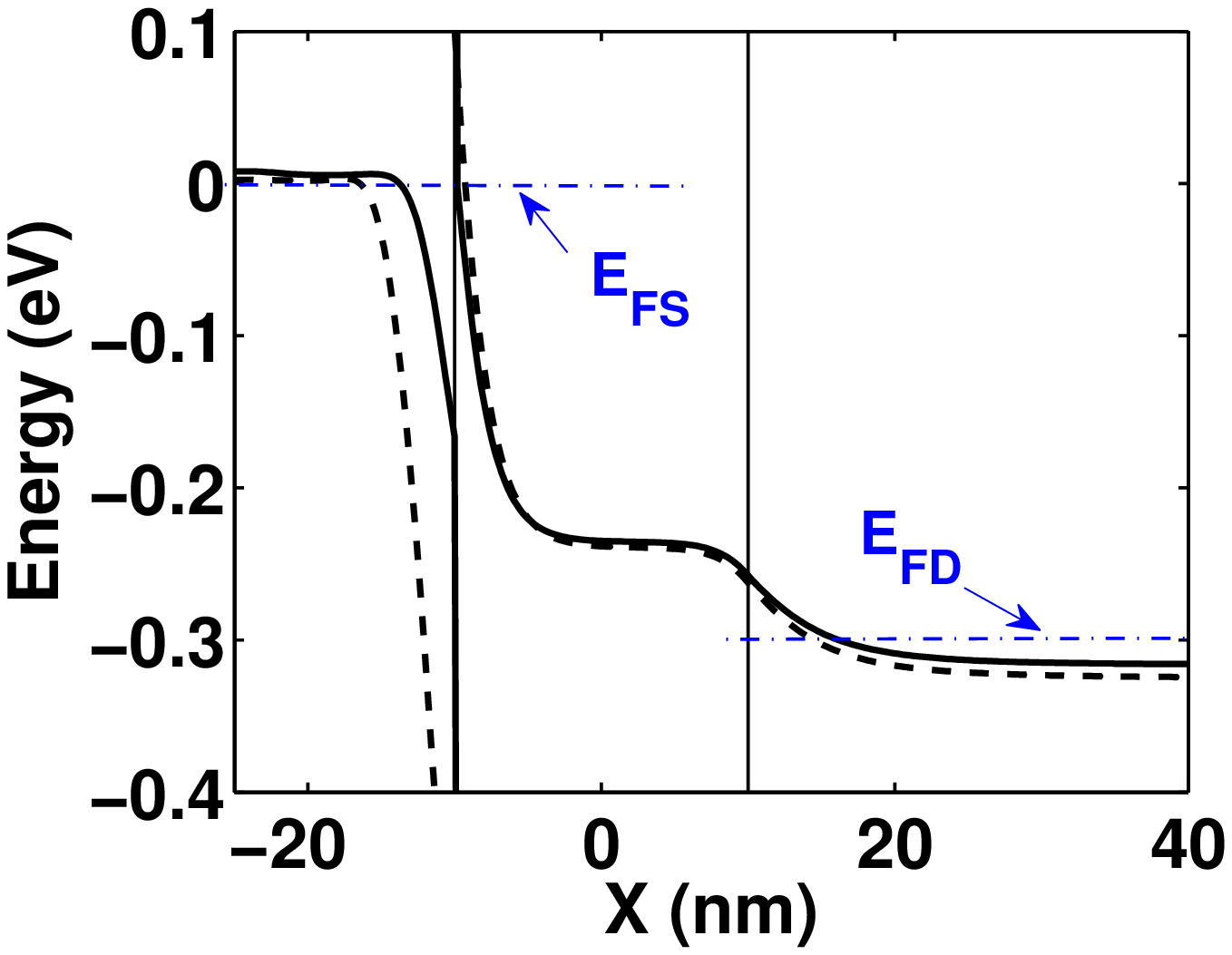}}
\subfigure[]{\includegraphics[width=5.8cm]{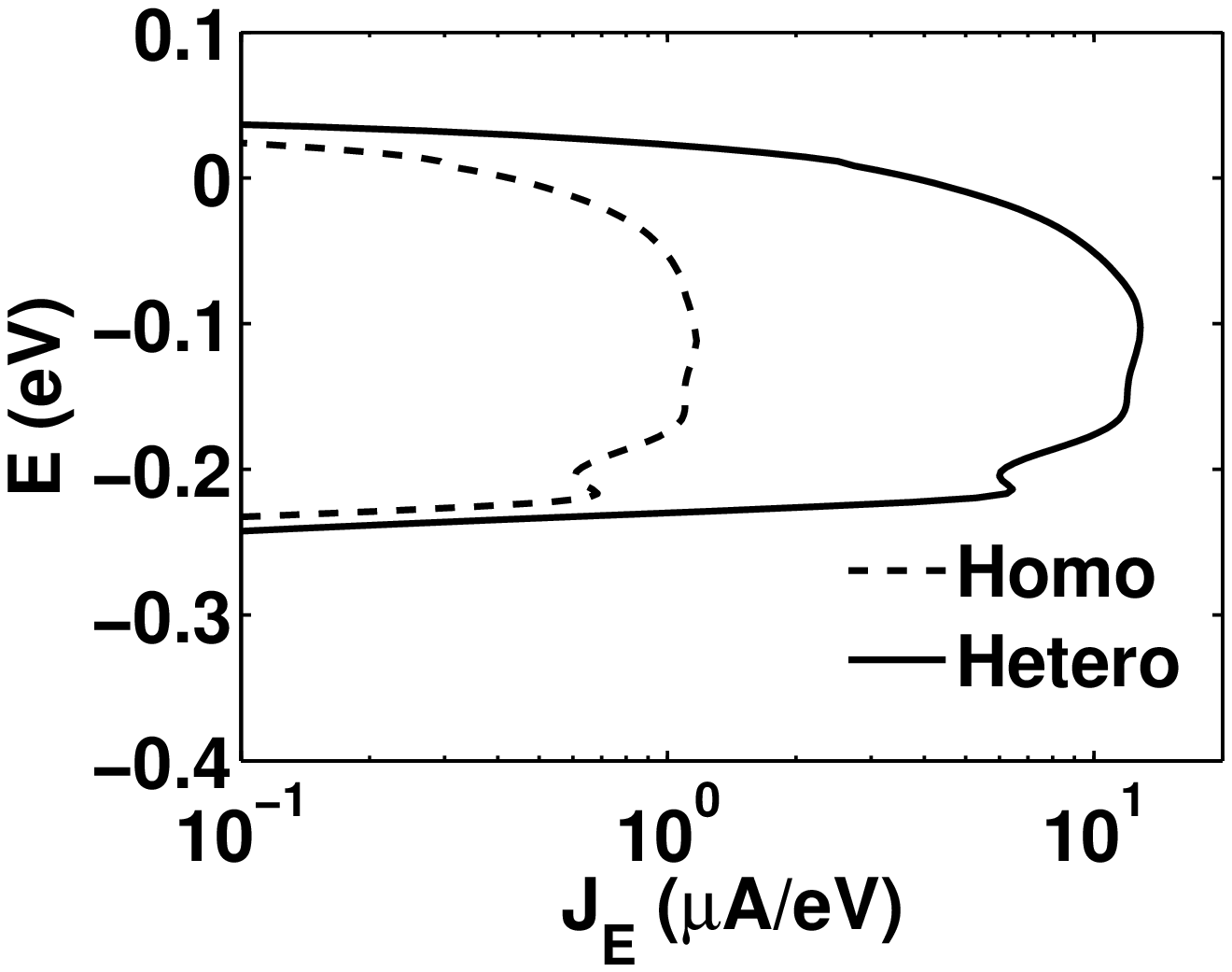}}
\caption{(a) $I_{DS}$-$V_{GS}$ curves (at $V_{DS}=0.3V$) and (b) $I_{ON}$-$I_{OFF}$ (at $V_{DD}=0.3V$) of the $5\rm{nm}\times5\rm{nm}$ cross section GaSb/InAs heterojunction TFETs in the [100], [110], and [111] orientations, compared with the homojunction cases. (c) Potential profiles (at $V_{DS}=0.3V$ and $V_{GS}=0.45V$) and (b) current spectra (at $V_{DS}=0.3V$ and $V_{GS}=0.45V$), of the 5nm x 5nm cross section GaSb/InAs heterojunction TFETs in the [100] orientation, compared with the homojunction case.}
\label{fig_hetero_100}
\end{figure}

As shown in Fig. \ref{fig_hetero_100} (a) and (b), the GaSb/InAs heterojunction TFETs significantly improve $I_{ON}$ for all HP, LOP, and LSTP applications. [111] orientation gives the best $I_{ON}$ for both HP and LOP applications, while [100] gives the best $I_{ON}$ for LSTP application. In Fig. \ref{fig_hetero_100} (c) and (d), we compare the band diagrams and current spectra of the GaSb/InAs heterojunction TFET with the InAs homojunction TFET. It is clear that the smaller tunneling height and distance of the heterojunction TFET, in particular at the GaSb side, leads to around $10\times$ larger tunneling current.

\subsection{Improvements of Heterojunction TFETs}
Employing the schemes for improving $I_{ON}$ of homojunction TFETs, we get source-pocket heterojunction TFETs or strained heterojunction TFETs, which are expected to deliver even larger $I_{ON}$.

\begin{figure}[htbp] \centering
\subfigure[]{\includegraphics[width=5.8cm]{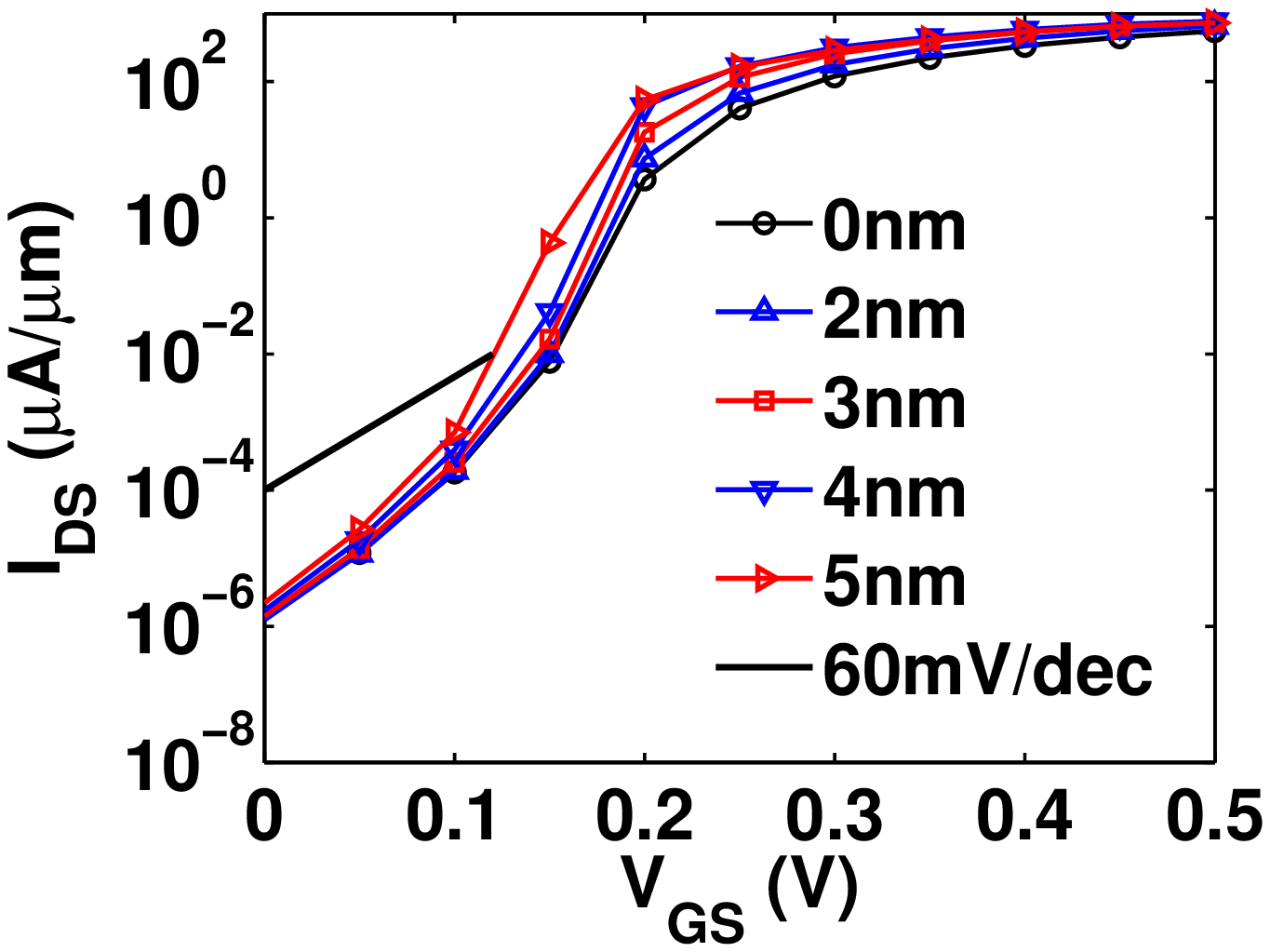}}
\subfigure[]{\includegraphics[width=5.8cm]{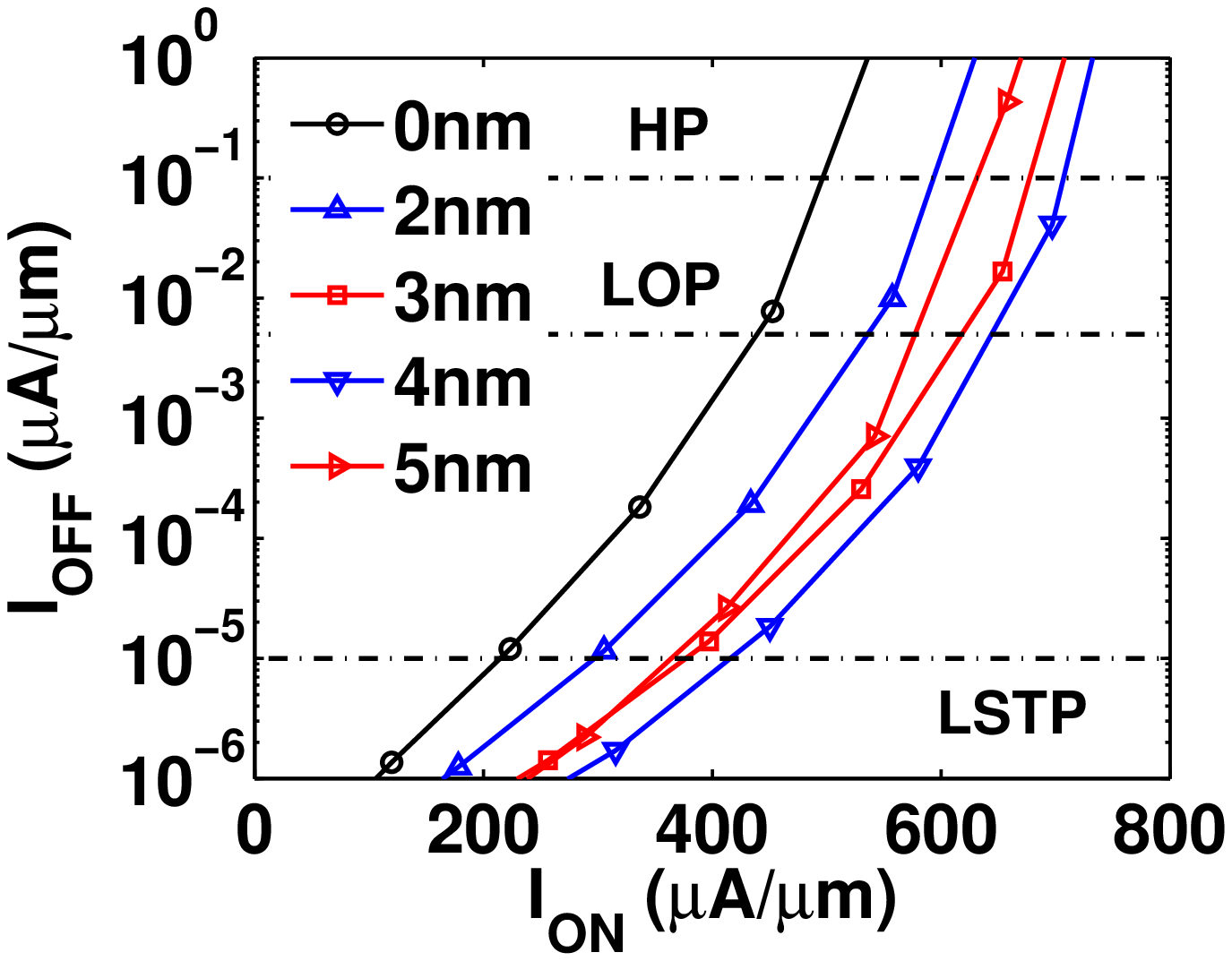}}
\caption{(a) $I_{DS}$-$V_{GS}$ curves (at $V_{DS}=0.3V$) and (b) $I_{ON}$-$I_{OFF}$ (at $V_{DD}=0.3V$), of the $5\rm{nm}\times5\rm{nm}$ cross section source-pocket GaSb/InAs heterojunction TFETs in the [100] orientation, compared with no pocket case. The doping density of the pocket is $N_{sp}=5\times10^{19}\rm{cm}^{-3}$. Pocket lengths of 2nm to 5nm are considered.}
\label{fig_hetero_100_sp}
\end{figure}

Indeed, as shown in Fig. \ref{fig_hetero_100_sp} (a) and (b), the source pocket can improve $I_{ON}$ for all HP, LOP, and LSTP applications, by up to $200\mu A/\mu m$. The optimal pocket length is found to be around 4nm, beyond which $I_{ON}$ will drop. The physics is similar to the source-pocket homojunction TFETs and will not be repeated here.

\begin{figure}[htbp] \centering
\subfigure[]{\includegraphics[width=5.8cm]{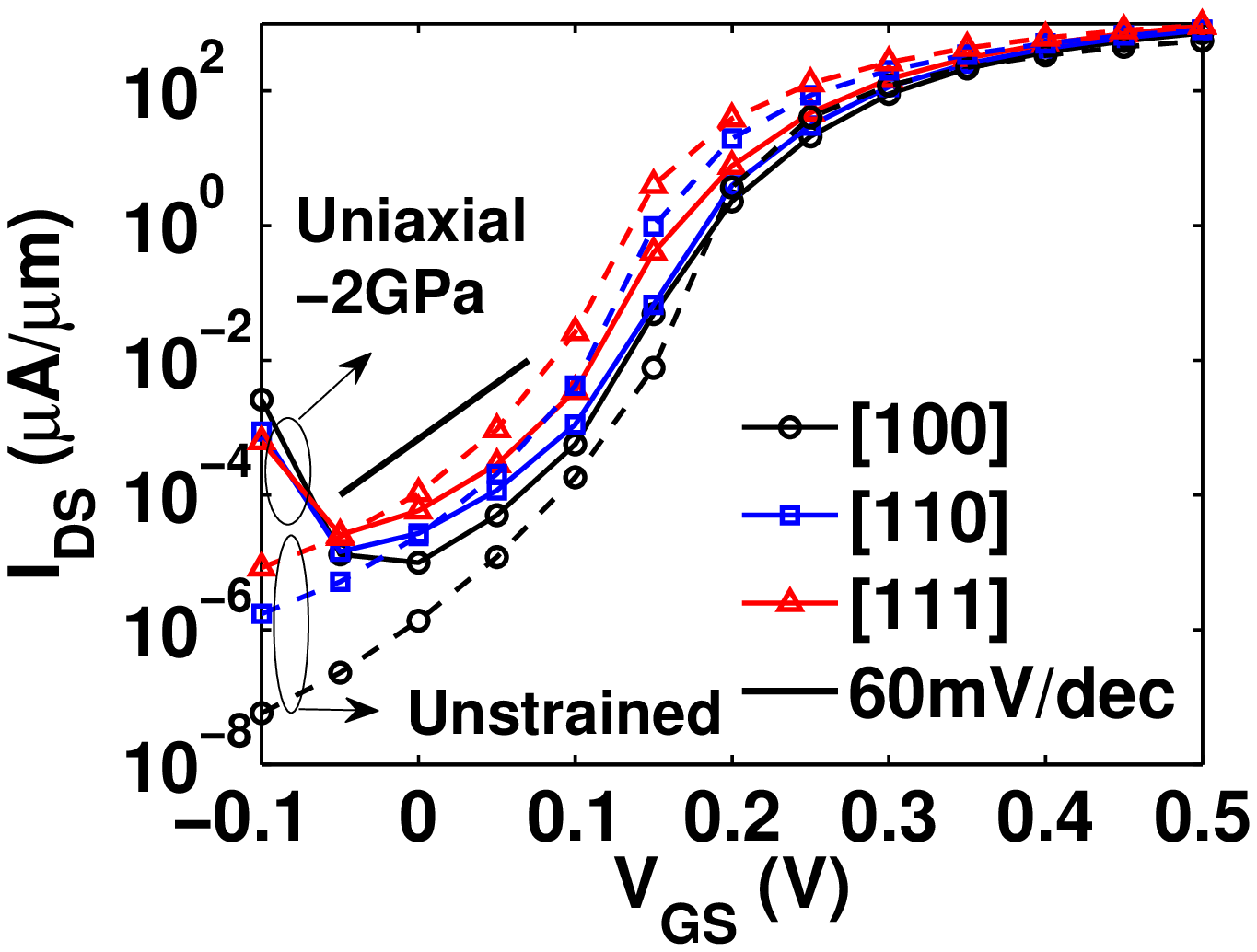}}
\subfigure[]{\includegraphics[width=5.8cm]{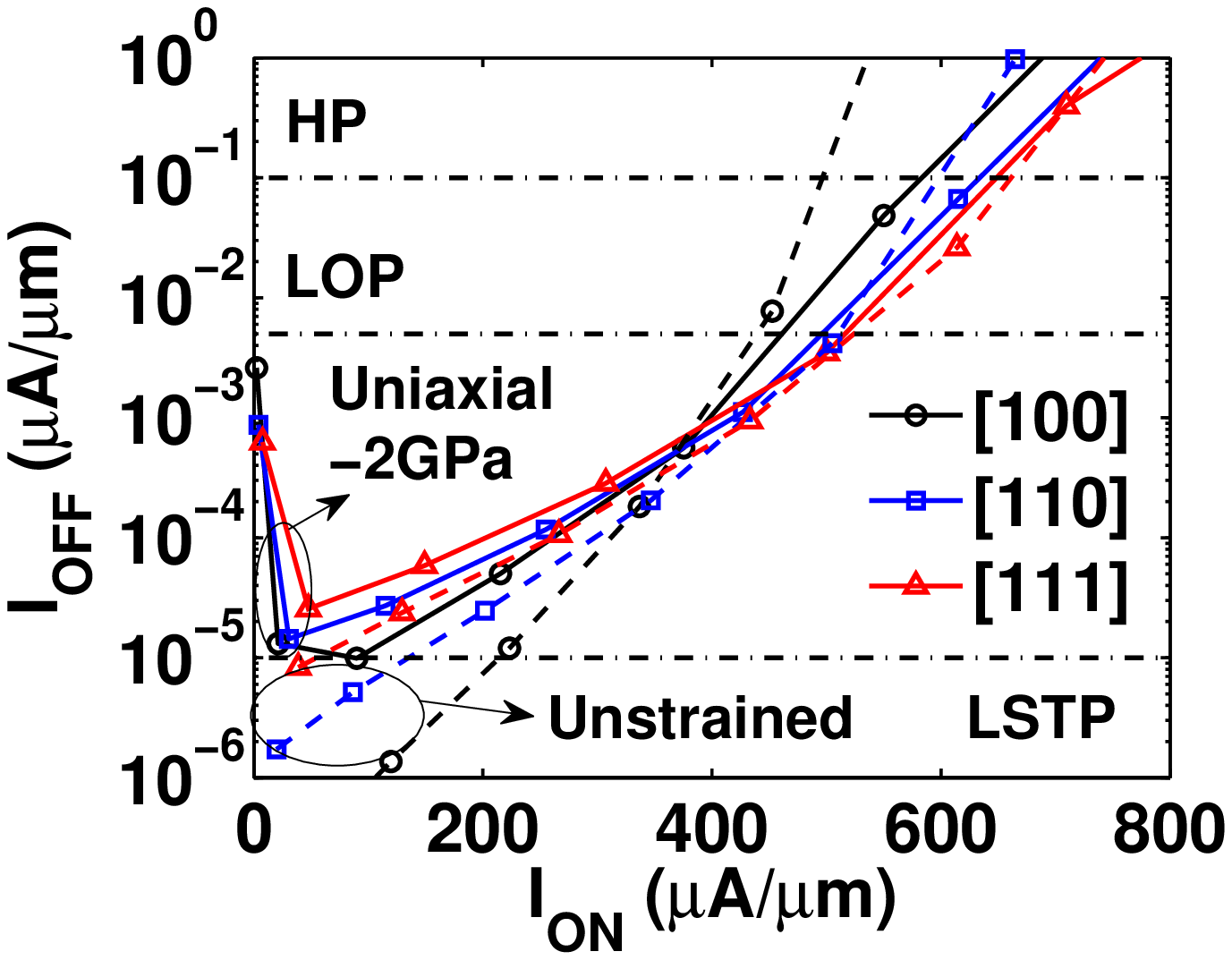}}
\caption{(a) $I_{DS}$-$V_{GS}$ curves (at $V_{DS}=0.3V$) and (b) $I_{ON}$-$I_{OFF}$ (at $V_{DD}=0.3V$), of the $5\rm{nm}\times5\rm{nm}$ cross section heterojunction TFETs in the [100], [110], and [111] orientations, with uniaxial compressive stress applied along the transport direction, in comparison with unstrained cases.}
\label{fig_hetero_strained}
\end{figure}

However, as shown in Fig. \ref{fig_hetero_strained} (a) and (b), uniaxial compressive stress only slightly improves $I_{ON}$ of [100] orientation for HP and LOP applications (and [110] orientation for HP application). In the [111] orientation, the stress even degrades $I_{ON}$. Again, uniaxial tensile stress leads to increased effective masses and thus degraded $I_{ON}$ (no shown here). The physics is similar to the strained homojunction TFETs and will not be repeated here.

\section{Conclusions}
To efficiently simulate III-V nanowire based TFETs, a reduced-order $\bm{k}\cdot\bm{p}$ NEGF method is developed. Through comparison with TB method, the $\bm{k}\cdot\bm{p}$ method is shown to be able to describe quite well the band structures of very small nanowires. By introducing a spurious band elimination process, the reduced-order $\bm{k}\cdot\bm{p}$ models can be constructed for reproducing the original band structures in an energy window near the band gap. The reduced models can also accurately capture the $I$-$V$ characteristics of homojunction and heterojunction TFETs within a short simulation time.

InAs TFETs with different cross sections and channel orientations are compared, it is found that [111] direction has the best cross section scaling ability. Various performance boosters are studied. It is found that embedding source pockets can improve the ``on" current due to the enhanced band bending at the source-to-channel junction, but this effect will saturate with increasing pocket length. Uniaxial compressive stress can also be used to boost the ``on" current, which is found to be more effective in the [100] orientation than in the [110] and [111] orientations. Adopting GaSb/InAs heterojunction achieves a much larger ``on" current due to the staggered-gap band alignment. Incorporating source pockets with proper pocket length into the heterojunction TFET is shown to further enhance the ``on" current.

However, there is a large gap between theoretical projections and experiments \cite{lu2014tunnel}. In experiments, the device performances are usually degraded by nonidealities such as phonon/dopant-induced band tails, defect-assisted tunneling, interface roughness and traps \cite{lu2014tunnel,avci2015tunnel}. To model these nonidealities, the $\bm{k}\cdot\bm{p}$ Hamiltonian needs to be modified properly to account for these defects and the transport solver needs to be extended to incorporate various scattering events due to impurity, alloy, phonon, and surface roughness. These will be done in the future.

\section*{Acknowledgment}
The use of nanoHUB.org computational resources operated by the Network for Computational Nanotechnology funded by the US National Science Foundation under Grant Nos. EEC-0228390, EEC-1227110, EEC-0228390, EEC-0634750, OCI-0438246, OCI-0832623 and OCI-0721680 is gratefully acknowledged.

\end{document}